\documentclass[sigconf, nonacm]{acmart}

\AtBeginDocument{%
  }

\usepackage{listings}
\usepackage{tabularx}
\usepackage{makecell}
\usepackage{array}
\usepackage{caption}
\usepackage{booktabs}
\usepackage{needspace}

\newcommand{\apptabletitle}[1]{%
  \par\medskip\noindent
  \begin{minipage}{\columnwidth}
  \captionof*{table}{#1}\par\smallskip
  \begingroup\fontsize{10pt}{14.4pt}\selectfont
}

\def\endapptable{%
  \endgroup
  \end{minipage}
  \par\medskip
}

\usepackage{enumitem}

\begin{document}

\title[Beyond Single-Vulnerability Evaluation: Closing the Engineering Decision Gap]{Beyond Single-Vulnerability Evaluation: Closing the Engineering Decision Gap Between C Retrofits and Native Safety}


\author{Andrew Laramore}
\email{andrew.laramore2@my.utsa.edu}
\orcid{0009-0005-4545-5767}
\affiliation{%
 \institution{The University of Texas at San Antonio}
 \city{San Antonio}
 \state{Texas}
 \country{USA}
}

\author{Joseph Spracklen}
\email{joe.spracklen@my.utsa.edu}
\affiliation{%
 \institution{The University of Texas at San Antonio}
 \city{San Antonio}
 \state{Texas}
 \country{USA}
}

\author{Murtuza Jadliwala}
\email{Murtuza.Jadliwala@utsa.edu}
\orcid{0000-0001-9316-1943}
\affiliation{%
 \institution{The University of Texas at San Antonio}
 \city{San Antonio}
 \state{Texas}
 \country{USA}
}

\renewcommand{\shortauthors}{Laramore, Spracklen, and Jadliwala}

\begin{abstract}



While decades of research have produced numerous retrofitted memory-safety protections for C, these mechanisms are almost exclusively evaluated in isolation, targeting specific vulnerability classes. This siloed evaluation paradigm leaves practitioners without a clear understanding of the cumulative performance costs, interoperability conflicts, and protection gaps that arise when layering defenses to achieve comprehensive safety. This paper presents a new evaluation paradigm that benchmarks natively memory-safe languages like Rust and Go against compounded C retrofits. Using standardized cross-language tasks, we evaluate the performance and protection tradeoffs of state-of-the-art mechanisms when deployed in combination. Our results demonstrate that layered C defenses incur compounding and workload-dependent performance penalties, can suffer from fundamental architectural incompatibilities, and fall short of the protection scope provided by native memory-safe languages. These findings expose a critical engineering decision gap where the true cost of backporting safety to C remains hidden from practitioners. We argue for a fundamental shift in memory-safety research: moving away from isolated evaluation toward holistic, comparative frameworks that inform the high-stakes choice between retrofitting legacy codebases and migrating to modern, safe languages.

\end{abstract}



\maketitle

\section{Introduction}
\label{sec:intro}

C and C++ remain critically important in modern software infrastructure, and both languages consistently rank among the top ten most popular on GitHub \cite{staff_octoverse_2024}. These languages underpin foundational systems such as the Linux and Windows NT kernels, Google Chrome, ffmpeg and countless device drivers that are actively maintained and developed today. Although these technologies drive contemporary computing, they remain vulnerable to persistent memory safety flaws inherent to C and C++. 
The 2024 Common Weakness Enumeration (CWE) Top 25 explicitly highlights this enduring challenge, listing "Out-of-bounds read" (CWE-125) as the sixth most critical vulnerability and "Use-after-free" (CWE-416) as the eighth \cite{CWE}.

This prevalence underscores a fundamental tension: despite decades of engineering effort, memory-unsafe languages continue to dominate performance-critical domains while contributing significantly to security breaches. Traditional mitigation techniques, such as Address Space Layout Randomization (ASLR) \cite{paxASLR}, Control Flow Integrity (CFI) \cite{clang_cfi}, and compiler-based checks such as \texttt{-fstack-protector} \cite{gcc_instrumentation_options} address specific attack vectors, but fail to provide holistic memory safety. Crucially, these protections operate in isolation, creating a fragmented defense landscape where adversaries can exploit gaps between disjointed mechanisms.

Concurrently, memory-safe languages such as Rust and Go have demonstrated that eliminating entire vulnerability classes (e.g., dangling pointers and buffer overflows) is achievable without sacrificing performance. However, wholesale migration from C/C++ ecosystems is often impractical due to legacy codebases, toolchain dependencies, and domain-specific performance constraints. This raises the question: \emph{Can synergistic integration of state-of-the-art memory protections for C achieve memory safety guarantees comparable to native memory-safe languages, while preserving compatibility and performance?}

We investigate whether a unified framework that combines these techniques can bridge the safety gap between C and memory-safe alternatives such as Rust and Go. Evaluating C/C++ memory safety retrofits against baseline C alone creates a false sense of progress. Without benchmarking combinatorial defense stacks against native memory-safe languages' inherent safety guarantees and performance profile, practitioners cannot determine whether retrofitting legacy code is more cost-effective than strategic migration. 
Native memory-safe languages, particularly Rust, establish a practical baseline for acceptable security guarantees and performance overhead, often delivering comprehensive memory safety with negligible measurable slowdown\cite{SHANG2025100351}. For retrofit C defenses to be a viable alternative, they must therefore demonstrate comparable protection breadth and predictable performance costs, rather than merely outperforming unprotected C. Absent such a comparison, claims of progress in C/C++ memory safety risk optimizing incomplete defenses whose cumulative cost and coverage remain poorly understood.

Taken together, these observations expose a fundamental security gap: practitioners lack empirical guidance on whether layered C/C++ defenses can serve as a practical alternative to migration to memory-safe languages. Addressing this gap requires a systematic evaluation of protection coverage and performance costs across languages. Given that motivation, this research addresses three critical questions:
\begin{enumerate}
    \item Can a combination of C/C++ memory safety mechanisms achieve Rust-like coverage without prohibitive overhead?
    \item Does the composition of these mechanisms introduce safety regressions relative to the benchmarks established in their respective original papers?

    \item What performance penalties arise from layered memory safety defenses versus native safe languages?
\end{enumerate}
Through a thorough evaluation of runtime performance and stability using the Checked C compiler\cite{checkedCGithub}, HWASAN compiler-based sanitizer \cite{AndroidHWAddressSanitizer}, and the memory allocators MarkUs \cite{9152661} and FFMalloc \cite{263880}, both individually and in combination with established benchmarks, 
we reveal the following key findings: 

\begin{enumerate}
    \item \textbf{Safety Completeness Gap}: Even when combined, state-of-the-art (SOTA) retrofits, specifically Checked C paired with security-focused allocators like MarkUs or FFMalloc, fail to match Rust's formal completeness, covering 5 out of 6 memory safety categories. WHile these 5 categories account for 99.8\% of CVEs in the evaluated CWEs, the remaining uncovered class highlights a fundamental gap between retrofitting and native memory safety.
    \item \textbf{Performance Superiority}: Rust consistently outperforms optimized C retrofits in 7 out of 9 benchmarks. While retrofitting aims to bridge the security gap, we observed significant performance penalties ranging from 1.5x to 4.5x slowdowns compared to Rust, demonstrating that native memory safety is often substantially more efficient than cumulative C protection layers.
    \item \textbf{Interoperability and Maintenance Failures}: We identified critical interoperability failures that undermine the practical viability of C retrofits. For example, combining HWASAN with Checked C triggered immediate runtime crashes due to conflicting memory handling. These failures reduce the breadth of protection and significantly increase maintenance overhead, making such combinations difficult to deploy in production environments.


\end{enumerate}


\section{Background}
\label{sec:background}

The history of memory errors is nearly as long as the history of computing itself. 
A 1972 USAF-funded security analysis identified `Scavenging' (uninitialized memory exploitation) and `Implied Sharing' (memory isolation failures) as active attack vectors, with technical descriptions aligning with modern concerns like use-after-free and cross-process memory leaks. The document noted the performance overhead of countermeasures, revealing the early tension between efficiency and memory integrity \cite{usafComputerSecurityStudy}. The rise of C and C++, with arbitrary pointer arithmetic, exacerbated memory safety errors. The 1988 Morris worm became the first large-scale memory-safety incident, exploiting a buffer overflow 
and resulting in a conviction \cite{dressler_cases_2007}. Initial mitigations were operating system features: the non-executable stack was proposed for Linux in 1997, followed by canaries in 1999 and ASLR in 2001 \cite{10.1007/978-3-642-33338-5_5}. The C and C++ languages also implemented minor changes, such as replacing functions such as \texttt{gets()} and \texttt{strcpy()} with bounds-checked alternatives such as \texttt{fgets()}, \texttt{gets\_s()} or \texttt{strlcpy()}. C++ introduced smart pointers for automatic deallocation to combat use-after-free errors. Although these changes help, they are insufficient for making C/C++ memory safe, prompting research into more robust measures.

\subsection{The Addition of New Systems-Level Languages}
C has long dominated systems programming, underpinning foundational infrastructure including the Linux \cite{kbuild-gnu11} and Windows NT kernels \cite{ntLanguage} as well as widely deployed webservers such as Apache httpd and Nginx\cite{webMarketShare}. This dominance reflects C's unmatched performance and control, but also entrenches its well-known memory safety risks.
Beginning in 2013, several languages emerged explicitly to challenge C/C++ in systems-level domains, including Go, Rust, D, and Nim 
each aiming to provide memory safety guarantees without sacrificing performance \cite{dobbs2013}. Among these, Rust has seen the most significant adoption. Released in 2015, Rust 1.0 introduced a novel ownership and borrowing model that enforces memory safety at compile time while preserving low-level control comparable to C \cite{rust1.0}.
Since it's release, Rust has moved beyond experimental use and into production systems contexts. It has been adopted for operating systems components and device drivers \cite{highleveldrivers}, and has enabled the development of fully memory-safe software such as the Redox operating system \cite{redoxHome}. This growing deployment signals a broader shift in systems engineering: rather than retrofitting safety onto memory-unsafe languages, developers are increasingly willing to adopt languages that enforce memory safety by construction.

\subsection{Rust Memory Safety Mechanisms}
Rust enforces memory safety through a statically checked ownership and borrowing model that prevents spatial and temporal memory erros at compile time. 
The borrowing method provides temporary access via immutable (\&T) or exclusive mutable (\&mut T) references. The borrow checker validates reference lifetimes, implementing an affine type system that prevents use-after-free, double-free, and data races without runtime overhead. Safety guarantees exclude logical errors and integer overflows in release builds. Unsafe blocks enable raw pointer operations, FFI, and inline assembly through programmer-verified invariants, necessary for systems programming while maintaining localized safety boundaries. Rust's ownership-based memory safety design establishes it as a practical baseline for memory safety, demonstrating that guaranteed protection can be achieved with minimal performance cost. Accordingly, Rust provides a natural point of comparison for evaluating whether layered C/C++ retrofits can deliver similar safety and performance.

\subsection{Go Memory Safety Mechanisms}
Go ensures memory safety via automatic garbage collection \cite{go_gc_guide}, strong static typing, runtime checks, and controlled pointers \cite{go_spec}. Its concurrent mark-sweep garbage collector tracks object references to reclaim unused memory, preventing leaks and use-after-free errors. Compile-time type enforcement eliminates memory corruption bugs by catching type mismatches early. Runtime checks trigger panics for unsafe operations like out-of-bounds array access or null pointer dereferencing. Although the unsafe package permits low-level pointer manipulation, it explicitly warns against bypassing safety guarantees. These mechanisms create a secure yet efficient memory model suitable for systems programming, reflecting Go's philosophy of balancing simplicity with performance. Together, these mechanisms position Go as an alternative memory-safe baseline, enabling direct comparison between garbage-collected safety guarantees and the performance and composability of retrofitted C/C++ defenses evaluated in this work.

\subsection{Programming Language Performance}
Cross-language performance comparison is notoriously difficult, and most existing evaluations have been informal, ad-hoc, and largely disconnected from security considerations. The Computer Programming Language Benchmarks Game \cite{benchmarkGames} provides one of the most widely referenced comparisons by measuring equivalent tasks in languages, consistently showing that C, C++, and Rust have closely matched performance profiles. Independent practitioner studies, such as Niklas Heer’s /dev/night analyses, reach similar conclusions, reporting only marginal performance differences between Rust and C across diverse workloads \cite{nHeerGit}. Although these results suggest that memory safety need not come at a substantial performance cost, they stop short of evaluating how retrofitted safety mechanisms in C compare to native safety guarantees in modern languages. 
\section{Problem Definition}
\label{sec:problem}
To compare state-of-the-art memory safety in C with that of natively memory-safe languages, we first characterize the classes of memory safety vulnerabilities and the extent to which each language addresses them.
Typically, C++ memory safety protections need to be explicitly enabled using a blacklist approach towards memory safety, but this requires developers to read each available protection mechanism and explicitly enable that mechanism one by one ("blacklisting" that memory safety bug). Forgetting or choosing not to explicitly enable protections available through standard tooling often leaves vulnerabilities that would otherwise have been avoided. Native memory safe languages operate with a whitelisting approach to memory safety bugs where the compiler assumes the program must be memory safe and requires the developer to explicitly \emph{disable} protections.

\subsection{\label{sec:adding_c_safety}Adding Safety to C}
Microsoft Research introduced Checked C in 2015, which enhances the C programming language with static and dynamic verification mechanisms to identify and prevent frequent coding mistakes, including buffer overflows and memory access beyond allocated boundaries \cite{checkedCGithub}. Checked C is meant to be a set of checks at both the compile time and the runtime that enable spatial safety for the C codebases \cite{tarditi2018checked} to protect against errors such as buffer overflows. It works on both regular C codebases and those that compile into LLVM \cite{llvmCheckC}, and is designed to be added to the C code incrementally over time to increase adoption. Despite this work, Checked C does not protect against all safety concerns. Notably, Microsoft states that Checked C is not designed to handle use-after-free errors \cite{checkedCGithub}. Separate research efforts have attempted to address memory safety issues through specialized allocators with distinct protection mechanisms. Fast-Forward Allocation \cite{263880} prevents use-after-free exploits by implementing one-time memory allocation, ensuring that each address is permanently retired after deallocation to eliminate memory reuse vulnerabilities. Freeguard \cite{Freeguard:silvestro} offers a high-performance secure heap allocator that mitigates multiple heap-related attacks through randomized memory layout and guard pages. MarkUs \cite{9152661} counters use-after-free errors by quarantining freed memory blocks and delaying reallocation until it is confirmed that no dangling pointers reference the reclaimed space. In all cases, these features add overhead to the runtime performance of the C programs to which they are applied. The most performant, Fast-Forward Allocation, added a performance overhead (runtime) of 2.3\% and a memory overhead of 61\% for SPEC CPU2006 benchmarks, but added negligible overhead to the performance of ChakraCore and Nginx during testing of real-world applications.

\subsection{Protection Breadth}
Attempts to make C memory safe tend to target specific known memory safety problems. While this targeted approach is required to maintain a reasonable scope of work and get a detailed analysis on the level of protection, it also means no single solution provides complete memory safety. 
Each solution only solves one problem, or at most one category, of memory safety. For security-critical applications, this makes C and C++ inferior to native memory-safe languages. Prior works have evaluated individual mechanisms in isolation and proven they are effective at mitigating specific vulnerabilities. However, such isolated evaluations do not reveal how protections interact when combined, not whether their cumulative coverage and performance costs justify retrofitting compared to migration. Combinatorial evaluation therefor becomes essential to address the engineering decision gap: \emph{Is a layered C/C++ defense stack more cost-effective than adopting Rust or Go?}

\subsection{Research Questions}
In line with the above, our work attempts to address the following research questions:

\begin{itemize}
    \item \textbf{RQ1: Can combinations of existing C/C++ memory safety mechanisms achieve coverage comparable to Rust’s protection scope?} Current state-of-the-art C/C++ protections (Section \ref{sec:adding_c_safety}) typically target isolated memory safety flaws (e.g., Checked C for spatial errors and MarkUs for temporal errors), but remain fragmented in scope. Crucially, ``comprehensive coverage'' here is operationally defined as matching native memory-safe languages' systematic prevention of spatial and temporal memory safety errors. Without confirming whether combinatorial approaches can collectively cover these categories before investing in performance/efficacy testing, we risk optimizing solutions for inherently incomplete protection models.
    \item \textbf{RQ2: Can individual safety methods be combined to give more robust protection to C programs?} Due to state-of-the-art protections being spread across many different solutions, getting the same protection provided by native memory safe languages will require combining these solutions to address the various weaknesses that may be present. The protections provided by these solutions should complement each other rather than interfere. To confirm whether a combination of these protections is plausible, it needs to be tested that their existing functionality is not adversely impacted when working in combination with other protections.
    \item \textbf{RQ3: What performance penalties arise from deploying individual memory safety mechanisms, layered combinations of mechanisms, and different protection stacks, relative to natively memory-safe languages?} Individual protections incur performance overheads, but the cost of combining multiple mechanisms cannot be inferred from their isolated effects and must be evaluated directly. These combined overheads are then compared against native memory-safe languages to determine whether retrofitted defenses impose similar or fundamentally different performance costs.
\end{itemize}
\section{Methodology}
\label{sec:method}

\subsection{Threat Model}
This work formalizes a threat model for memory safety protections in C programs, focusing on defenses against spatial and temporal memory violations under the constraint that adversaries possess only the ability to craft malicious input data while lacking privileges to modify executable binaries, runtime environments, or underlying hardware. We consider an adversary whose objective is to achieve arbitrary code execution or information disclosure via memory corruption vulnerabilities (e.g., buffer overflows, use-after-free, or type confusion), operating within the standard attack surface of unprivileged user-space applications. Crucially, the threat model assumes that the adversary cannot subvert the compilation toolchain, tamper with runtime libraries, or exploit side channels beyond the memory safety boundaries enforced by the protection mechanisms. This scope explicitly excludes threats involving physical access, kernel exploits, or compiler backdoors, thereby isolating the evaluation to software-only memory corruption vectors originating from untrusted input.

\subsection{Protection Mechanisms}

\begin{table*}[t]
  \caption{Comparison of memory safety mechanisms across vulnerability types. 
           $\oplus$~indicates complete protection, $\ominus$~partial protection, 
           $\bigcirc$~no protection. CVE counts from NVD API~\cite{nvd_api_2025}.}
  \label{fig:claim-protections-table}
  \small
  \centering
  \begin{tabular}{@{}lrcccccc@{}}
    \toprule
    Vulnerability (CWE) & CVE Count & Rust & Go & Checked C & HWASAN & MarkUs & FFMalloc \\
    \midrule
    Buffer Overflow (120) & 2,795 & $\oplus$ & $\oplus$ & $\oplus$ & $\oplus$ & $\bigcirc$ & $\bigcirc$ \\
    Buffer Overread (126) & 74 & $\oplus$ & $\oplus$ & $\oplus$ & $\oplus$ & $\bigcirc$ & $\bigcirc$ \\
    Use-after-free (416) & 5,121 & $\oplus$ & $\oplus$ & $\bigcirc$ & $\oplus$ & $\oplus$ & $\oplus$ \\
    Double Free (415) & 574 & $\oplus$ & $\oplus$ & $\bigcirc$ & $\oplus$ & $\oplus$ & $\oplus$ \\
    Uninitialized Var (457) & 19 & $\oplus$ & $\oplus$ & $\bigcirc$ & $\oplus$ & $\bigcirc$ & $\ominus$ \\
    Null Pointer Deref (476) & 3,723 & $\oplus$ & $\oplus$ & $\oplus$ & $\oplus$ & $\bigcirc$ & $\bigcirc$ \\
    \bottomrule
  \end{tabular}
  
\end{table*}

\textbf{Checked C} \cite{tarditi2018checked} is a Microsoft Research extension to the C language designed to enforce spatial memory safety while maintaining backward compatibility with existing C code. Its core innovation lies in introducing three distinct checked pointer types with explicit bounds declarations, enabling both static verification and optimized runtime checks:
\begin{itemize}[leftmargin=*]
    \item \texttt{\_Ptr<T>}: This type is used for single-object pointers where dereferencing occurs without arithmetic. The compiler inserts null checks before dereference. Pointer arithmetic is not performed on these pointers, as they are intended solely for dereference operations (e.g., output parameters such as \texttt{\_Ptr<int>} out in functions).
    \item \texttt{\_Array\_ptr<T>}: This type supports pointer arithmetic for arrays. Programmers declare bounds expressions (e.g., \texttt{count(n)} and \texttt{byte\_count(s*n)}) specifying valid memory ranges. Bounds expressions consist of non-modifying C expressions and can involve variables, parameters, and struct field members. The compiler either statically proves accesses are safe or inserts runtime bounds checks before dereference. Bounds are stored separately from the pointer itself (avoiding ``fat pointers'').
    \item \texttt{\_Nt\_array\_ptr<T>}: This type extends \texttt{\_Array\_ptr<T>} for null terminated arrays (e.g., strings). Types without explicit bounds default to bounds of \texttt{count(0)}. It enforces zero-termination invariants by restricting writes beyond the current bounds unless the terminating null is verified.
\end{itemize}

\noindent Checked C enables incremental adoption through two mechanisms:

\begin{itemize}[leftmargin=*]
    \item \textbf{Checked regions:} Annotated via \texttt{\#pragma}, function prototypes, or blocks such as \texttt{\_Checked \{ ... \}}, these regions enforce strict spatial safety. Within the checked regions: (1) explicit casts from unchecked to checked pointers are disallowed, (2) reads/writes to unchecked pointers are restricted to ensure spatial safety, and (3) all pointer operations undergo mandatory null/bounds checks.
    \item \textbf{Bounds-safe interfaces:} Allow unchecked legacy code (e.g., standard libraries) and as-yet unconverted code to interoperate safely with checked regions. For instance, \texttt{fwrite} can be annotated as \texttt{void *p  byte\_count(s*n)}, enabling the compiler to treat \texttt{p} as \texttt{\_Array\_ptr<void>} in checked contexts while preserving C compatibility elsewhere.
\end{itemize}

\noindent Checks are inserted only when static verification fails. All valid C remains valid Checked C, as unchecked pointers (\texttt{T*}) coexist with checked types, enabling gradual conversion—developers can start with isolated checked regions and expand them incrementally.\\

\noindent \textbf{Hardware-assisted AddressSanitizer (HWASAN)} \cite{Hardware-assisted} uses ARM\allowbreak v8's Top-Byte Ignore (TBI) to detect memory errors (buffer overflows, use-after-free, stack use-after-return) with 10-35\% RAM overhead \cite{AndroidHWAddressSanitizer}, significantly lower than ASAN. It embeds 8-bit tags in the pointer top bytes and shadow memory, validating matches during memory access. 
Collisions cause a 0.39\% per-access error miss probability, but eliminate ASAN's redzones/quarantine. HWASAN reliably detects stack use-after-return and supports ARM64 (full TBI) and x86\_64 (limited page table aliasing).
\\
\textbf{MarkUs} \cite{9152661} is a prototype memory allocator that prevents use-after-free (UAF) vulnerabilities in C through a two-phase memory reclamation process decoupling programmer-initiated freeing from actual memory reuse. When \texttt{free()} is called, MarkUs places the freed object onto a quarantine list rather than immediately returning it to free lists. Verification occurs through periodic ``marking'' inspired by garbage collection but optimized for C/C++ constraints. MarkUs initiates conservative graph traversal from root pointers, treating any word within heap bounds as a potential pointer, and recursively identifying all live objects accessible from program roots, flagging quarantined memory still reachable via dangling pointers. This marking tolerates false positives, prioritizing safety over precision. After marking, MarkUs walks the quarantine list: unmarked objects are confirmed unreachable and migrated to size-categorized free lists for future allocation, while marked objects remain under quarantine for re-evaluation. This process ensures that UAF exploits become unexploitable by guaranteeing freed memory cannot be reallocated while any live pointer could reference it, with reallocation occurring strictly after empirical validation of pointer absence via marking, and allowing early physical memory reclamation for large objects while retaining metadata under quarantine.
\\
\textbf{FFMalloc} \cite{263880} is an experimental memory allocator that eliminates UAF vulnerabilities in C programs through one-time allocation, ensuring that each virtual memory address is assigned to exactly one allocation request during execution and never reused, though freed memory regions are eventually returned to the operating system when sufficient contiguous pages accumulate. This prevents attackers from reclaiming freed memory to manipulate the contents, removing exploitation of dangling pointers through content control. To maintain performance, FFMalloc combines two specialized allocators: a binning allocator for requests up to 2048 bytes prevents small persistent allocations from blocking page release, while a continuous allocator for larger requests minimizes alignment-related waste. Memory is organized into per-core 4MB pools to reduce contention, with freed pages batch-released to the kernel via \texttt{munmap} only after contiguous blocks of at least eight pages accumulate. Address tracking uses a three-level tree structure mapping addresses to pool metadata, with status information stored differently per allocator type: the continuous allocator encodes states in the least significant bits of 8-byte aligned addresses, while the binning allocator uses per-page bitmaps. This metadata system enforces the one-time allocation guarantee, while inherently enabling detection of double-free and invalid-free conditions.\\

Table \ref{fig:claim-protections-table} provides a visual summary of the claimed protections for the various protection mechanisms to illustrate the coverage they provide and the intent when combining these mechanisms.


\subsection{Protection Validation}
We validation the functional correctness of each protection mechanism with respect to its claimed security guarantees, following the evaluation methodology described in the original works. For Checked C, we evaluate the reference programs provided by the authors \cite{checkedc-samples}, which demonstrate protection against spatial memory errors such as buffer overflows, out-of-bounds accesses, and null pointer dereferences when code is properly annotated. For FFMalloc, we rely on the exploitation and validation examples included in its original evaluation, which demonstrate prevention of temporal memory errors including use-after-free and double-free vulnerabilities. MarkUs does not provide a standalone validation suite but because its protections closely mirror FFMalloc, and FFMalloc explicitly cites MarkUs, we evaluate MarkUs using the same test cases. To assess whether these mechanisms remain effective when combined, we consider a protection stack valid if each constituent mechanism preserves its expected behavior under composition. This validation criterion aligns with prior work evaluating the layering of security mechanisms, which treats preservation of individual guarantees under combination as the baseline requirement for practical deployment.

\subsection{Performance Evaluation}
To create an initial comparison, we selected experimental features to create combinations: HWASAN, MarkUs, and FFMalloc. These features were tested individually and in various configurations with standard compiler protections. The features can be seen in Table \ref{sec:mod_map_defs}. This allowed us to measure the performance trade-offs required for more robust memory security.
Native compiler protections were included to provide a comparison point to how a program might run in production today, and added to the combinations to show what performance might look like in a production system with added memory safety mechanisms.

\subsection{Platform}
The benchmarks were conducted on an Apple M1 MacBook Air \cite{apple_macbook_air_m1_2020}. Apple Silicon was chosen as the desired platform due to its desktop-class performance and the availability of ARM 8.5 instructions for testing. ARM architecture was selected as the primary platform for three reasons: (1) it represents the only architecture with full HWASAN support, enabling evaluation of hardware-assisted memory safety; (2) ARM's increasing adoption in cloud computing (AWS Graviton, Azure Cobalt) makes it relevant for future production deployments; and (3) it provides a consistent platform for comparing relative performance impacts across protection mechanisms. While x86\_64 remains dominant in current deployments, the relative performance penalties observed here provide valuable insights for any architecture. To run the benchmark programs, an instance of Linux was required to minimize possible compatibility issues. At the time of testing, the only usable Linux distribution available for Apple Silicon was Asahi Linux \cite{asahilinux2025}, specifically version 40 with kernel version 6.12.1. 

\subsection{Software}
Addressing existing memory safety issues in C is crucial, and any proposed mitigations should be evaluated for their impact on overall program performance, if they are to be considered feasible. To effectively benchmark these memory safety mitigations, it is essential to establish a baseline of expected performance across various tasks. The Computer Language Benchmark Games (CLBG) \cite{benchmarkGames}, hosted by the Debian project, served as an initial reference for benchmarking due to their popular methodology in comparing the performance of different programs and different programming languages. The CLBG is conducted as a public competition among different programming languages, featuring a range of predefined tasks. Programmers are challenged to write solutions in any language they choose, aiming for the lowest runtime, with the fastest solution being recognized as the ``winner'' for that task. Notably, C and Rust frequently compete for the title of the fastest language, making them excellent foundations for this analysis. 
We also used \emph{hyperfine} to manage benchmark execution and measurement, in accordance with CLBG recommendations \cite{benchmarksgame_bencher}. 

\subsection{Benchmarking}
We utilize all nine benchmarking tasks featured in CLBG for our study:
\begin{enumerate}
    \item \emph{binarytrees} - Constructs perfect binary trees in their entirety prior to any garbage collection of tree nodes.
    \item \emph{pidigits} - Employs arbitrary precision arithmetic alongside a consistent step-by-step, single-threaded algorithm to generate digits of Pi.
    \item \emph{regex-redux} - Manipulates FASTA format data using identical regex patterns and actions.
    \item \emph{fannkuch-redux} - Implements a custom algorithm as delineated in ``Performing Lisp Analysis of the FANNKUCH Benchmark'' \cite{10.1145/382109.382124}
    \item \emph{n-body} - Models the orbits of Jovian planets utilizing a straightforward symplectic integrator.
    \item \emph{spectral-norm} - Computes the spectral norm of an infinite matrix \( A \) with specified entries.
    \item \emph{mandelbrot} - Plots the Mandelbrot set [-1.5-i,0.5+i] on an N-by-N bitmap
    \item \emph{fasta} - Generates DNA sequences by replicating a given sequence and through weighted random selection from two alphabets.
    \item \emph{k-nucleotide} - Utilizes a built-in or library hash table to aggregate count values, allowing for key lookups and updates to the corresponding counts within the hash table.
\end{enumerate}

Our investigation evaluates all available candidates written in Rust, Go, and C for the tasks described above. During this assessment, certain benchmark candidates were observed to utilize x86\_64 hardware acceleration, leading to their exclusion from the testing, as the evaluation was conducted on an AArch64 platform. To facilitate the execution of tests, a JSON configuration file, accompanied by a Python script, was developed to automate the process. The JSON configuration encompassed each candidate code across the different programming languages, along with their compilation prerequisites as delineated in the original CLBG. This configuration served as the control group for performance comparisons. 
Subsequently, a series of modifications were identified to alter the execution parameters for the C language candidates:

\begin{itemize}
    \item \textbf{Compiler Selection:} Candidates were compiled using gcc, clang, or checked-c, with the control group using gcc. Both gcc and clang are highly regarded within the C development community, with clang offering features that are particularly relevant to this study, such as Hardware AddressSanitizer (HWASAN). Checked C is a fork of clang 12 and required some slight modifications to the programs. The Checked C 3c tool \cite{10.1145/3527322} was used to automatically annotate the existing C code to enable the extra features that Checked C provides. However, not all benchmark candidates worked with the default configuration. Of the 33 total benchmark candidates, 14 (42\%) failed to compile with Checked C annotations and were excluded. All failing candidates utilized OpenMP for multi-threading parallelization. This consistent failure pattern suggests limitations in the 3c annotation tool's handling of OpenMP pragmas or fundamental incompatibilities between Checked C's type system and OpenMP's threading model-a determination beyond the scope of this study. This limitation restricts generalizability to programs compatible with Checked C's current alpha implementation. Those programs that did not compile with annotations were omitted from the performance analysis.
    \item \textbf{Compiler Hardening Flags:} The study incorporated recommended compiler hardening flags, as outlined in the ``Compiler Options Hardening Guide for C and C++'' provided by OpenSSF \cite{Foundation}. The options included either a basic selection or an extended set of production-use flags. These flags enable compiler protections that influence the behavior of the final binary, guided by OpenSSF's recommendations for robust security.
    \item \textbf{AddressSanitizer and Hardware-assisted AddressSanitizer:} The configurations allowed for the enabling of Address Sanitization in the compiler, or alternatively, the activation of ARM's HWASAN for hardware-assisted ASAN, utilizing the top-byte ignore feature. ASAN serves as a memory debugging tool in C compilers, significantly increasing both memory usage and CPU overhead, but facilitating the resolution of obscure memory issues. Although this overhead limits ASAN's practicality to debugging scenarios, its capabilities hold potential merit for enhancing security. HWASAN, conversely, effectively leverages ARM's top-byte-ignore feature to mitigate ASAN's memory overhead, representing a potential advancement toward employing ASAN-like capabilities in production environments.
    \item \textbf{Memory Allocation Alternatives:} The study considered two proof-of-concept replacements for the standard \texttt{malloc} function, namely MarkUs and FFMalloc, both of which are designed to address use-after-free vulnerabilities.
\end{itemize}

Each modification set was selectively incorporated into the control group configuration. For instance, a configuration could utilize clang, integrate the tl;dr hardening flags, exclude ASAN, and implement FFMalloc. However, it was not permissible to combine both ASAN and HWASAN within the same configuration due to their inherent conflicts as modifications of the same type. Details on the different modifications tested in this study can be found in Appendix \ref{sec:mod_map_defs}.

\section{Results}
\subsection{RQ1: Comparison of claimed memory protections}

\subsubsection{Checked C}
Checked C focuses on protections on bounds checking and null pointer checking with the addition of new pointer and array types. The added types will run checks when used and cause program stoppage when attempting to access memory that is out of bounds of a checked variable or upon dereferencing a null pointer  \cite{tarditi2018checked}. Use of Checked C along with it's new types should address buffer overflows, buffer overreads, and null pointer dereference problems. Checked C does not address temporal memory safety issues; however, the addition of these protections has been suggested by the authors of Checked C \cite{noauthor_use-after-free_nodate}.

\subsubsection{HWASAN}
HWASAN is intended to be a lower resource intensive alternative to ASAN. ASAN and HWASAN are designed to debug memory problems in C programs and thus cause failure for all of the memory safety flaws. HWASAN does, however, come with some caveats \cite{Hardware-assisted}:

\begin{itemize}
    \item HWASAN is limited to use on AArch64 architectures due to its use of ARM's hardware address tagging. Intel CPUs on x86\_64 can use HWASAN in a limited capacity so long as the CPU supports Intel's Linear Address Masking.
    \item Detection of buffer overflows or use-after-free is probabilistic with a 6.25\% chance of missing a bug with four-bit tags and 0.39\% chance with eight-bit tags \cite{Hardware-assisted}
    \item Can cause interference with other programs that may make use of high bits in an address.
    \item As a diagnostic tool, HWASAN is unsuitable for production security because its debugging features may require disabling or bypassing standard security mitigations to facilitate error reporting.
\end{itemize}

HWASAN was included despite its ARM-specific design and current unsuitability for production software to evaluate whether hardware-assisted approaches could provide production-viable alternatives to ASAN. While ASAN explicitly states that it is not suitable for production, as does HWASAN, HWASAN promises lower overhead. Testing revealed fundamental compatibility issues that suggest that hardware-assisted sanitizers require further development before they can be deployed in production. Despite these caveats, testing the combination of memory tagging with other memory safety features using HWASAN will be helpful as a preliminary look at what a solution like ARM MTE may be able to accomplish.

\subsubsection{MarkUs}

Despite protecting from use-after-free vulnerabilities, MarkUs does not completely protect against double-free issues \cite{9152661}. Although the target vulnerability for MarkUs is use-after-free, the MarkUs paper \cite{9152661} mentions that an attacker may be able to use an existing double free vulnerability to edit the quarantine list, but editing the list would not result in any useful privilege escalation, thereby giving some protection for double-free as well \cite{9152661}. We chose to count the weak protection since the double-free would not be usefully exploitable, despite it technically still being possible according to the original study.

\subsubsection{FFMalloc}
FFMalloc is an alternative memory allocator that uses Fast-Forward allocation to fully prevent use-after-free and double-free vulnerabilities. 

FFMalloc provides only partial mitigation for the Uninitialized Variables category; while it successfully handles invalid-free errors, it does not detect the use of uninitialized pointers.

\subsubsection{Summary}

As demonstrated in Table \ref{fig:claim-protections-table}, current state-of-the-art solutions, excluding HWASAN, fail to provide a combination of protections that matches the comprehensive safety guarantees of Rust. Combining Checked C with either FFMalloc or MarkUs covered only five of the six categories of memory safety issues. 
That combination still greatly increases memory safety to give both spatial and temporal memory safety, even if incomplete in comparison. Based on vulnerability CWE classification, 
12287 CVEs are potentially mitigated, covering 99.8\% of the CVEs in the analyzed CWE categories. The benefit of using mechanisms like Checked C, FFMalloc, and MarkUs is the ability to retrofit existing code with these features to avoid the time and effort required for a complete rewrite. However, it should be noted that should one choose to incorporate Checked C, there are changes required to ensure the use of the custom types offered by Checked C. This analysis acknowledges two critical limitations: (1) CVE records reflect reported vulnerabilities rather than absolute exploit prevalence, and (2) coverage claims derive from theoretical protection scope (e.g., Checked C for spatial errors, MarkUs for temporal errors) rather than empirical validation against each CVE. Crucially, we frame these findings as indicative of vulnerability type coverage, not definitive prevention counts, aligning with security community standards that emphasize vulnerability class elimination over raw CVE metrics. Our threshold for meaningful progress ($\geq$5/6 categories) remains justified by the concentration of high-impact risks in the covered categories, while acknowledging that individual CVE mitigation would require case-specific validation. \textbf{The fundamental takeaway of RQ1 is that a protection scope comparable to Rust is indeed achievable through a combination of C retrofits; by mitigating 99.8\% of analyzed vulnerability classes, these combinations prove to be a practically significant alternative that justifies a shift from isolated tool testing toward the holistic evaluation of integrated defenses. As of 2026 there is no established method for this holistic evaluation!}

\subsection{RQ2: Effectiveness of the Combination of Safety Methods}
\subsubsection{Methodology}
Combining these methods of memory safety is only useful if the protections hold in combination. To validate whether protections still hold in combination, we used samples directly from the referenced literature. To test the memory protections of Checked C, sample programs are used from the Checked C GitHub repository and behavior is compared before and after adding HWASAN, MarkUs, and FFMalloc. To test for protection against use-after-free vulnerabilities, the Capture-the-Flag tests used to test FFMalloc are compiled using combinations of Checked C, HWASAN, MarkUs, and FFMalloc. The programs are initially exploitable with no protections enabled, so exploitation should not be possible after introducing temporal memory safety protections. 
The vulnerabilities are tested with the same combinations and results noted in Table 4.3.

\subsubsection{Results}

\begin{table}[ht!]
\caption{A filled circle ($\oplus$) signifies the expected behavior was unchanged, and the program received the expected exit code(s) noted under the exit code column with the additional modification. An empty circle ($\bigcirc$) indicates that the program's behavior changed after adding the mechanism in combination.}
\label{fig:checked-c-table}
\begin{tabular}{|| c | c c c ||} 
 \hline
 Checked C & HWASAN & MarkUs & FFMalloc \\ [0.5ex] 
 \hline\hline
 avoid-warning.c \cite{checkedc-samples} & $\bigcirc$ & $\oplus$ & $\oplus$ \\
 echo-args.c \cite{checkedc-samples} & $\bigcirc$ & $\oplus$ & $\oplus$ \\
 echo-args-buggy.c \cite{checkedc-samples} & $\bigcirc$ & $\oplus$ & $\oplus$ \\
 exit-on-failure.c \cite{checkedc-samples} & $\bigcirc$ & $\oplus$ & $\oplus$ \\
 hello-world.c \cite{checkedc-samples} & $\bigcirc$ & $\oplus$ & $\oplus$ \\
 \hline
\end{tabular}
\end{table}

\begin{table}[ht]
  \caption{Vulnerability test results comparing behavior during use-after-free exploit. A filled circle ($\oplus$) confirms expected behavior. Our tests validated that all combined protections resulted in expected behavior.}
  \label{tab:vuln_tests}
  \small  
  \centering
  \begin{tabular}{@{}lcccc@{}}
    \toprule
    Test Case & Checked C & HWASAN & MarkUs & FFMalloc \\
    \midrule
    ghostparty~\cite{pwnablePwnabletw} & $\oplus$ & $\oplus$ & $\oplus$ & $\oplus$ \\
    uaf~\cite{pwnableKr} & $\oplus$ & $\oplus$ & $\oplus$ & $\oplus$ \\
    Python Issue 24613~\cite{pythonIssue24613} & $\oplus$ & $\oplus$ & $\oplus$ & $\oplus$ \\
    Python Issue 39421~\cite{Issue39421} & $\oplus$ & $\oplus$ & $\oplus$ & $\oplus$ \\
    \bottomrule
  \end{tabular}
\label{fig:allocator-table}
\end{table}

The experimental evaluation reveals critical insights regarding the interoperability of memory safety frameworks with Checked C. When executing the standardized Checked C example suite (Table \ref{fig:checked-c-table}), HWASAN exhibited complete incompatibility with the Checked C compiler toolchain, manifesting as immediate failure upon attempting to allocate memory. This failure occurs because HWASAN's instrumentation pass conflicts with Checked C's extended type system semantics, which alter pointer representation in ways that violate HWASAN's memory tagging assumptions.

Conversely, runtime memory allocators demonstrated seamless integration with Checked C's compile-time protections. Both MarkUs (a metadata-enriched allocator) and FFMalloc (a fence-based allocator) maintained identical behavioral profiles when processing Checked C-annotated binaries compared to their operation with standard C code. This compatibility was empirically validated through two canonical test cases:

\begin{itemize}
    \item The echo-args-buggy.c example, which intentionally triggers an off-by-one error, consistently generated an exception regardless of allocator implementation
    \item The exit-on-failure.c test case, designed to showcase Checked C's bounds checking, produced identical termination behavior under all allocator configurations
\end{itemize}

These results confirm that Checked C's bounds enforcement operates orthogonally to heap memory management layers, as the allocator swap neither suppressed nor altered the spatial safety violations detected by Checked C's type system.

Further validation through Capture-the-Flag security challenges demonstrated robust compatibility across protection layers. When compiled with Checked C annotations (automatically inserted via the 3c conversion tool), all test binaries maintained full exploit resistance under MarkUs and FFMalloc. Notably, the protections remained effective even when Checked C annotations were applied to legacy codebases originally developed without spatial safety considerations.

Architectural constraints necessitated methodological adaptation for Python Issue \#24613 testing. The absence of aarch64-compatible HWASAN tooling prevented direct evaluation on Apple Silicon platforms. Consequently, we migrated testing to x86\_64 infrastructure, where both MarkUs and FFMalloc successfully mitigated exploitation attempts even when Python was compiled with Checked C annotations.

\subsubsection{Summary}
The empirical analysis demonstrates that allocator-based memory safety mechanisms exhibit full interoperability with Checked C, as evidenced by Table \ref{fig:allocator-table}. However, a critical exception to this interoperability principle emerges in the interaction between HWASAN and the Checked C extension. The tests reveal that HWASAN consistently fails during runtime validation when processing Checked C-annotated codebases, incorrectly producing segmentation faults in all cases. This incompatibility stems from fundamental architectural conflicts: Checked C introduces extended pointer types (\texttt{\_Ptr<T>}, \texttt{\_Array\_ptr<T>}) with compiler-enforced spatial and temporal safety guarantees, while HWASAN relies on precise memory tagging aligned with standard C type semantics. Specifically, Checked C's custom type management system modifies pointer representation and metadata handling in ways that violate HWASAN's assumptions about memory layout expectations. This incompatibility represents a significant barrier to comprehensive adoption of memory safety in legacy code modernization efforts using memory sanitization functionality. Three possible directions to resolve this conflict are:

\begin{itemize}
    \item Development of a unified metadata schema that accommodates Checked C's extended type annotations within HWASAN's tagging framework
    \item Creation of a translation layer in LLVM's middle-end to reconcile Checked C's type transformations with HWASAN's instrumentation passes
    \item Formal verification of the combined system using separation logic to prove the absence of semantic conflicts
\end{itemize}

Such work would enable synergistic deployment of compile-time (Checked C) and runtime (HWASAN) protections, closing a critical gap in the memory safety toolchain.

\begin{figure*}[t]
    \centering
    \begin{minipage}[t]{0.48\textwidth}
        \centering
        \includegraphics[width=\columnwidth]{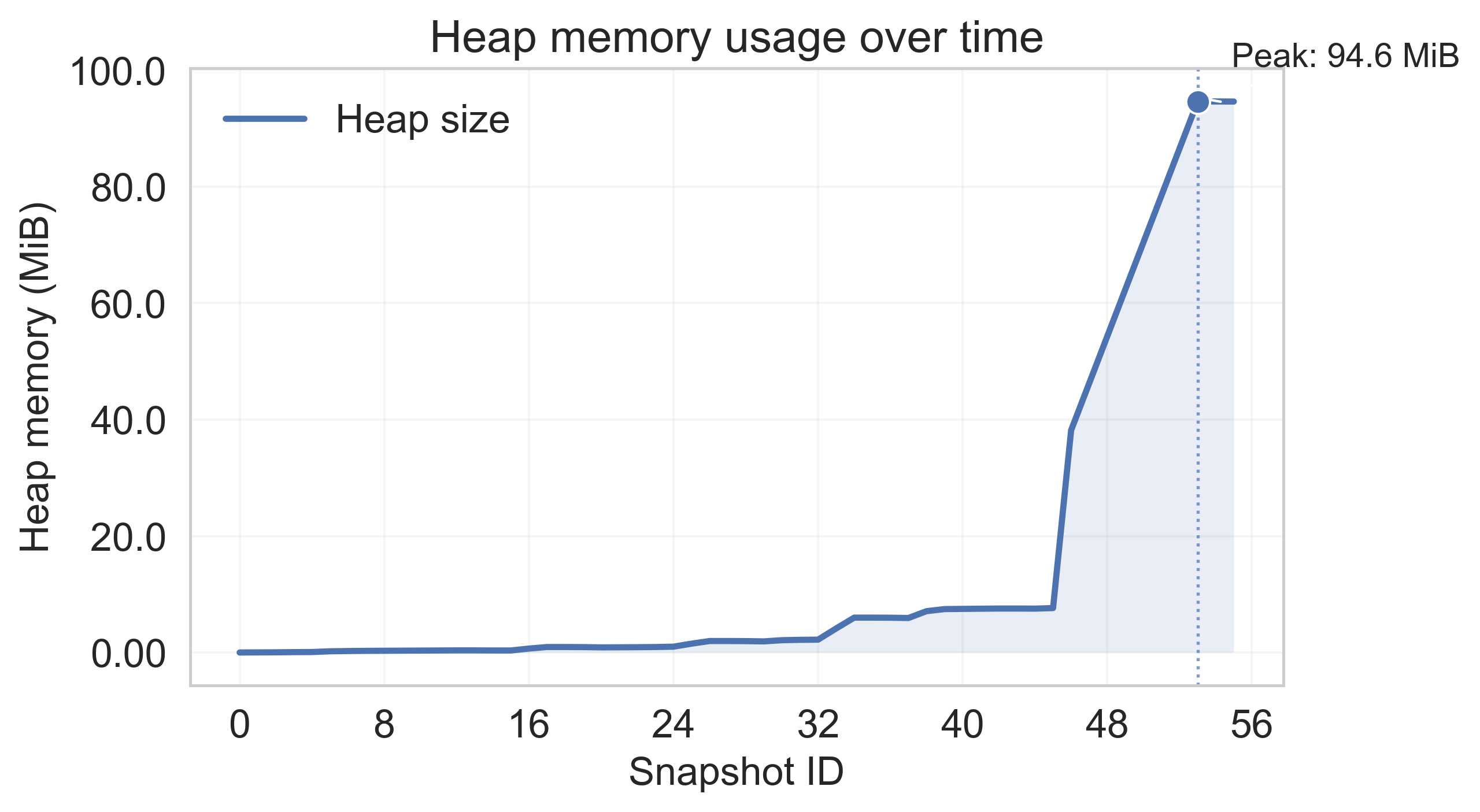}
        \caption{Massif memory usage graph for mandelbrot displayed using Massif Visualizer}
        \label{fig:massif_mandelbrot}
    \end{minipage}\hfill
    \begin{minipage}[t]{0.48\textwidth}
        \centering
        \includegraphics[width=\columnwidth]{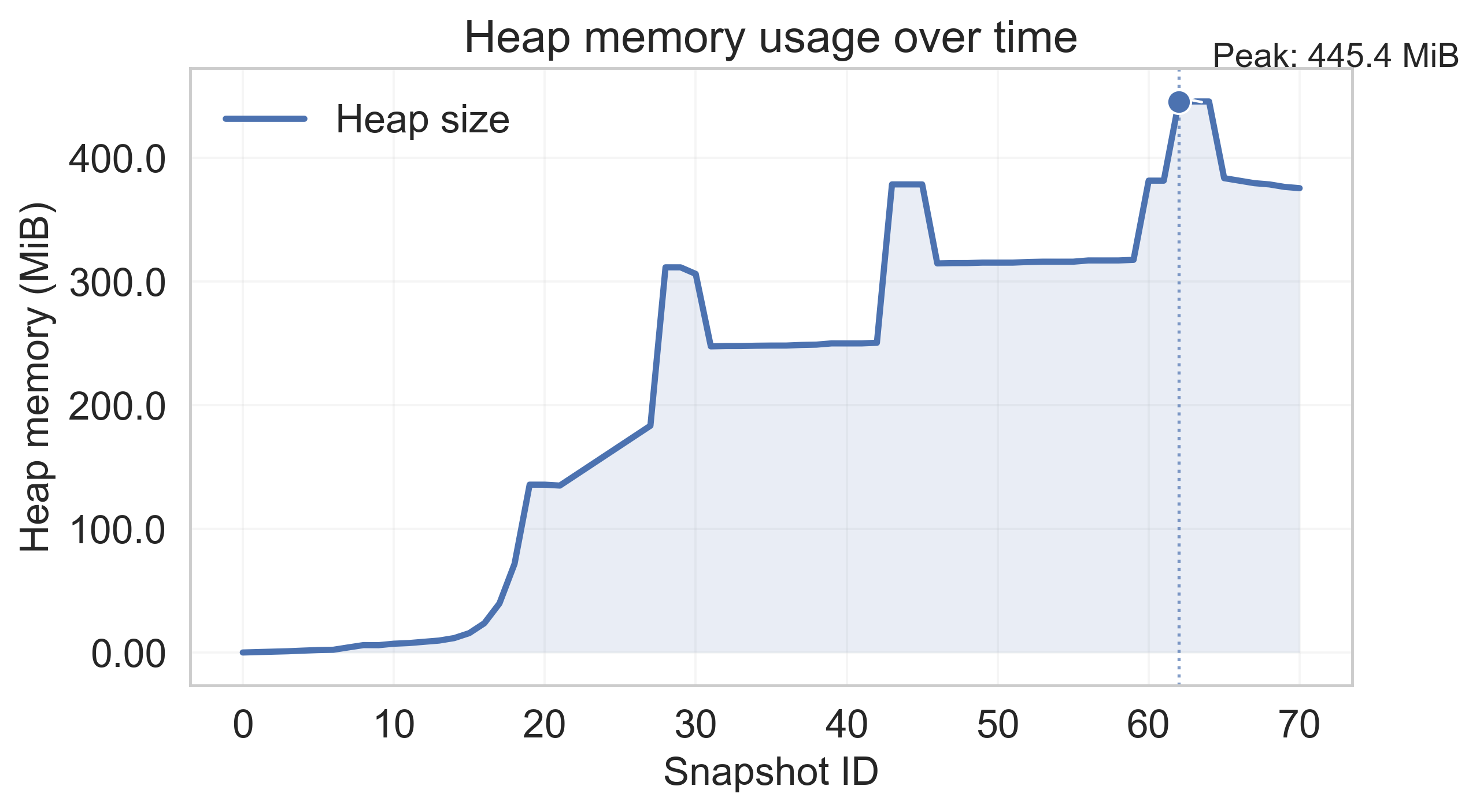}
        \caption{Massif memory usage graph for k-nucleotide displayed using Massif Visualizer}
        \label{fig:massif_knucleotide}
    \end{minipage}
\end{figure*}

\subsection{RQ3: Performance Evaluation}
A total of 2,316 benchmark tests were conducted, including control baselines, with each modification combination tested over 30 independent runs to support statistical comparison. 
Normality of execution-time distributionswas assessed using the Shapiro-Wilk test \cite{10.1093/biomet/52.3-4.591}. At a significance level of 0.05, 740 sample sets did not exhibit evidence of non-normality, while 1,576 deviated significantly from a normal distribution. 
Given this prevalence of non-normality, median values were used as measures of central tendency and non-parametric statistical tests were employed. 
To compare execution times between Rust and C implementations, we applied the Mann-Whitney U test \cite{10.1214/aoms/1177730491}, testing the null hypothesis that the two languages produce execution-time distributions with equal central tendency.
To compare unmodified C baselines against C binaries compiled with memory-safety mechanisms, we used the Wilcoxon signed-rank test \cite{c4091bd3-d888-3152-8886-c284bf66a93a}, testing the null hypothesis that applying a given protection mechanism does not change execution time relative to the baseline.
Across all comparisons, 24,903 test pairs yielded sufficient evidence to reject their respective null hypotheses at $\alpha = 0.05$, while 895 comparisons showed no evidence of a statistically significant performance difference.

\subsubsection{Stability}
The application of various modification sets did not consistently lead to successful executions. Many of the errors encountered were attributed to memory mismanagement, as identified by ASAN and HWASAN. In the case of ``binarytrees.c'', the function \texttt{NewTreeNode()} allocates memory using \texttt{malloc} to create a new \texttt{treeNode}; however, this memory is not deallocated, resulting in a memory leak upon program termination. Similar issues arise in other instances, such as in fasta.gcc-7.gcc at line 146. Discovering memory leaks of this nature is not unexpected, given the context of the programs being assessed. These programs are small, simple, and designed to complete specific tasks for a benchmarking competition. They execute once and can depend on the operating system to reclaim any memory not otherwise deallocated. Although the memory leak does not significantly impact the tasks at hand under the benchmark parameters, it underscores the utility of ASAN in debugging. It suggests that a more performant ASAN implementation could provide greater protection in production environments, considering that other modifications did not detect the leak. The error code charts feature data points from ASAN and HWASAN that completed successfully but ultimately exited due to memory not being deallocated.

Once again, another significant source of errors stemmed from the combination of ASAN/HWASAN with Checked C. When compiled with both, the binaries failed to allocate memory and terminated immediately. The failure occurred at line 54 of sanitizer\_\allowbreak common.cpp in the Checked C compiler, with error code 22 indicating an "Invalid Argument." This behavior was consistently observed across the benchmarks, and given that the Checked C code is considered an alpha implementation, it seems reasonable to speculate that this may be a bug.

Finally, the error figures in Appendix \ref{fig:error_count_fannkuchredux} when split by compiler show that \texttt{clang} failed more often than \texttt{gcc}. While the two compilers are technically different, it was a bit surprising to see \texttt{clang} having more errors than \texttt{gcc}. Further investigation found that all errors have to do with HWASAN. A core dump was triggered on all benchmarks using \texttt{clang} with HWASAN after running \texttt{atoi()} with argv given as an argument. An example is shown in Figure \ref{fig:gdb_debug_output}. 

\begin{figure*}[t]
\noindent
\begin{lstlisting}[basicstyle=\scriptsize\ttfamily, frame=lines]
Program received signal SIGSEGV, Segmentation fault.

0x0000fffff7b75814 in __GI_____strtol_l_internal ... at ../stdlib/strtol_l.c:304
304       while (ISSPACE (*s))

#0  0x0000fffff7b75814 in __GI_____strtol_l_internal ... at ../stdlib/strtol_l.c:304
#1  0x0000fffff7b7575c [PAC] in __strtol ... at ../stdlib/strtol.c:117
#2  0x00000000004670c0 in atoi (__nptr=0x0) at /usr/include/stdlib.h:483
#3  main (argc=<optimized out>, argv=<optimized out>) at fannkuchredux.gcc-5.gcc:30
\end{lstlisting}
    \caption{ GDB output from debugging (ellipses indicate truncated text for space)}
  \label{fig:gdb_debug_output}
\end{figure*}




With significantly more failures, it shows that HWASAN, while more performant, may not be ready for production use (at least when using \texttt{clang}).

\subsubsection{Workload Classification}

Differences in performance are not consistent across the tests. The fundamental difference lies in each program's memory access patterns and computational characteristics. Mandelbrot's execution is predominantly CPU-bound with minimal heap activity, where approximately 92\% of its peak memory usage (87 MB out of 94.6 MB) consists of thread stacks for OpenMP parallelization rather than dynamic heap allocations (Figure \ref{fig:massif_mandelbrot}). This workload profile means that the program spends most of its time performing computations rather than memory operations, making it relatively insensitive to the overhead introduced by different security mechanisms. Fasta performs similarly with relatively little change across various combinations of memory protections.

Conversely, k-nucleotide demonstrates relatively intensive heap activity with frequent allocation/deallocation cycles, where 57\% of its peak memory usage (268 MB out of 467 MB) consists of program data allocated through standard malloc operations, including significant realloc activity accounting for 130 MB of memory usage (Figure \ref{fig:massif_knucleotide}).

This memory-bound workload can be hampered depending on the mechanism of protection. For example, this directly conflicts with FFMalloc's design principle, where "any given virtual memory address is only returned to the calling application once" \cite{263880}, forcing new allocations for operations that would normally reuse memory in standard allocators. These contrasting behaviors highlight a critical insight: \emph{the performance penalty of memory safety protections is not inherent to the protection mechanism itself, but rather emerges from the interaction between the protection strategy and the application's memory usage patterns}. As noted in FFMalloc research \cite{263880}, it could more aggresively release pages, but this tradeoff only becomes performance-critical for applications like k-nucleotide that frequently reuse memory locations. For CPU-bound applications such as mandelbrot with minimal memory reuse patterns, security benefits can be obtained with negligible performance impact.


\subsubsection{Individual Protection Mechanisms}
\begin{figure*}[t]
    \centering
    \begin{minipage}[t]{0.48\textwidth}
        \centering
        \includegraphics[width=\columnwidth]{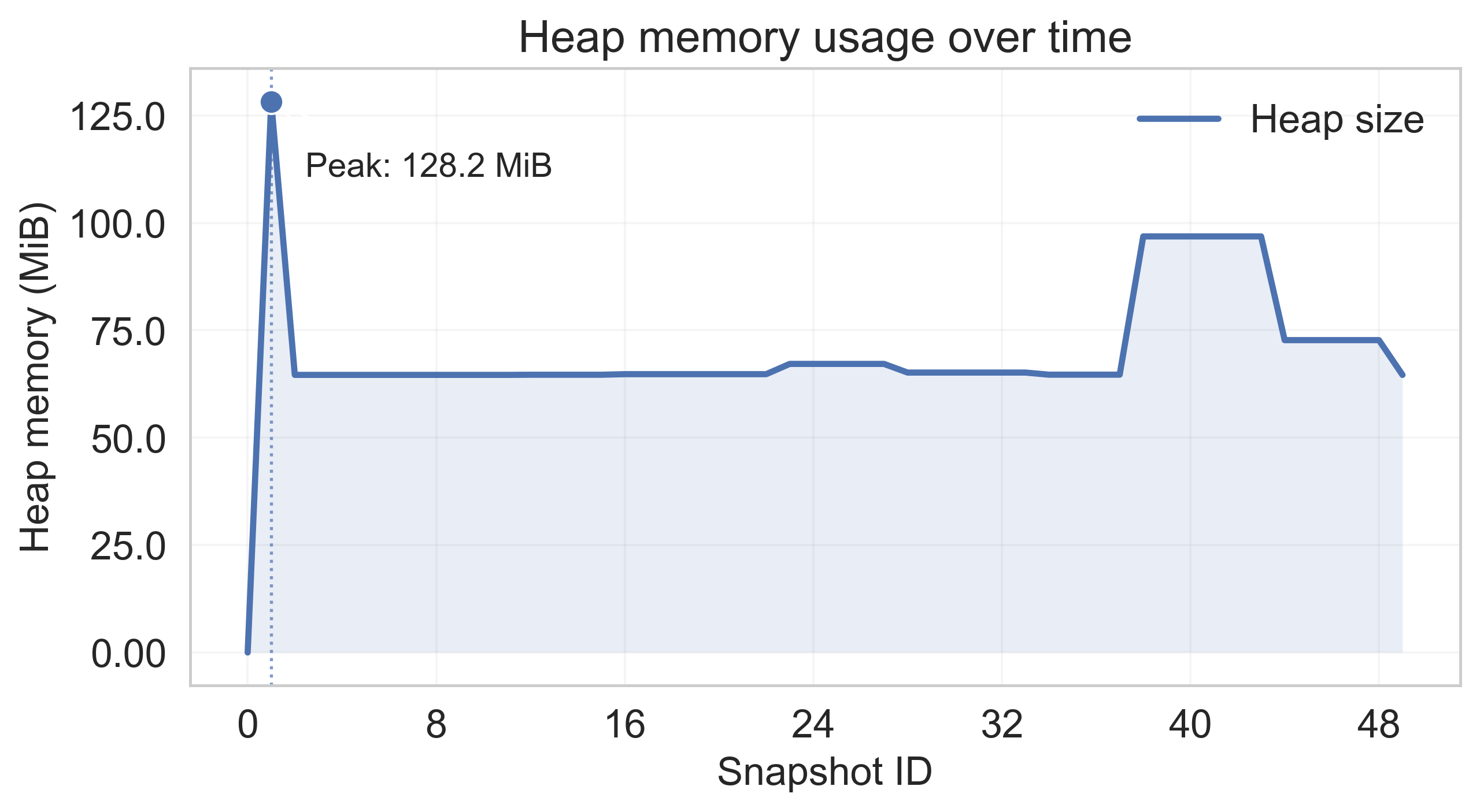}
        \caption{Massif heap memory usage graph for \texttt{binarytrees.gcc-3.gcc} displayed using Massif Visualizer}
        \label{fig:bin3-heap}
    \end{minipage}\hfill
    \begin{minipage}[t]{0.48\textwidth}
        \centering
        \includegraphics[width=\columnwidth]{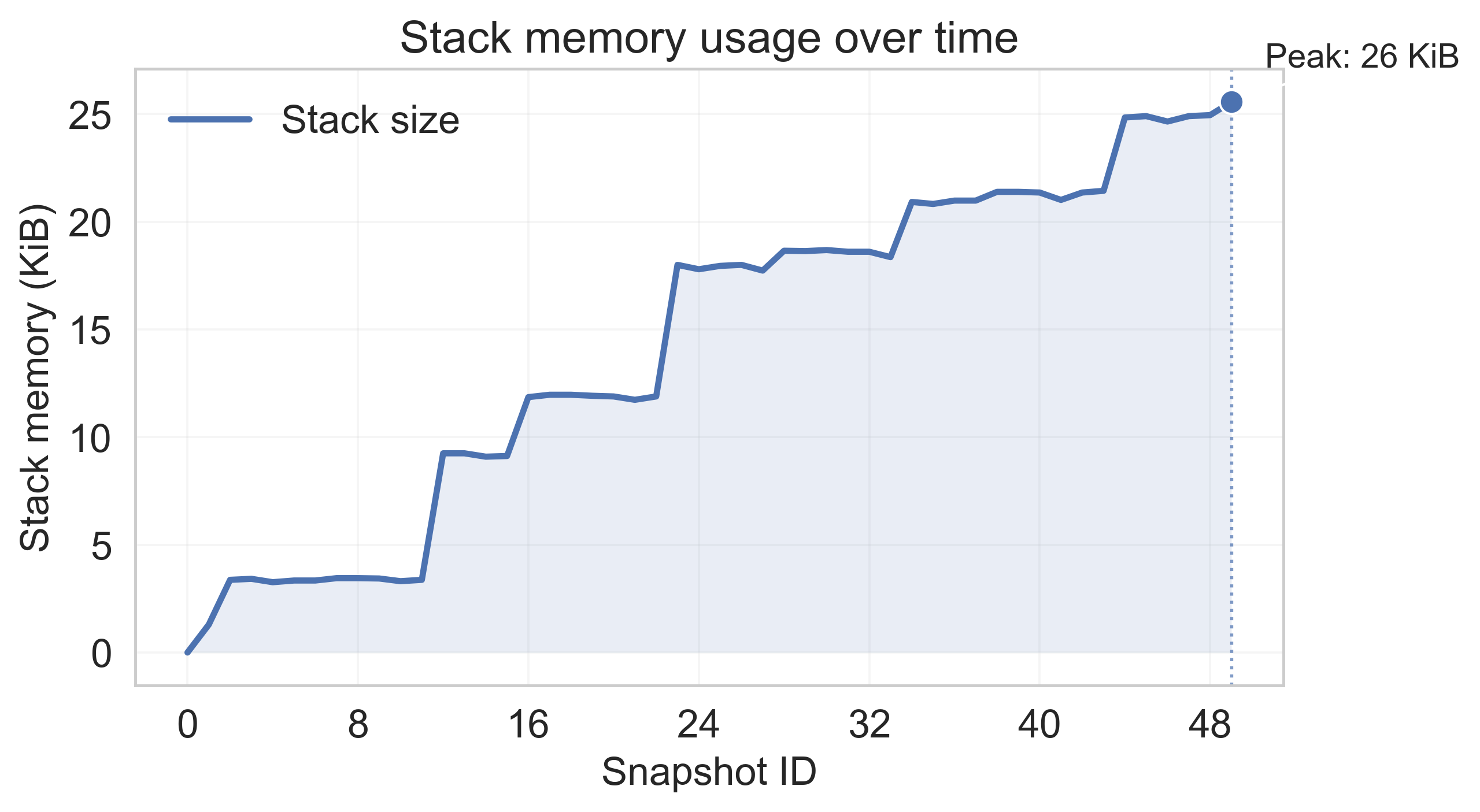}
        \caption{Massif stack memory usage graph for \texttt{binarytrees.gcc-3.gcc} displayed using Massif Visualizer}
        \label{fig:bin3-stack}
    \end{minipage}
\end{figure*}

\begin{figure*}[t]
    \centering
    \begin{minipage}[t]{0.48\textwidth}
        \centering
        \includegraphics[width=\columnwidth]{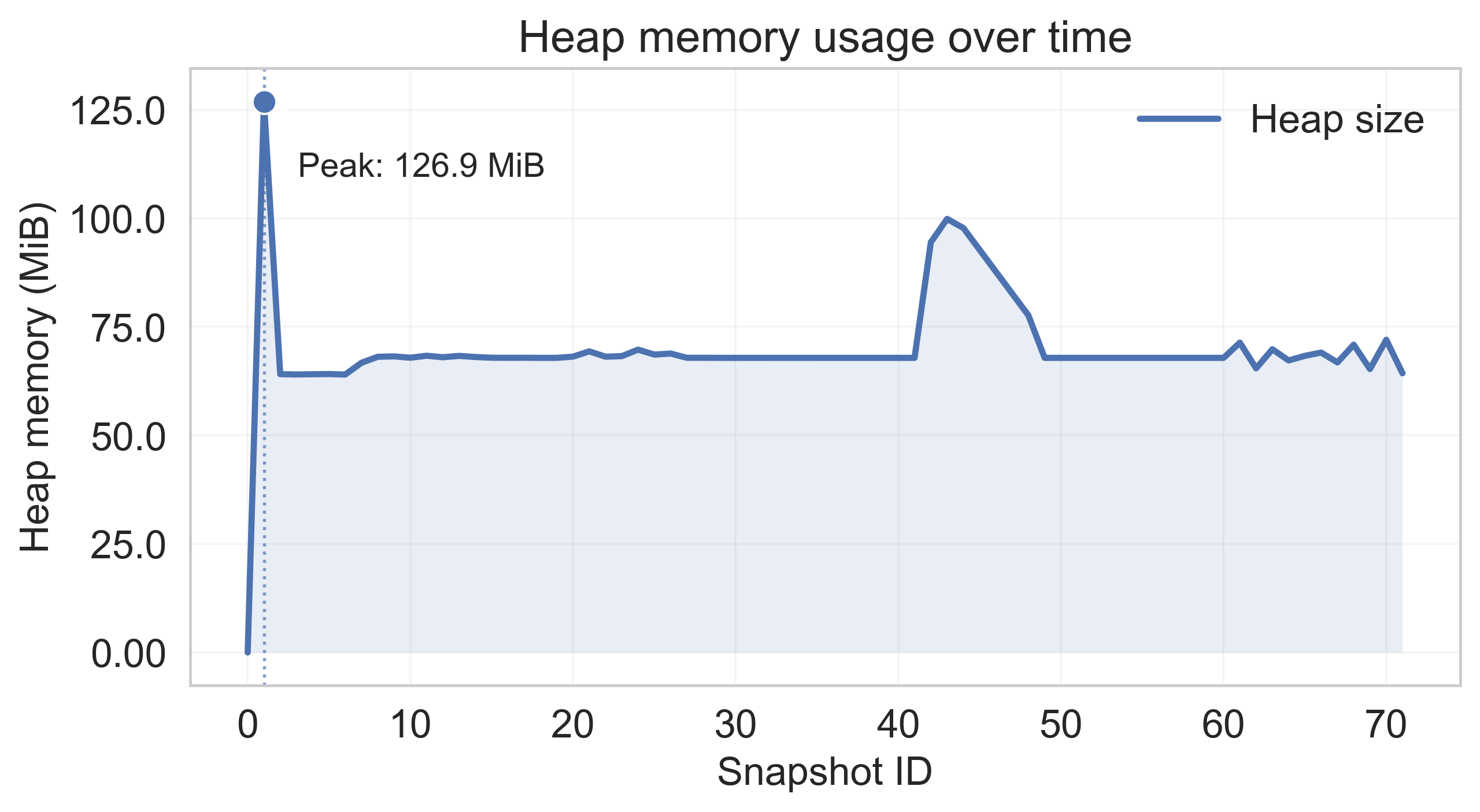}
        \captionof{figure}{Massif heap memory usage graph for \texttt{binarytrees.gcc-5.gcc} displayed using Massif Visualizer}
        \label{fig:bin5-heap}
    \end{minipage}\hfill
    \begin{minipage}[t]{0.48\textwidth}
        \centering
        \includegraphics[width=\columnwidth]{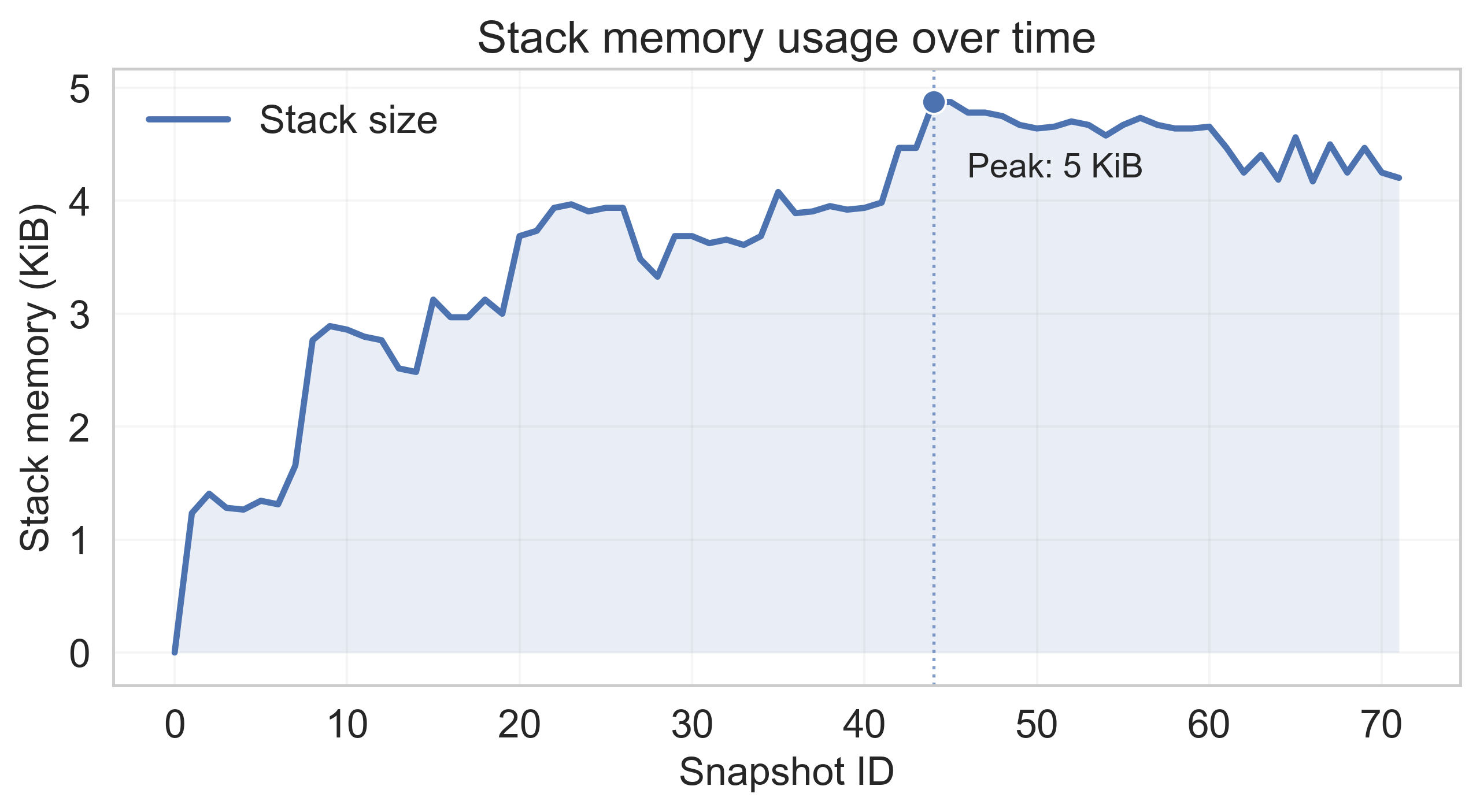}
        \captionof{figure}{Massif stack memory usage graph for \texttt{binarytrees.gcc-5.gcc} displayed using Massif Visualizer}
        \label{fig:bin5-stack}
    \end{minipage}
\end{figure*}

\textbf{(HW)ASAN} 
Empirical analysis reveals that while HWASAN generally offers comparable or better runtime than ASAN in many scenarios, certain workload characteristics can reverse this expectation. When evaluating two implementations of the same task, binarytrees.gcc-3.gcc and binarytrees.gcc-5.gcc under identical conditions on Asahi Linux running on Apple Silicon hardware, it was observed that binarytrees.gcc-3.gcc exhibited a significant runtime penalty with HWASAN compared to ASAN with the HWASAN median time being 42\% slower than the ASAN median time, while binarytrees.gcc-5.gcc the HWASAN median time was 36\% faster than the ASAN median time. Massif profiling with --time-unit=B revealed that binarytrees.gcc-5.gcc had 64 snapshots recording stack changes (Figure \ref{fig:bin5-stack}) and 71 snapshots recording heap changes (Figure \ref{fig:bin5-heap}) while binarytrees.gcc-3.gcc had 42 snapshots recording stack changes (Figure \ref{fig:bin3-stack}) and 10 snapshots recording heap changes (Figure \ref{fig:bin3-heap}).

 This evidence strongly suggests that HWASAN's runtime characteristics are particularly sensitive to programs with a higher ratio of stack operations. The technical explanation lies in the fundamental differences between the stack protection mechanisms of the two sanitizers: while ASAN primarily relies on redzone padding with compiler instrumentation, HWASAN leverages the top-byte-ignore feature of aarch64 to store metadata in the top byte of pointers \cite{Hardware-assisted}. This hardware-assisted approach enables HWASAN to detect additional error classes including stack-use-after-return that ASAN may miss \cite{AddressSanitizer}, but requires additional operations during each stack frame management event-tag generation, storage, and consistency checking. However, this methodology is much more efficient than ASAN's when applied to the heap.

 Consequently, in workloads with intensive stack allocation ratios like binarytrees.gcc-3.gcc, the overhead of these additional operations on the stack results in runtimes that are worse than ASAN. While ASAN also monitors the stack, it utilizes a less computationally demanding approach of redzone poisoning and shadow memory bit-flipping to detect overflows. In contrast, HWASAN implements a more intensive stack protection protocol. While this methodology enables HWASAN to detect complex error classes, it introduces a significant performance penalty in environments characterized by high-frequency stack frame management.
 This finding demonstrates that memory sanitizer selection must consider not only the target architecture and the desired error coverage, but also the specific memory access patterns of the application if the feature is intended to be implemented in a production capacity, as the use of HWASAN does not necessarily translate into a runtime advantage despite lower RAM overhead (10\%-35\%) \cite{Hardware-assisted}.

\textbf{Memory Allocators}
FFMalloc and MarkUs exhibited workload-dependent performance characteristics that diverged from their published benchmarks. Although the original papers reported overheads of 2.3\% (FFMalloc\cite{263880}) and 5\% (MarkUs\cite{9152661}) on SPEC CPU2006, our testing revealed penalties ranging from negligible (<5\%) for CPU-bound workloads to substantial (>70\%) for allocation-intensive programs such as k-nucleotide. This discrepancy likely stems from differences in workload characteristics: SPEC CPU2006 emphasizes computational tasks with predictable memory patterns, while the CLBG benchmarks include allocation-intensive scenarios that stress the allocators' quarantine mechanisms. The k-nucleotide benchmark's extensive use of realloc operations particularly conflicts with FFMalloc's forward-only allocation strategy, as each reallocation requires entirely new address space rather than in-place expansion.

\textbf{Compiler Selection}
Analysis of mean runtimes across successful protection combinations reveals nuanced compiler performance characteristics that challenge conventional expectations about compiler age and optimization. Due to the high failure rate of Checked C configurations (absent entirely from knucleotide and spectralnorm benchmarks), mean values provide more representative comparisons than medians for successful runs. Surprisingly, Checked C (based on \texttt{clang}-12) outperformed modern compilers in several benchmarks, despite its outdated foundation. For compute-intensive workloads like fannkuchredux (21.5s vs 25.1s \texttt{clang}, 32.1s \texttt{gcc}) and pidigits (427ms vs 653ms \texttt{clang}, 798ms \texttt{gcc}), Checked C demonstrated superior optimization. However, Checked C showed significant degradation in others, particularly regexredux (2.56s vs 1.17-1.34s) and mandelbrot when compared to modern \texttt{clang} (14.2s vs 4.7s). The absence of Checked C results for knucleotide and spectralnorm, due to compilation or runtime failures, underscores the fragility of the Checked C toolchain rather than performance concerns. These findings suggest that Checked C's performance penalties are not uniform but highly workload-dependent, with the primary limitation being compatibility rather than raw performance. The selective success of Checked C configurations creates a survivorship bias in the performance data: only simpler, less-protected configurations succeed, potentially explaining the unexpectedly competitive performance in successful runs.

\subsection{RQ3: Memory Protections in Combination}
\textbf{Non-linear Scaling of Protection Mechanisms}
The empirical data from Tables \ref{sec:mean-tables} and \ref{sec:median-tables} demonstrate conclusively that combining memory protection mechanisms produces non-linear performance degradation patterns that cannot be predicted from individual component analysis.

Examining the knucleotide benchmark runtime data reveals striking non-linearity. The baseline configuration shows a median runtime of 1.93s (Tables \ref{sec:median-rust}). When FFMalloc is added individually [0,0,0,0,1,0], the median drops to 1.86s—actually improving performance by 3.7\%. Several combinations of memory allocator and OpenSSF flags stay below the runtime of the control by as much as 5.4\%. However, when FFMalloc combines with ASAN [0,0,1,0,1,0], the median jumps to 3.99s, representing a 69.4\% degradation from baseline. If the effects were linear, we would expect approximately 65.7\% overhead (69.4\% from ASAN alone minus the 3.7\% improvement from FFMalloc), but instead observe the full 69.4\%—demonstrating that FFMalloc's optimization completely disappears when combined with ASAN.
The binarytrees benchmark provides another compelling example of non-additive overhead. From Tables \ref{sec:mean-no-rust}, the baseline shows a 4.43s mean runtime. ASAN alone adds an overhead of 106.2\%. The combination of OpenSSF production flags along with FFMalloc adds an overhead of 9.4\%. Combine all of those to enable OpenSSF production flags, ASAN, and FFMalloc, and the overhead compared to control is 106.78\%, negligibly different than ASAN alone. 
The mean runtime analysis (Tables \ref{sec:mean-no-rust}) further confirms non-linearity. For fannkuchredux, the OpenSSF production flags alone add minimal overhead, while HWASAN alone adds 78.7\% overhead. Yet their combination shows 79.0\% overhead—virtually identical to HWASAN alone, indicating the production flags become negligible when combined with HWASAN's instrumentation.

\textbf{Comparing Runtimes to Rust and Go}
The experimental objective was to achieve memory-safety guarantees comparable to natively memory-safe languages while addressing equivalent classes of memory vulnerabilities within C programs. Performance evaluation across nine standardized benchmarks revealed Rust's computational efficiency relative to the implemented mitigation techniques. As documented in Tables \ref{sec:compiler-median}, Rust demonstrated the lowest median runtime in seven of nine benchmarks. Meanwhile, Go had a lower median runtime than at least one C compiler in six out of nine benchmarks. Similarly, Tables \ref{sec:mean-compiler} indicates Rust achieved the lowest mean runtime in eight cases, whereas Go had a lower mean runtime than at least one C compiler in four out of nine benchmarks. A notable exception occurred in minimum runtime analysis (Tables \ref{sec:compiler-min}), where Rust ranked lowest in only four benchmarks and Go only outperformed at least one C compiler in two out of nine benchmarks. These results indicate that existing methodologies for mitigating memory-safety violations in C introduce measurable runtime overhead, resulting in Rust's predominant performance advantage across the majority of evaluated metrics. This trade-off between safety enforcement and execution efficiency warrants consideration in systems programming contexts requiring stringent memory protection. While Go generally exhibits lower runtime performance compared to Rust, the language offers greater approachability for novice programmers through features such as automatic memory management, built-in concurrency primitives, and an extensive package ecosystem. Excluding statistical outliers (binarytrees, regexredux), median Go runtimes remain within 17-400\% of Rust and C implementations. Although these performance differences are quantitatively significant, in practical application contexts, such differences are often acceptable and can be mitigated by the substantially longer latency inherent in human interaction cycles.

\section{Limitations}
This work's primary contribution lies not in prescribing specific memory safety configurations for C, but in advocating for a paradigm shift in how the academic community approaches memory safety research for C. Specifically, it argues that future work must:

\begin{itemize}
    \item Systematically compare memory safety modifications for C against the comprehensive protection models of native memory-safe languages like Rust and Go, and
    \item Design and evaluate compound memory safety mechanisms that collectively approximate the robustness of native solutions.
\end{itemize}

By demonstrating this dual approach through empirical analysis of combined protections and cross-language benchmarking, this paper serves as a methodological exemplar for advancing C memory safety research beyond isolated mechanism evaluation. The intent is not to recommend specific implementation strategies, but to establish a necessary framework for honest comparative analysis that accounts for the modern programming language landscape. This conceptual contribution precedes the technical limitations of the current implementation, which focus on narrower empirical constraints.

Given the intent of this paper, several limitations within this study should be kept in mind when reviewing the results of these performance benchmarks. First, Apple Silicon's unified memory architecture (UMA) fundamentally differs from traditional x86\_64 systems by integrating the CPU, GPU, and neural engine into a single system-on-a-chip (SoC) with shared high-bandwidth LPDDR5X memory. This design eliminates data copying between discrete components, as all processors access the same memory pool via ultra-fast on-package interconnects. Memory-bound applications may appear exceptionally efficient on Apple hardware but reveal significant slowdowns on DIMM-based systems where bandwidth constraints dominate. Developers must isolate compute-bound metrics from memory-dependent operations when cross-platform benchmarking, as Apple's architectural advantages in memory hierarchy can mask underlying algorithmic inefficiencies that only manifest on conventional hardware.

Second, benchmarks on Asahi Linux require cautious interpretation due to its experimental reverse-engineered approach, which relies on community-driven trial-and-error to interface with Apple Silicon. Without Apple's documentation, developers reconstruct hardware behavior through incomplete driver implementations \cite{asahilinux2025}. Key subsystems like memory management, GPU acceleration, and power efficiency often operate suboptimally compared to macOS or mature x86\_64 Linux distributions. Critical optimizations (e.g., cache coherence, thermal throttling) remain partially reverse-engineered, causing benchmarks to underrepresent hardware potential. Performance tests may reflect software-layer inefficiencies rather than hardware capabilities, skewing comparisons with well-documented platforms. Results lack consistency across versions and platforms due to rapid, unstable development cycles. Updates can introduce significant performance shifts as drivers evolve from approximations to accurate control, rendering longitudinal comparisons unreliable. Unlike x86\_64 Linux distributions with decades of vendor-collaborative integration, Asahi’s inferred behavior creates transient performance profiles. Apple Silicon’s unified memory architecture, which Asahi utilizes less effectively than macOS’s native optimizations, exacerbates discrepancies in memory-intensive workloads.

Lastly, academic tools like Checked C, FFMalloc, and MarkUs represent valuable but inherently limited research prototypes designed to explore specific memory safety or allocation concepts—not production-grade solutions. Checked C is a proof-of-concept with critical gaps: it lacks full compiler integration beyond experimental \texttt{clang} forks (now obsolete), breaks compatibility with standard C libraries, and this paper demonstrated it has breaking incompatibilities, such as HWASAN feature production-grade compilers don't have issues with, as well as 42\% of the benchmark candidates which failed to compile. MarkUs and FFMalloc are both very young compared to production memory allocators present in libraries like glibc or musl, and have only been tested in limited and very controlled circumstances. Much more testing would be required to begin including them in production workloads.
\section{Ethical Considerations}

This research evaluates the efficacy of C retrofits and native safety mechanisms through the analysis of memory safety vulnerabilities. We have considered the ethical implications of this work in accordance with the ACM CCS 2026 \cite{ccs2026cfp} guidelines and the USENIX Security Ethics Policy \cite{usenixethics}.

\subsection{Vulnerability Analysis and Disclosure}
Our evaluation utilizes publicly available CVEs and established benchmark suites to assess security defenses. No new vulnerabilities were discovered in production software during this study. Had any previously unknown vulnerabilities been identified, we would have followed the standard responsible disclosure process, notifying the relevant vendors and maintainers before public dissemination.

\subsection{Impact and Dual Use}
While our findings highlight specific limitations and "engineering gaps" in existing security retrofits, we believe the defensive benefits of this research outweigh the risks. By providing a systematic framework for comparing C retrofits against native safety (e.g., Rust), we empower engineers to make informed security decisions. We do not provide exploit code or actionable information that would simplify the creation of new attacks; rather, we focus on the architectural and performance trade-offs of defensive mechanisms.

\subsection{Human Subjects and Data}
This study does not involve human subjects, nor does it utilize private or sensitive user data. All experiments were conducted in controlled environments using open-source software and synthetic or historical vulnerability data. Consequently, Institutional Review Board (IRB) approval was not required for this work.

\section{Generative AI Usage}
Generative AI (Claude Opus 4.2 \cite{anthropic2025claudeopus}, Grammarly \cite{grammarlyAi}, and Google Gemini 3 \cite{google_gemini_2026}) was systematically used to refine grammar, academic tone, and technical phrasing throughout this paper. All AI-generated suggestions underwent rigorous manual verification for accuracy in programming language concepts and memory safety principles and followed university guidelines on the use of generative AI in academia. 
The author retains full scholarly ownership of all analytical content, with AI serving strictly as an editing tool under human oversight \cite{Hosseini_Resnik_Holmes_2023}.

\bibliographystyle{ACM-Reference-Format}
\bibliography{references}

\appendix

\section{Open Science}
\label{sec:open-science}
To support the transparency and reproducibility of our findings, we provide the complete benchmarking framework and datasets used in this study. The artifacts include:

\begin{enumerate}
    \item \textbf{Automated Benchmarking Framework:} Includes \texttt{cbench.py}, and \texttt{benchmark.py} used to orchestrate the build and execution of compounding defenses.
    \item \textbf{Benchmark Task Suite:} The complete source code for the nine benchmark tasks taken from the Computer Language Benchmark Games. The dataset for running the knucleotide benchmarks was compressed to \texttt{knucleotide-input25000\allowbreak 000.txt.zstd} using the \texttt{zstd} compression tool to fit within Github's 100 megabyte limit. The file \texttt{knucleotide-input1\allowbreak 000.txt} is available as a toy example without compression.
    \item \textbf{Analysis and Visualization Scripts:} Python scripts (including \texttt{dataframe.py} and \texttt{shapiro.py}) used to process raw telemetry and generate the statistical analyses in Section 6.
    \item \textbf{Replication Environment:} A \texttt{Dockerfile} and configuration scripts that define the exact compiler toolchain and library versions used for our evaluation.
    \item \textbf{Raw Experimental Data:} The full dataset of performance and security metrics generated during our evaluation in \texttt{results}.
\end{enumerate}

These artifacts are publicly available at the following URL:\\

\url{https://github.com/red-avalanche/compounding-defenses-evaluation}
 (commit \texttt{59411c3f9e254738bed66a48cfff0241daf83377})

The repository for the forked \texttt{ffmalloc}, \texttt{ffmalloc-16k}, is available at:\\
\url{https://github.com/red-avalanche/ffmalloc-16k}
 (commit \texttt{9661ccacf9ace0ffcae3071f9e74249cadd8738c})

No artifacts have been withheld.
\section{\label{sec:mod_map_defs}Mod Map Definitions}
The combinations of various mechanisms are described using "mod maps" or Modification Maps. Resembling bitmaps, the mod maps had six binary digits each representing a set  of compiler flags (specifics of the each shown in section \ref{sec:compiler-flags}). The mod maps are as follows:
\begin{description}
    \item[{[}0, 0, 0, 0, 0, 0{]}] - Control group (original compiler settings from benchmark games used)
    \item[{[}1, 0, 0, 0, 0, 0{]}] - OpenSSF tl;dr
    \item[{[}0, 1, 0, 0, 0, 0{]}] - OpenSSF Production
    \item[{[}0, 0, 1, 0, 0, 0{]}] - Memory Sanitizer ASAN
    \item[{[}0, 0, 0, 1, 0, 0{]}] - Memory Sanitizer HWASAN
    \item[{[}0, 0, 0, 0, 1, 0{]}] - Memory Allocator FFMalloc
    \item[{[}0, 0, 0, 0, 0, 1{]}] - Memory Allocator MarkUs
\end{description}
When the benchmarks were conducted all test samples had either one or multiple mechanisms enabled as an addition to the original compiler settings from the benchmark games. Which mechanisms were enabled are described with these mod maps. Notes on the some patterns of mod maps:
\begin{itemize}
    \item OpenSSF tl;dr is a subset or OpenSSF Production. All mod maps with OpenSSF will have either one or both digits enabled. The production flags are not enabled without the tl;dr flags.
    \item Only one memory sanitizer can be enabled at a time. They are two different methods of achieving the same end goal so there are no benchmarks where they are both enabled.
    \item Only one memory allocator can be enabled at a time. They use two different methods of allocating memory and both replace the malloc() function present in the standard library (glibc for Asahi). There are no benchmarks where they are both enabled.
\end{itemize}

\section{\label{sec:compiler-flags}Compiler Flags Provided}

\clearpage

\begin{table*}[t]
\centering
\caption*{Compiler Flags Provided}
\begin{tabular}{|| c | c ||}
 \hline
 OpenSSF & \\ [0.5ex] 
 \hline\hline
 tl;dr & \makecell{-U\_FORTIFY\_SOURCE\\ -D\_FORTIFY\_SOURCE=3\\ -D\_GLIBCXX\_ASSERTIONS\\ -fstrict-flex-arrays=3\\ -fstack-clash-protection\\ -fstack-protector-strong\\ -Wl,-z,nodlopen\\ -Wl,-z,noexecstack\\ -Wl,-z,relro\\ -Wl,-z,now\\ -Wl,--as-needed\\ -Wl,--no-copy-dt-needed-entries\\ -fPIE\\ -pie} \\
 \hline
Production & tl;dr + \makecell{-fno-delete-null-pointer-checks\\
          -fno-strict-overflow\\
          -fno-strict-aliasing\\
          -ftrivial-auto-var-init=zero} \\
 \hline\hline
Memory Sanitize & \\
\hline\hline
ASAN & \makecell{-L/usr/lib/gcc/aarch64-redhat-linux/14\\ -fsanitize=address \\ -fsanitize-recover=all\\ -static-libasan} \\
\hline
HWASAN & \makecell{-L/usr/lib/gcc/aarch64-redhat-linux/14\\ -fsanitize=hwaddress\\ -fsanitize-recover=all}\\
\hline\hline
Malloc & \\
\hline\hline
FFMalloc (single thread) & \makecell{-L/usr/local/lib/ffmalloc\\ -l:libffmallocnpst.so\\ -Wl,-rpath=/usr/local/lib/ffmalloc} \\
\hline
FFMalloc (multi thread) & \makecell{-L/usr/local/lib/ffmalloc\\ -l:libffmallocnpmt.so\\ -Wl,-rpath=/usr/local/lib/ffmalloc} \\
\hline
MarkUs & \makecell{-L/usr/local/lib/markus-allocator/lib\\ -I /usr/local/lib/markus-allocator/include\\ -l:libgc.so\\ -l:libgccpp.so\\ -Wl,-rpath-link,/usr/lib64}\\
\hline\hline
\end{tabular}
\end{table*}

\clearpage


\section{\label{sec:median-tables}Median Runtime Tables}
\subsection{\label{sec:median-rust}By mod\_map including Rust and Go}

\par\medskip
\apptabletitle{{\large binarytrees Median Durations}}
\fontsize{10.0pt}{14.4pt}\selectfont
\begin{tabular*}{\linewidth}{@{\extracolsep{\fill}}clr}
\toprule
mod\_map & time\_as\_duration & \%\_difference \\ 
\midrule\addlinespace[2.5pt]
{[}1, 1, 1, 0, 1, 0] & 2s 32861102ns & 70.744 \\
{[}0, 0, 1, 0, 0, 0] & 2s 49840569ns & 70.015 \\
{[}1, 0, 1, 0, 0, 1] & 2s 79303898ns & 68.76 \\
{[}1, 1, 1, 0, 0, 0] & 2s 130858809ns & 66.591 \\
{[}1, 0, 0, 1, 0, 0] & 2s 268153184ns & 60.982 \\
{[}1, 0, 0, 1, 0, 1] & 2s 333344353ns & 58.401 \\
{[}1, 1, 1, 0, 0, 1] & 2s 340136279ns & 58.135 \\
{[}1, 0, 0, 1, 1, 0] & 2s 370944604ns & 56.936 \\
{[}1, 1, 0, 1, 0, 1] & 2s 389668825ns & 56.212 \\
{[}1, 1, 0, 1, 0, 0] & 2s 437078371ns & 54.398 \\
{[}1, 1, 0, 1, 1, 0] & 2s 458268465ns & 53.595 \\
{[}0, 0, 1, 0, 1, 0] & 2s 492526755ns & 52.308 \\
{[}0, 0, 0, 1, 1, 0] & 2s 512938441ns & 51.547 \\
{[}0, 0, 0, 1, 0, 1] & 2s 535197836ns & 50.723 \\
{[}0, 0, 0, 1, 0, 0] & 2s 696562207ns & 44.906 \\
{[}1, 0, 1, 0, 0, 0] & 3s 107707060ns & 31.236 \\
{[}1, 0, 1, 0, 1, 0] & 3s 143455294ns & 30.119 \\
{[}0, 0, 1, 0, 0, 1] & 3s 393016507ns & 22.613 \\
{[}0, 0, 0, 0, 1, 0] & 3s 891049580ns & 9.0079 \\
{[}1, 0, 0, 0, 1, 0] & 4s 213404207ns & 1.0547 \\
{[}0, 0, 0, 0, 0, 1] & 4s 224625004ns & 0.78878 \\
{[}0, 0, 0, 0, 0, 0] & 4s 258080924ns & 0.0 \\
{[}1, 1, 0, 0, 0, 1] & 4s 286361933ns & 0.66197 \\
{[}1, 0, 0, 0, 0, 0] & 4s 289112387ns & 0.72613 \\
{[}1, 1, 0, 0, 1, 0] & 4s 313419616ns & 1.2912 \\
{[}1, 0, 0, 0, 0, 1] & 4s 381562379ns & 2.8585 \\
{[}1, 1, 0, 0, 0, 0] & 4s 443435317ns & 4.2603 \\
\bottomrule
\end{tabular*}
\endapptable

\par\medskip
\apptabletitle{{\large fannkuchredux Median Durations}}
\fontsize{10.0pt}{14.4pt}\selectfont

\begin{tabular*}{\linewidth}{@{\extracolsep{\fill}}clr}
\toprule
mod\_map & time\_as\_duration & \%\_difference \\ 
\midrule\addlinespace[2.5pt]
{[}0, 0, 0, 0, 0, 0] & 7s 716126922ns & 0.0 \\
{[}0, 0, 0, 0, 0, 1] & 23s 293798556ns & 100.47 \\
{[}0, 0, 0, 0, 1, 0] & 23s 353293507ns & 100.66 \\
{[}1, 0, 0, 0, 1, 0] & 23s 454567830ns & 100.98 \\
{[}1, 0, 0, 0, 0, 0] & 23s 455147355ns & 100.98 \\
{[}1, 0, 0, 0, 0, 1] & 23s 498811897ns & 101.12 \\
{[}1, 1, 0, 0, 0, 1] & 23s 586171255ns & 101.4 \\
{[}1, 1, 0, 0, 0, 0] & 23s 596873781ns & 101.43 \\
{[}1, 1, 0, 0, 1, 0] & 23s 600597766ns & 101.44 \\
{[}1, 0, 1, 0, 0, 1] & 27s 543939175ns & 112.47 \\
{[}0, 0, 1, 0, 0, 0] & 27s 683255040ns & 112.81 \\
{[}0, 0, 1, 0, 0, 1] & 27s 927590301ns & 113.41 \\
{[}1, 0, 1, 0, 0, 0] & 27s 928904639ns & 113.41 \\
{[}0, 0, 1, 0, 1, 0] & 28s 11496459ns & 113.61 \\
{[}1, 0, 1, 0, 1, 0] & 28s 217245161ns & 114.11 \\
{[}1, 1, 1, 0, 1, 0] & 29s 121002594ns & 116.21 \\
{[}1, 1, 1, 0, 0, 0] & 29s 494527037ns & 117.05 \\
{[}1, 1, 1, 0, 0, 1] & 30s 150331007ns & 118.49 \\
{[}0, 0, 0, 1, 0, 0] & 38s 407972725ns & 133.08 \\
{[}0, 0, 0, 1, 0, 1] & 38s 453165545ns & 133.15 \\
{[}0, 0, 0, 1, 1, 0] & 38s 526122665ns & 133.25 \\
{[}1, 0, 0, 1, 0, 1] & 39s 483923375ns & 134.61 \\
{[}1, 0, 0, 1, 0, 0] & 39s 712817557ns & 134.92 \\
{[}1, 0, 0, 1, 1, 0] & 39s 719572170ns & 134.93 \\
{[}1, 1, 0, 1, 0, 0] & 40s 291480631ns & 135.71 \\
{[}1, 1, 0, 1, 1, 0] & 40s 493011462ns & 135.98 \\
{[}1, 1, 0, 1, 0, 1] & 40s 576456760ns & 136.09 \\
\bottomrule
\end{tabular*}
\endapptable

\par\medskip
\apptabletitle{
{\large fasta Median Durations}
} 

\fontsize{10.0pt}{14.4pt}\selectfont

\begin{tabular*}{\linewidth}{@{\extracolsep{\fill}}clr}
\toprule
mod\_map & time\_as\_duration & \%\_difference \\ 
\midrule\addlinespace[2.5pt]
{[}1, 1, 0, 0, 0, 1] & 998396639ns & 39.334 \\
{[}1, 1, 0, 0, 1, 0] & 1s 2969940ns & 38.894 \\
{[}1, 0, 0, 0, 1, 0] & 1s 18738831ns & 37.391 \\
{[}1, 0, 0, 0, 0, 0] & 1s 19313116ns & 37.337 \\
{[}0, 0, 0, 0, 1, 0] & 1s 24363481ns & 36.859 \\
{[}0, 0, 0, 0, 0, 1] & 1s 29410536ns & 36.384 \\
{[}1, 1, 0, 0, 0, 0] & 1s 32781673ns & 36.068 \\
{[}1, 0, 0, 0, 0, 1] & 1s 59855670ns & 33.559 \\
{[}0, 0, 0, 0, 0, 0] & 1s 487245716ns & 0.0 \\
{[}1, 0, 1, 0, 0, 1] & 1s 599325078ns & 7.2624 \\
{[}1, 1, 1, 0, 0, 1] & 1s 600811455ns & 7.3552 \\
{[}1, 0, 1, 0, 1, 0] & 1s 667635260ns & 11.436 \\
{[}1, 1, 1, 0, 0, 0] & 1s 710305901ns & 13.952 \\
{[}1, 1, 1, 0, 1, 0] & 1s 746504903ns & 16.035 \\
{[}1, 0, 1, 0, 0, 0] & 1s 822675943ns & 20.268 \\
{[}0, 0, 1, 0, 0, 1] & 1s 888435437ns & 23.769 \\
{[}0, 0, 1, 0, 1, 0] & 2s 57137469ns & 32.157 \\
{[}0, 0, 1, 0, 0, 0] & 2s 263212505ns & 41.38 \\
{[}1, 0, 0, 1, 0, 1] & 3s 440671842ns & 79.28 \\
{[}0, 0, 0, 1, 0, 0] & 3s 466964829ns & 79.921 \\
{[}1, 0, 0, 1, 1, 0] & 3s 483451708ns & 80.319 \\
{[}1, 0, 0, 1, 0, 0] & 3s 488201486ns & 80.433 \\
{[}0, 0, 0, 1, 0, 1] & 3s 839671787ns & 88.322 \\
{[}0, 0, 0, 1, 1, 0] & 3s 883711551ns & 89.238 \\
{[}1, 1, 0, 1, 1, 0] & 4s 301688552ns & 97.235 \\
{[}1, 1, 0, 1, 0, 1] & 4s 443246207ns & 99.688 \\
{[}1, 1, 0, 1, 0, 0] & 4s 581203933ns & 101.97 \\
\bottomrule
\end{tabular*}
\endapptable

\par\medskip
\apptabletitle{
{\large knucleotide Median Durations}
} 

\fontsize{10.0pt}{14.4pt}\selectfont

\begin{tabular*}{\linewidth}{@{\extracolsep{\fill}}clr}
\toprule
mod\_map & time\_as\_duration & \%\_difference \\ 
\midrule\addlinespace[2.5pt]
{[}1, 0, 0, 0, 1, 0] & 1s 829948135ns & 50.068 \\
{[}1, 1, 0, 0, 1, 0] & 1s 838519189ns & 49.629 \\
{[}1, 0, 0, 0, 0, 1] & 1s 858599369ns & 48.609 \\
{[}0, 0, 0, 0, 1, 0] & 1s 862534748ns & 48.41 \\
{[}0, 0, 0, 0, 0, 1] & 1s 890772899ns & 46.99 \\
{[}1, 1, 0, 0, 0, 1] & 1s 898015128ns & 46.629 \\
{[}1, 1, 0, 0, 0, 0] & 2s 77741974ns & 37.988 \\
{[}1, 0, 0, 0, 0, 0] & 2s 80159397ns & 37.876 \\
{[}0, 0, 0, 0, 0, 0] & 3s 52112662ns & 0.0 \\
{[}1, 1, 1, 0, 0, 0] & 3s 537135748ns & 14.722 \\
{[}0, 0, 1, 0, 0, 0] & 3s 714023788ns & 19.565 \\
{[}1, 0, 1, 0, 0, 0] & 3s 878621573ns & 23.851 \\
{[}0, 0, 1, 0, 1, 0] & 3s 988719588ns & 26.605 \\
{[}1, 0, 1, 0, 1, 0] & 4s 51464255ns & 28.137 \\
{[}0, 0, 1, 0, 0, 1] & 4s 101552056ns & 29.34 \\
{[}1, 1, 1, 0, 1, 0] & 4s 116309892ns & 29.691 \\
{[}1, 1, 1, 0, 0, 1] & 4s 161069662ns & 30.748 \\
{[}1, 0, 1, 0, 0, 1] & 4s 250337017ns & 32.817 \\
{[}1, 0, 0, 1, 1, 0] & 6s 395152534ns & 70.773 \\
{[}0, 0, 0, 1, 0, 0] & 6s 420811965ns & 71.123 \\
{[}1, 0, 0, 1, 0, 1] & 6s 565219855ns & 73.058 \\
{[}1, 1, 0, 1, 0, 0] & 6s 583466351ns & 73.298 \\
{[}1, 1, 0, 1, 0, 1] & 6s 597831592ns & 73.487 \\
{[}1, 1, 0, 1, 1, 0] & 6s 646750275ns & 74.125 \\
{[}0, 0, 0, 1, 0, 1] & 7s 147984203ns & 80.31 \\
{[}1, 0, 0, 1, 0, 0] & 7s 234202114ns & 81.314 \\
{[}0, 0, 0, 1, 1, 0] & 7s 289109294ns & 81.944 \\
\bottomrule
\end{tabular*}
\endapptable

\par\medskip
\apptabletitle{
{\large mandelbrot Median Durations}
} 

\fontsize{10.0pt}{14.4pt}\selectfont

\begin{tabular*}{\linewidth}{@{\extracolsep{\fill}}clr}
\toprule
mod\_map & time\_as\_duration & \%\_difference \\ 
\midrule\addlinespace[2.5pt]
{[}0, 0, 0, 0, 1, 0] & 1s 372407075ns & 80.66 \\
{[}1, 0, 0, 0, 0, 1] & 1s 393161897ns & 79.399 \\
{[}1, 0, 0, 0, 1, 0] & 1s 407514057ns & 78.534 \\
{[}0, 0, 0, 0, 0, 1] & 1s 443017492ns & 76.417 \\
{[}1, 1, 0, 0, 1, 0] & 1s 452841981ns & 75.836 \\
{[}1, 1, 0, 0, 0, 1] & 1s 481397848ns & 74.164 \\
{[}1, 0, 0, 0, 0, 0] & 1s 508760576ns & 72.58 \\
{[}1, 1, 0, 0, 0, 0] & 1s 515155776ns & 72.212 \\
{[}1, 1, 0, 1, 0, 0] & 2s 9994371ns & 46.494 \\
{[}0, 0, 0, 1, 0, 1] & 2s 286067276ns & 34.152 \\
{[}0, 0, 0, 1, 1, 0] & 2s 299822659ns & 33.569 \\
{[}0, 0, 0, 1, 0, 0] & 2s 345449350ns & 31.657 \\
{[}1, 0, 0, 1, 0, 0] & 2s 360442193ns & 31.035 \\
{[}1, 0, 0, 1, 1, 0] & 2s 379142550ns & 30.265 \\
{[}1, 1, 0, 1, 1, 0] & 2s 404482069ns & 29.229 \\
{[}1, 1, 0, 1, 0, 1] & 2s 413834285ns & 28.849 \\
{[}1, 0, 0, 1, 0, 1] & 2s 430611386ns & 28.17 \\
{[}0, 0, 0, 0, 0, 0] & 3s 227565390ns & 0.0 \\
{[}0, 0, 1, 0, 0, 0] & 7s 850678912ns & 83.463 \\
{[}0, 0, 1, 0, 0, 1] & 7s 867542604ns & 83.64 \\
{[}1, 0, 1, 0, 0, 0] & 7s 877854707ns & 83.748 \\
{[}0, 0, 1, 0, 1, 0] & 7s 902111548ns & 84.001 \\
{[}1, 0, 1, 0, 1, 0] & 8s 21473295ns & 85.232 \\
{[}1, 1, 1, 0, 1, 0] & 8s 504150490ns & 89.954 \\
{[}1, 0, 1, 0, 0, 1] & 8s 512217793ns & 90.03 \\
{[}1, 1, 1, 0, 0, 1] & 8s 531445517ns & 90.21 \\
{[}1, 1, 1, 0, 0, 0] & 8s 556567954ns & 90.444 \\
\bottomrule
\end{tabular*}
\endapptable

\par\medskip
\apptabletitle{
{\large nbody Median Durations}
} 

\fontsize{10.0pt}{14.4pt}\selectfont

\begin{tabular*}{\linewidth}{@{\extracolsep{\fill}}clr}
\toprule
mod\_map & time\_as\_duration & \%\_difference \\ 
\midrule\addlinespace[2.5pt]
{[}0, 0, 0, 0, 0, 0] & 2s 181703329ns & 0.0 \\
{[}0, 0, 0, 0, 1, 0] & 2s 328540828ns & 6.5113 \\
{[}0, 0, 0, 0, 0, 1] & 2s 330787495ns & 6.6076 \\
{[}1, 0, 0, 0, 0, 1] & 2s 347674287ns & 7.3286 \\
{[}1, 0, 0, 0, 1, 0] & 2s 349521825ns & 7.4072 \\
{[}1, 0, 0, 0, 0, 0] & 2s 356721997ns & 7.7127 \\
{[}1, 1, 0, 0, 0, 0] & 3s 332593609ns & 41.742 \\
{[}1, 1, 0, 0, 1, 0] & 3s 354073712ns & 42.356 \\
{[}1, 1, 0, 0, 0, 1] & 3s 389477337ns & 43.358 \\
{[}1, 0, 1, 0, 0, 1] & 3s 412999738ns & 44.016 \\
{[}1, 0, 1, 0, 1, 0] & 3s 578430510ns & 48.496 \\
{[}1, 0, 1, 0, 0, 0] & 3s 863110869ns & 55.631 \\
{[}1, 1, 1, 0, 0, 1] & 3s 997347153ns & 58.768 \\
{[}1, 1, 1, 0, 1, 0] & 4s 32422369ns & 59.565 \\
{[}0, 0, 1, 0, 0, 1] & 4s 148217041ns & 62.134 \\
{[}1, 1, 1, 0, 0, 0] & 4s 189969306ns & 63.037 \\
{[}0, 0, 1, 0, 0, 0] & 4s 206688380ns & 63.396 \\
{[}0, 0, 1, 0, 1, 0] & 4s 254183789ns & 64.404 \\
{[}1, 1, 0, 1, 1, 0] & 7s 222710208ns & 107.21 \\
{[}1, 1, 0, 1, 0, 1] & 7s 230657368ns & 107.28 \\
{[}0, 0, 0, 1, 1, 0] & 7s 236632517ns & 107.34 \\
{[}1, 1, 0, 1, 0, 0] & 7s 238125354ns & 107.36 \\
{[}0, 0, 0, 1, 0, 1] & 7s 248889319ns & 107.46 \\
{[}1, 0, 0, 1, 0, 1] & 7s 268065019ns & 107.65 \\
{[}1, 0, 0, 1, 1, 0] & 7s 269158871ns & 107.66 \\
{[}0, 0, 0, 1, 0, 0] & 7s 276286200ns & 107.73 \\
{[}1, 0, 0, 1, 0, 0] & 7s 297401433ns & 107.94 \\
\bottomrule
\end{tabular*}
\endapptable

\par\medskip
\apptabletitle{
{\large pidigits Median Durations}
} 

\fontsize{10.0pt}{14.4pt}\selectfont

\begin{tabular*}{\linewidth}{@{\extracolsep{\fill}}clr}
\toprule
mod\_map & time\_as\_duration & \%\_difference \\ 
\midrule\addlinespace[2.5pt]
{[}1, 0, 0, 0, 0, 0] & 381553778ns & 42.02 \\
{[}1, 0, 0, 0, 1, 0] & 383419608ns & 41.553 \\
{[}1, 0, 0, 0, 0, 1] & 388122005ns & 40.385 \\
{[}1, 1, 0, 0, 1, 0] & 397716155ns & 38.037 \\
{[}1, 1, 0, 0, 0, 0] & 460939768ns & 23.642 \\
{[}0, 0, 0, 0, 0, 1] & 526278898ns & 10.487 \\
{[}1, 1, 0, 0, 0, 1] & 561323987ns & 4.0498 \\
{[}0, 0, 0, 0, 0, 0] & 584525213ns & 0.0 \\
{[}0, 0, 0, 0, 1, 0] & 634471212ns & 8.1946 \\
{[}1, 0, 1, 0, 0, 1] & 699532269ns & 17.913 \\
{[}1, 0, 1, 0, 0, 0] & 713454343ns & 19.866 \\
{[}1, 1, 1, 0, 0, 0] & 740282575ns & 23.514 \\
{[}1, 1, 1, 0, 1, 0] & 740663350ns & 23.565 \\
{[}1, 1, 1, 0, 0, 1] & 747753456ns & 24.504 \\
{[}1, 0, 1, 0, 1, 0] & 773971333ns & 27.891 \\
{[}0, 0, 1, 0, 0, 0] & 779993970ns & 28.65 \\
{[}0, 0, 1, 0, 1, 0] & 785086581ns & 29.287 \\
{[}0, 0, 1, 0, 0, 1] & 789269493ns & 29.807 \\
{[}1, 0, 0, 1, 1, 0] & 902110904ns & 42.725 \\
{[}1, 1, 0, 1, 1, 0] & 904474342ns & 42.975 \\
{[}0, 0, 0, 1, 1, 0] & 909519325ns & 43.505 \\
{[}1, 1, 0, 1, 0, 1] & 917025782ns & 44.288 \\
{[}0, 0, 0, 1, 0, 1] & 919682564ns & 44.563 \\
{[}1, 0, 0, 1, 0, 1] & 920576343ns & 44.655 \\
{[}1, 0, 0, 1, 0, 0] & 1s 183110673ns & 67.727 \\
{[}0, 0, 0, 1, 0, 0] & 1s 184638125ns & 67.841 \\
{[}1, 1, 0, 1, 0, 0] & 1s 255696635ns & 72.945 \\
\bottomrule
\end{tabular*}
\endapptable

\par\medskip
\apptabletitle{
{\large regexredux Median Durations}
} 

\fontsize{10.0pt}{14.4pt}\selectfont

\begin{tabular*}{\linewidth}{@{\extracolsep{\fill}}clr}
\toprule
mod\_map & time\_as\_duration & \%\_difference \\ 
\midrule\addlinespace[2.5pt]
{[}0, 0, 1, 0, 0, 0] & 858411278ns & 10.208 \\
{[}1, 1, 1, 0, 1, 0] & 905870468ns & 4.835 \\
{[}1, 1, 1, 0, 0, 0] & 912055543ns & 4.1549 \\
{[}1, 0, 1, 0, 0, 0] & 925373857ns & 2.7057 \\
{[}0, 0, 1, 0, 1, 0] & 944552726ns & 0.65446 \\
{[}1, 1, 1, 0, 0, 1] & 945778246ns & 0.52475 \\
{[}0, 0, 1, 0, 0, 1] & 947677190ns & 0.32416 \\
{[}0, 0, 0, 0, 0, 0] & 950754267ns & 0.0 \\
{[}1, 0, 0, 1, 0, 0] & 969658660ns & 1.9687 \\
{[}1, 0, 0, 1, 0, 1] & 977405338ns & 2.7644 \\
{[}1, 0, 1, 0, 1, 0] & 1s 25465685ns & 7.561 \\
{[}1, 0, 0, 1, 1, 0] & 1s 54316078ns & 10.33 \\
{[}1, 0, 1, 0, 0, 1] & 1s 56256779ns & 10.513 \\
{[}0, 0, 0, 1, 0, 0] & 1s 121015496ns & 16.436 \\
{[}0, 0, 0, 1, 0, 1] & 1s 164709239ns & 20.228 \\
{[}0, 0, 0, 1, 1, 0] & 1s 170492119ns & 20.718 \\
{[}1, 1, 0, 1, 0, 0] & 1s 200109273ns & 23.187 \\
{[}1, 0, 0, 0, 1, 0] & 1s 207429959ns & 23.786 \\
{[}1, 1, 0, 1, 1, 0] & 1s 210736573ns & 24.056 \\
{[}1, 1, 0, 1, 0, 1] & 1s 266543323ns & 28.484 \\
{[}1, 0, 0, 0, 0, 1] & 1s 324648411ns & 32.864 \\
{[}1, 1, 0, 0, 1, 0] & 1s 356042980ns & 35.139 \\
{[}0, 0, 0, 0, 1, 0] & 1s 374394155ns & 36.44 \\
{[}0, 0, 0, 0, 0, 1] & 1s 393703246ns & 37.787 \\
{[}1, 1, 0, 0, 0, 1] & 1s 418592773ns & 39.491 \\
{[}1, 1, 0, 0, 0, 0] & 1s 524932749ns & 46.385 \\
{[}1, 0, 0, 0, 0, 0] & 1s 581160950ns & 49.797 \\
\bottomrule
\end{tabular*}
\endapptable

\par\medskip
\apptabletitle{
{\large spectralnorm Median Durations}
} 

\fontsize{10.0pt}{14.4pt}\selectfont

\begin{tabular*}{\linewidth}{@{\extracolsep{\fill}}clr}
\toprule
mod\_map & time\_as\_duration & \%\_difference \\ 
\midrule\addlinespace[2.5pt]
{[}0, 0, 0, 0, 0, 0] & 308246580ns & 0.0 \\
{[}1, 0, 1, 0, 0, 0] & 311494232ns & 1.0482 \\
{[}0, 0, 1, 0, 0, 0] & 317158343ns & 2.85 \\
{[}1, 0, 1, 0, 0, 1] & 328716001ns & 6.4274 \\
{[}1, 1, 1, 0, 0, 0] & 337458389ns & 9.0481 \\
{[}1, 0, 1, 0, 1, 0] & 342899642ns & 10.644 \\
{[}1, 1, 1, 0, 0, 1] & 343479043ns & 10.812 \\
{[}0, 0, 1, 0, 0, 1] & 346147126ns & 11.584 \\
{[}1, 0, 0, 0, 0, 1] & 350465804ns & 12.819 \\
{[}0, 0, 0, 0, 1, 0] & 352857345ns & 13.496 \\
{[}1, 0, 0, 0, 1, 0] & 353096263ns & 13.563 \\
{[}1, 0, 0, 0, 0, 0] & 355705622ns & 14.296 \\
{[}0, 0, 0, 0, 0, 1] & 359260899ns & 15.285 \\
{[}1, 1, 1, 0, 1, 0] & 364807555ns & 16.807 \\
{[}0, 0, 1, 0, 1, 0] & 374396822ns & 19.381 \\
{[}1, 1, 0, 0, 0, 1] & 384178648ns & 21.932 \\
{[}1, 1, 0, 0, 1, 0] & 387803137ns & 22.86 \\
{[}1, 1, 0, 0, 0, 0] & 405966248ns & 27.364 \\
{[}1, 1, 0, 1, 0, 1] & 585031608ns & 61.971 \\
{[}0, 0, 0, 1, 0, 1] & 633501678ns & 69.075 \\
{[}0, 0, 0, 1, 0, 0] & 650040817ns & 71.334 \\
{[}1, 0, 0, 1, 0, 1] & 650556592ns & 71.404 \\
{[}1, 1, 0, 1, 0, 0] & 655668970ns & 72.086 \\
{[}1, 1, 0, 1, 1, 0] & 680693910ns & 75.323 \\
{[}1, 0, 0, 1, 0, 0] & 680977303ns & 75.358 \\
{[}1, 0, 0, 1, 1, 0] & 685844156ns & 75.969 \\
{[}0, 0, 0, 1, 1, 0] & 694538396ns & 77.044 \\
\bottomrule
\end{tabular*}
\endapptable

\subsection{\label{median-no-rust}By mod\_map excluding Rust and Go}
\par\medskip
\apptabletitle{
{\large binarytrees Median Durations} \\
{\small Only C programs shown}
}

\fontsize{10.0pt}{14.4pt}\selectfont

\begin{tabular*}{\linewidth}{@{\extracolsep{\fill}}clr}
\toprule
mod\_map & time\_as\_duration & \%\_difference \\
\midrule\addlinespace[2.5pt]
{[}1, 1, 1, 0, 1, 0] & 2s 32861102ns & 21.721 \\
{[}0, 0, 1, 0, 0, 0] & 2s 49840569ns & 20.899 \\
{[}1, 0, 1, 0, 0, 1] & 2s 79303898ns & 19.486 \\
{[}1, 1, 1, 0, 0, 0] & 2s 130858809ns & 17.057 \\
{[}1, 0, 0, 1, 0, 0] & 2s 268153184ns & 10.844 \\
{[}1, 0, 0, 1, 0, 1] & 2s 333344353ns & 8.0167 \\
{[}1, 1, 1, 0, 0, 1] & 2s 340136279ns & 7.7265 \\
{[}1, 0, 0, 1, 1, 0] & 2s 370944604ns & 6.4202 \\
{[}1, 1, 0, 1, 0, 1] & 2s 389668825ns & 5.6343 \\
{[}1, 1, 0, 1, 0, 0] & 2s 437078371ns & 3.6709 \\
{[}1, 1, 0, 1, 1, 0] & 2s 458268465ns & 2.8054 \\
{[}0, 0, 1, 0, 1, 0] & 2s 492526755ns & 1.4216 \\
{[}0, 0, 0, 1, 1, 0] & 2s 512938441ns & 0.60601 \\
{[}0, 0, 0, 0, 0, 0] & 2s 528213589ns & 0.0 \\
{[}0, 0, 0, 1, 0, 1] & 2s 535197836ns & 0.27586 \\
{[}0, 0, 0, 1, 0, 0] & 2s 696562207ns & 6.4443 \\
{[}1, 0, 1, 0, 0, 0] & 3s 107707060ns & 20.564 \\
{[}1, 0, 1, 0, 1, 0] & 3s 143455294ns & 21.695 \\
{[}0, 0, 1, 0, 0, 1] & 3s 393016507ns & 29.21 \\
{[}0, 0, 0, 0, 1, 0] & 3s 891049580ns & 42.461 \\
{[}1, 0, 0, 0, 1, 0] & 4s 213404207ns & 49.994 \\
{[}0, 0, 0, 0, 0, 1] & 4s 224625004ns & 50.243 \\
{[}1, 1, 0, 0, 0, 1] & 4s 286361933ns & 51.6 \\
{[}1, 0, 0, 0, 0, 0] & 4s 289112387ns & 51.66 \\
{[}1, 1, 0, 0, 1, 0] & 4s 313419616ns & 52.187 \\
{[}1, 0, 0, 0, 0, 1] & 4s 381562379ns & 53.644 \\
{[}1, 1, 0, 0, 0, 0] & 4s 443435317ns & 54.943 \\
\bottomrule
\end{tabular*}
\endapptable

\par\medskip
\apptabletitle{
{\large fannkuchredux Median Durations} \\
{\small Only C programs shown}
}

\fontsize{10.0pt}{14.4pt}\selectfont

\begin{tabular*}{\linewidth}{@{\extracolsep{\fill}}clr}
\toprule
mod\_map & time\_as\_duration & \%\_difference \\
\midrule\addlinespace[2.5pt]
{[}0, 0, 0, 0, 0, 0] & 20s 852956479ns & 0.0 \\
{[}0, 0, 0, 0, 0, 1] & 23s 293798556ns & 11.058 \\
{[}0, 0, 0, 0, 1, 0] & 23s 353293507ns & 11.312 \\
{[}1, 0, 0, 0, 1, 0] & 23s 454567830ns & 11.743 \\
{[}1, 0, 0, 0, 0, 0] & 23s 455147355ns & 11.746 \\
{[}1, 0, 0, 0, 0, 1] & 23s 498811897ns & 11.931 \\
{[}1, 1, 0, 0, 0, 1] & 23s 586171255ns & 12.301 \\
{[}1, 1, 0, 0, 0, 0] & 23s 596873781ns & 12.346 \\
{[}1, 1, 0, 0, 1, 0] & 23s 600597766ns & 12.362 \\
{[}1, 0, 1, 0, 0, 1] & 27s 543939175ns & 27.65 \\
{[}0, 0, 1, 0, 0, 0] & 27s 683255040ns & 28.145 \\
{[}0, 0, 1, 0, 0, 1] & 27s 927590301ns & 29.006 \\
{[}1, 0, 1, 0, 0, 0] & 27s 928904639ns & 29.011 \\
{[}0, 0, 1, 0, 1, 0] & 28s 11496459ns & 29.3 \\
{[}1, 0, 1, 0, 1, 0] & 28s 217245161ns & 30.015 \\
{[}1, 1, 1, 0, 1, 0] & 29s 121002594ns & 33.089 \\
{[}1, 1, 1, 0, 0, 0] & 29s 494527037ns & 34.328 \\
{[}1, 1, 1, 0, 0, 1] & 30s 150331007ns & 36.458 \\
{[}0, 0, 0, 1, 0, 0] & 38s 407972725ns & 59.247 \\
{[}0, 0, 0, 1, 0, 1] & 38s 453165545ns & 59.354 \\
{[}0, 0, 0, 1, 1, 0] & 38s 526122665ns & 59.527 \\
{[}1, 0, 0, 1, 0, 1] & 39s 483923375ns & 61.756 \\
{[}1, 0, 0, 1, 0, 0] & 39s 712817557ns & 62.279 \\
{[}1, 0, 0, 1, 1, 0] & 39s 719572170ns & 62.294 \\
{[}1, 1, 0, 1, 0, 0] & 40s 291480631ns & 63.582 \\
{[}1, 1, 0, 1, 1, 0] & 40s 493011462ns & 64.03 \\
{[}1, 1, 0, 1, 0, 1] & 40s 576456760ns & 64.215 \\
\bottomrule
\end{tabular*}
\endapptable

\par\medskip
\apptabletitle{
{\large fasta Median Durations} \\
{\small Only C programs shown}
}

\fontsize{10.0pt}{14.4pt}\selectfont

\begin{tabular*}{\linewidth}{@{\extracolsep{\fill}}clr}
\toprule
mod\_map & time\_as\_duration & \%\_difference \\
\midrule\addlinespace[2.5pt]
{[}1, 1, 0, 0, 0, 1] & 998396639ns & 53.116 \\
{[}1, 1, 0, 0, 1, 0] & 1s 2969940ns & 52.691 \\
{[}1, 0, 0, 0, 1, 0] & 1s 18738831ns & 51.237 \\
{[}1, 0, 0, 0, 0, 0] & 1s 19313116ns & 51.184 \\
{[}0, 0, 0, 0, 1, 0] & 1s 24363481ns & 50.722 \\
{[}0, 0, 0, 0, 0, 1] & 1s 29410536ns & 50.262 \\
{[}1, 1, 0, 0, 0, 0] & 1s 32781673ns & 49.955 \\
{[}1, 0, 0, 0, 0, 1] & 1s 59855670ns & 47.521 \\
{[}1, 0, 1, 0, 0, 1] & 1s 599325078ns & 7.2989 \\
{[}1, 1, 1, 0, 0, 1] & 1s 600811455ns & 7.2061 \\
{[}1, 0, 1, 0, 1, 0] & 1s 667635260ns & 3.1194 \\
{[}1, 1, 1, 0, 0, 0] & 1s 710305901ns & 0.5931 \\
{[}0, 0, 0, 0, 0, 0] & 1s 720479643ns & 0.0 \\
{[}1, 1, 1, 0, 1, 0] & 1s 746504903ns & 1.5013 \\
{[}1, 0, 1, 0, 0, 0] & 1s 822675943ns & 5.7686 \\
{[}0, 0, 1, 0, 0, 1] & 1s 888435437ns & 9.3078 \\
{[}0, 0, 1, 0, 1, 0] & 2s 57137469ns & 17.824 \\
{[}0, 0, 1, 0, 0, 0] & 2s 263212505ns & 27.248 \\
{[}1, 0, 0, 1, 0, 1] & 3s 440671842ns & 66.659 \\
{[}0, 0, 0, 1, 0, 0] & 3s 466964829ns & 67.335 \\
{[}1, 0, 0, 1, 1, 0] & 3s 483451708ns & 67.755 \\
{[}1, 0, 0, 1, 0, 0] & 3s 488201486ns & 67.876 \\
{[}0, 0, 0, 1, 0, 1] & 3s 839671787ns & 76.228 \\
{[}0, 0, 0, 1, 1, 0] & 3s 883711551ns & 77.201 \\
{[}1, 1, 0, 1, 1, 0] & 4s 301688552ns & 85.724 \\
{[}1, 1, 0, 1, 0, 1] & 4s 443246207ns & 88.348 \\
{[}1, 1, 0, 1, 0, 0] & 4s 581203933ns & 90.792 \\
\bottomrule
\end{tabular*}
\endapptable

\par\medskip
\apptabletitle{
{\large knucleotide Median Durations} \\
{\small Only C programs shown}
}

\fontsize{10.0pt}{14.4pt}\selectfont

\begin{tabular*}{\linewidth}{@{\extracolsep{\fill}}clr}
\toprule
mod\_map & time\_as\_duration & \%\_difference \\
\midrule\addlinespace[2.5pt]
{[}1, 0, 0, 0, 1, 0] & 1s 829948135ns & 5.4741 \\
{[}1, 1, 0, 0, 1, 0] & 1s 838519189ns & 5.0072 \\
{[}1, 0, 0, 0, 0, 1] & 1s 858599369ns & 3.9215 \\
{[}0, 0, 0, 0, 1, 0] & 1s 862534748ns & 3.7101 \\
{[}0, 0, 0, 0, 0, 1] & 1s 890772899ns & 2.2057 \\
{[}1, 1, 0, 0, 0, 1] & 1s 898015128ns & 1.8234 \\
{[}0, 0, 0, 0, 0, 0] & 1s 932941501ns & 0.0 \\
{[}1, 1, 0, 0, 0, 0] & 2s 77741974ns & 7.2207 \\
{[}1, 0, 0, 0, 0, 0] & 2s 80159397ns & 7.3369 \\
{[}1, 1, 1, 0, 0, 0] & 3s 537135748ns & 58.653 \\
{[}0, 0, 1, 0, 0, 0] & 3s 714023788ns & 63.081 \\
{[}1, 0, 1, 0, 0, 0] & 3s 878621573ns & 66.959 \\
{[}0, 0, 1, 0, 1, 0] & 3s 988719588ns & 69.432 \\
{[}1, 0, 1, 0, 1, 0] & 4s 51464255ns & 70.801 \\
{[}0, 0, 1, 0, 0, 1] & 4s 101552056ns & 71.874 \\
{[}1, 1, 1, 0, 1, 0] & 4s 116309892ns & 72.186 \\
{[}1, 1, 1, 0, 0, 1] & 4s 161069662ns & 73.125 \\
{[}1, 0, 1, 0, 0, 1] & 4s 250337017ns & 74.957 \\
{[}1, 0, 0, 1, 1, 0] & 6s 395152534ns & 107.16 \\
{[}0, 0, 0, 1, 0, 0] & 6s 420811965ns & 107.45 \\
{[}1, 0, 0, 1, 0, 1] & 6s 565219855ns & 109.02 \\
{[}1, 1, 0, 1, 0, 0] & 6s 583466351ns & 109.21 \\
{[}1, 1, 0, 1, 0, 1] & 6s 597831592ns & 109.37 \\
{[}1, 1, 0, 1, 1, 0] & 6s 646750275ns & 109.88 \\
{[}0, 0, 0, 1, 0, 1] & 7s 147984203ns & 114.86 \\
{[}1, 0, 0, 1, 0, 0] & 7s 234202114ns & 115.66 \\
{[}0, 0, 0, 1, 1, 0] & 7s 289109294ns & 116.16 \\
\bottomrule
\end{tabular*}
\endapptable

\par\medskip
\apptabletitle{
{\large mandelbrot Median Durations} \\
{\small Only C programs shown}
}

\fontsize{10.0pt}{14.4pt}\selectfont

\begin{tabular*}{\linewidth}{@{\extracolsep{\fill}}clr}
\toprule
mod\_map & time\_as\_duration & \%\_difference \\
\midrule\addlinespace[2.5pt]
{[}0, 0, 0, 0, 1, 0] & 1s 372407075ns & 0.26066 \\
{[}0, 0, 0, 0, 0, 0] & 1s 375989431ns & 0.0 \\
{[}1, 0, 0, 0, 0, 1] & 1s 393161897ns & 1.2402 \\
{[}1, 0, 0, 0, 1, 0] & 1s 407514057ns & 2.2651 \\
{[}0, 0, 0, 0, 0, 1] & 1s 443017492ns & 4.7554 \\
{[}1, 1, 0, 0, 1, 0] & 1s 452841981ns & 5.4335 \\
{[}1, 1, 0, 0, 0, 1] & 1s 481397848ns & 7.3779 \\
{[}1, 0, 0, 0, 0, 0] & 1s 508760576ns & 9.205 \\
{[}1, 1, 0, 0, 0, 0] & 1s 515155776ns & 9.6271 \\
{[}1, 1, 0, 1, 0, 0] & 2s 9994371ns & 37.449 \\
{[}0, 0, 0, 1, 0, 1] & 2s 286067276ns & 49.703 \\
{[}0, 0, 0, 1, 1, 0] & 2s 299822659ns & 50.266 \\
{[}0, 0, 0, 1, 0, 0] & 2s 345449350ns & 52.101 \\
{[}1, 0, 0, 1, 0, 0] & 2s 360442193ns & 52.695 \\
{[}1, 0, 0, 1, 1, 0] & 2s 379142550ns & 53.428 \\
{[}1, 1, 0, 1, 1, 0] & 2s 404482069ns & 54.411 \\
{[}1, 1, 0, 1, 0, 1] & 2s 413834285ns & 54.77 \\
{[}1, 0, 0, 1, 0, 1] & 2s 430611386ns & 55.41 \\
{[}0, 0, 1, 0, 0, 0] & 7s 850678912ns & 140.35 \\
{[}0, 0, 1, 0, 0, 1] & 7s 867542604ns & 140.46 \\
{[}1, 0, 1, 0, 0, 0] & 7s 877854707ns & 140.52 \\
{[}0, 0, 1, 0, 1, 0] & 7s 902111548ns & 140.68 \\
{[}1, 0, 1, 0, 1, 0] & 8s 21473295ns & 141.43 \\
{[}1, 1, 1, 0, 1, 0] & 8s 504150490ns & 144.29 \\
{[}1, 0, 1, 0, 0, 1] & 8s 512217793ns & 144.34 \\
{[}1, 1, 1, 0, 0, 1] & 8s 531445517ns & 144.45 \\
{[}1, 1, 1, 0, 0, 0] & 8s 556567954ns & 144.59 \\
\bottomrule
\end{tabular*}
\endapptable

\par\medskip
\apptabletitle{
{\large nbody Median Durations} \\
{\small Only C programs shown}
}

\fontsize{10.0pt}{14.4pt}\selectfont

\begin{tabular*}{\linewidth}{@{\extracolsep{\fill}}clr}
\toprule
mod\_map & time\_as\_duration & \%\_difference \\
\midrule\addlinespace[2.5pt]
{[}0, 0, 0, 0, 0, 0] & 2s 50706612ns & 0.0 \\
{[}0, 0, 0, 0, 1, 0] & 2s 328540828ns & 12.689 \\
{[}0, 0, 0, 0, 0, 1] & 2s 330787495ns & 12.785 \\
{[}1, 0, 0, 0, 0, 1] & 2s 347674287ns & 13.504 \\
{[}1, 0, 0, 0, 1, 0] & 2s 349521825ns & 13.582 \\
{[}1, 0, 0, 0, 0, 0] & 2s 356721997ns & 13.886 \\
{[}1, 1, 0, 0, 0, 0] & 3s 332593609ns & 47.625 \\
{[}1, 1, 0, 0, 1, 0] & 3s 354073712ns & 48.23 \\
{[}1, 1, 0, 0, 0, 1] & 3s 389477337ns & 49.218 \\
{[}1, 0, 1, 0, 0, 1] & 3s 412999738ns & 49.867 \\
{[}1, 0, 1, 0, 1, 0] & 3s 578430510ns & 54.279 \\
{[}1, 0, 1, 0, 0, 0] & 3s 863110869ns & 61.294 \\
{[}1, 1, 1, 0, 0, 1] & 3s 997347153ns & 64.372 \\
{[}1, 1, 1, 0, 1, 0] & 4s 32422369ns & 65.155 \\
{[}0, 0, 1, 0, 0, 1] & 4s 148217041ns & 67.673 \\
{[}1, 1, 1, 0, 0, 0] & 4s 189969306ns & 68.559 \\
{[}0, 0, 1, 0, 0, 0] & 4s 206688380ns & 68.91 \\
{[}0, 0, 1, 0, 1, 0] & 4s 254183789ns & 69.897 \\
{[}1, 1, 0, 1, 1, 0] & 7s 222710208ns & 111.54 \\
{[}1, 1, 0, 1, 0, 1] & 7s 230657368ns & 111.62 \\
{[}0, 0, 0, 1, 1, 0] & 7s 236632517ns & 111.68 \\
{[}1, 1, 0, 1, 0, 0] & 7s 238125354ns & 111.69 \\
{[}0, 0, 0, 1, 0, 1] & 7s 248889319ns & 111.79 \\
{[}1, 0, 0, 1, 0, 1] & 7s 268065019ns & 111.98 \\
{[}1, 0, 0, 1, 1, 0] & 7s 269158871ns & 111.99 \\
{[}0, 0, 0, 1, 0, 0] & 7s 276286200ns & 112.05 \\
{[}1, 0, 0, 1, 0, 0] & 7s 297401433ns & 112.25 \\
\bottomrule
\end{tabular*}
\endapptable

\par\medskip
\apptabletitle{
{\large pidigits Median Durations} \\
{\small Only C programs shown}
}

\fontsize{10.0pt}{14.4pt}\selectfont

\begin{tabular*}{\linewidth}{@{\extracolsep{\fill}}clr}
\toprule
mod\_map & time\_as\_duration & \%\_difference \\
\midrule\addlinespace[2.5pt]
{[}1, 0, 0, 0, 0, 0] & 381553778ns & 30.511 \\
{[}1, 0, 0, 0, 1, 0] & 383419608ns & 30.034 \\
{[}1, 0, 0, 0, 0, 1] & 388122005ns & 28.842 \\
{[}1, 1, 0, 0, 1, 0] & 397716155ns & 26.447 \\
{[}1, 1, 0, 0, 0, 0] & 460939768ns & 11.836 \\
{[}0, 0, 0, 0, 0, 0] & 518926285ns & 0.0 \\
{[}0, 0, 0, 0, 0, 1] & 526278898ns & 1.4068 \\
{[}1, 1, 0, 0, 0, 1] & 561323987ns & 7.8495 \\
{[}0, 0, 0, 0, 1, 0] & 634471212ns & 20.036 \\
{[}1, 0, 1, 0, 0, 1] & 699532269ns & 29.645 \\
{[}1, 0, 1, 0, 0, 0] & 713454343ns & 31.569 \\
{[}1, 1, 1, 0, 0, 0] & 740282575ns & 35.158 \\
{[}1, 1, 1, 0, 1, 0] & 740663350ns & 35.208 \\
{[}1, 1, 1, 0, 0, 1] & 747753456ns & 36.13 \\
{[}1, 0, 1, 0, 1, 0] & 773971333ns & 39.453 \\
{[}0, 0, 1, 0, 0, 0] & 779993970ns & 40.198 \\
{[}0, 0, 1, 0, 1, 0] & 785086581ns & 40.822 \\
{[}0, 0, 1, 0, 0, 1] & 789269493ns & 41.331 \\
{[}1, 0, 0, 1, 1, 0] & 902110904ns & 53.93 \\
{[}1, 1, 0, 1, 1, 0] & 904474342ns & 54.173 \\
{[}0, 0, 0, 1, 1, 0] & 909519325ns & 54.688 \\
{[}1, 1, 0, 1, 0, 1] & 917025782ns & 55.447 \\
{[}0, 0, 0, 1, 0, 1] & 919682564ns & 55.714 \\
{[}1, 0, 0, 1, 0, 1] & 920576343ns & 55.804 \\
{[}1, 0, 0, 1, 0, 0] & 1s 183110673ns & 78.046 \\
{[}0, 0, 0, 1, 0, 0] & 1s 184638125ns & 78.155 \\
{[}1, 1, 0, 1, 0, 0] & 1s 255696635ns & 83.034 \\
\bottomrule
\end{tabular*}
\endapptable

\par\medskip
\apptabletitle{
{\large regexredux Median Durations} \\
{\small Only C programs shown}
}

\fontsize{10.0pt}{14.4pt}\selectfont

\begin{tabular*}{\linewidth}{@{\extracolsep{\fill}}clr}
\toprule
mod\_map & time\_as\_duration & \%\_difference \\
\midrule\addlinespace[2.5pt]
{[}0, 0, 1, 0, 0, 0] & 858411278ns & 43.722 \\
{[}1, 1, 1, 0, 1, 0] & 905870468ns & 38.569 \\
{[}1, 1, 1, 0, 0, 0] & 912055543ns & 37.913 \\
{[}1, 0, 1, 0, 0, 0] & 925373857ns & 36.514 \\
{[}0, 0, 1, 0, 1, 0] & 944552726ns & 34.527 \\
{[}1, 1, 1, 0, 0, 1] & 945778246ns & 34.401 \\
{[}0, 0, 1, 0, 0, 1] & 947677190ns & 34.207 \\
{[}1, 0, 0, 1, 0, 0] & 969658660ns & 31.976 \\
{[}1, 0, 0, 1, 0, 1] & 977405338ns & 31.201 \\
{[}1, 0, 1, 0, 1, 0] & 1s 25465685ns & 26.501 \\
{[}1, 0, 0, 1, 1, 0] & 1s 54316078ns & 23.77 \\
{[}1, 0, 1, 0, 0, 1] & 1s 56256779ns & 23.589 \\
{[}0, 0, 0, 1, 0, 0] & 1s 121015496ns & 17.702 \\
{[}0, 0, 0, 1, 0, 1] & 1s 164709239ns & 13.902 \\
{[}0, 0, 0, 1, 1, 0] & 1s 170492119ns & 13.409 \\
{[}1, 1, 0, 1, 0, 0] & 1s 200109273ns & 10.92 \\
{[}1, 0, 0, 0, 1, 0] & 1s 207429959ns & 10.314 \\
{[}1, 1, 0, 1, 1, 0] & 1s 210736573ns & 10.041 \\
{[}1, 1, 0, 1, 0, 1] & 1s 266543323ns & 5.5415 \\
{[}1, 0, 0, 0, 0, 1] & 1s 324648411ns & 1.0573 \\
{[}0, 0, 0, 0, 0, 0] & 1s 338728526ns & 0.0 \\
{[}1, 1, 0, 0, 1, 0] & 1s 356042980ns & 1.285 \\
{[}0, 0, 0, 0, 1, 0] & 1s 374394155ns & 2.6291 \\
{[}0, 0, 0, 0, 0, 1] & 1s 393703246ns & 4.0239 \\
{[}1, 1, 0, 0, 0, 1] & 1s 418592773ns & 5.7929 \\
{[}1, 1, 0, 0, 0, 0] & 1s 524932749ns & 13.005 \\
{[}1, 0, 0, 0, 0, 0] & 1s 581160950ns & 16.606 \\
\bottomrule
\end{tabular*}
\endapptable

\par\medskip
\apptabletitle{
{\large spectralnorm Median Durations} \\
{\small Only C programs shown}
}

\fontsize{10.0pt}{14.4pt}\selectfont

\begin{tabular*}{\linewidth}{@{\extracolsep{\fill}}clr}
\toprule
mod\_map & time\_as\_duration & \%\_difference \\
\midrule\addlinespace[2.5pt]
{[}1, 0, 1, 0, 0, 0] & 311494232ns & 2.227 \\
{[}0, 0, 1, 0, 0, 0] & 317158343ns & 0.42507 \\
{[}0, 0, 0, 0, 0, 0] & 318509350ns & 0.0 \\
{[}1, 0, 1, 0, 0, 1] & 328716001ns & 3.1541 \\
{[}1, 1, 1, 0, 0, 0] & 337458389ns & 5.7774 \\
{[}1, 0, 1, 0, 1, 0] & 342899642ns & 7.3752 \\
{[}1, 1, 1, 0, 0, 1] & 343479043ns & 7.5439 \\
{[}0, 0, 1, 0, 0, 1] & 346147126ns & 8.3165 \\
{[}1, 0, 0, 0, 0, 1] & 350465804ns & 9.5537 \\
{[}0, 0, 0, 0, 1, 0] & 352857345ns & 10.232 \\
{[}1, 0, 0, 0, 1, 0] & 353096263ns & 10.3 \\
{[}1, 0, 0, 0, 0, 0] & 355705622ns & 11.034 \\
{[}0, 0, 0, 0, 0, 1] & 359260899ns & 12.025 \\
{[}1, 1, 1, 0, 1, 0] & 364807555ns & 13.551 \\
{[}0, 0, 1, 0, 1, 0] & 374396822ns & 16.131 \\
{[}1, 1, 0, 0, 0, 1] & 384178648ns & 18.691 \\
{[}1, 1, 0, 0, 1, 0] & 387803137ns & 19.621 \\
{[}1, 1, 0, 0, 0, 0] & 405966248ns & 24.144 \\
{[}1, 1, 0, 1, 0, 1] & 585031608ns & 58.995 \\
{[}0, 0, 0, 1, 0, 1] & 633501678ns & 66.174 \\
{[}0, 0, 0, 1, 0, 0] & 650040817ns & 68.459 \\
{[}1, 0, 0, 1, 0, 1] & 650556592ns & 68.529 \\
{[}1, 1, 0, 1, 0, 0] & 655668970ns & 69.219 \\
{[}1, 1, 0, 1, 1, 0] & 680693910ns & 72.495 \\
{[}1, 0, 0, 1, 0, 0] & 680977303ns & 72.531 \\
{[}1, 0, 0, 1, 1, 0] & 685844156ns & 73.149 \\
{[}0, 0, 0, 1, 1, 0] & 694538396ns & 74.237 \\
\bottomrule
\end{tabular*}
\endapptable

\subsection{\label{sec:compiler-median}By compiler}
\par\medskip
\apptabletitle{
{\large binarytrees Compiler Median Durations}
} 

\fontsize{12.0pt}{14.4pt}\selectfont

\begin{tabular*}{\linewidth}{@{\extracolsep{\fill}}cl}
\toprule
lang & time\_as\_duration \\ 
\midrule\addlinespace[2.5pt]
go & 14s 283967536ns \\
rust & 631295242ns \\
gcc & 2s 644193388ns \\
clang & 2s 748777936ns \\
checked-c & 8s 336653050ns \\
\bottomrule
\end{tabular*}
\endapptable

\par\medskip
\apptabletitle{
{\large fannkuchredux Compiler Median Durations}
} 

\fontsize{10.0pt}{14.4pt}\selectfont

\begin{tabular*}{\linewidth}{@{\extracolsep{\fill}}cl}
\toprule
lang & time\_as\_duration \\ 
\midrule\addlinespace[2.5pt]
go & 7s 634097602ns \\
rust & 6s 398731996ns \\
gcc & 28s 840526586ns \\
clang & 26s 904184203ns \\
checked-c & 23s 464719732ns \\
\bottomrule
\end{tabular*}
\endapptable

\par\medskip
\apptabletitle{
{\large fasta Compiler Median Durations}
} 

\fontsize{10.0pt}{14.4pt}\selectfont

\begin{tabular*}{\linewidth}{@{\extracolsep{\fill}}cl}
\toprule
lang & time\_as\_duration \\ 
\midrule\addlinespace[2.5pt]
go & 1s 409343499ns \\
rust & 1s 199237012ns \\
gcc & 2s 623590185ns \\
clang & 1s 17638204ns \\
checked-c & 1s 27606116ns \\
\bottomrule
\end{tabular*}
\endapptable

\par\medskip
\apptabletitle{
{\large knucleotide Compiler Median Durations}
} 

\fontsize{10.0pt}{14.4pt}\selectfont

\begin{tabular*}{\linewidth}{@{\extracolsep{\fill}}cl}
\toprule
lang & time\_as\_duration \\ 
\midrule\addlinespace[2.5pt]
go & 3s 125889378ns \\
rust & 2s 317144819ns \\
gcc & 3s 267298147ns \\
clang & 4s 23158505ns \\
\bottomrule
\end{tabular*}
\endapptable

\par\medskip
\apptabletitle{
{\large mandelbrot Compiler Median Durations}
} 

\fontsize{10.0pt}{14.4pt}\selectfont

\begin{tabular*}{\linewidth}{@{\extracolsep{\fill}}cl}
\toprule
lang & time\_as\_duration \\ 
\midrule\addlinespace[2.5pt]
go & 4s 218654939ns \\
rust & 1s 247961384ns \\
gcc & 2s 558363988ns \\
clang & 1s 91177433ns \\
checked-c & 14s 109587824ns \\
\bottomrule
\end{tabular*}
\endapptable

\par\medskip
\apptabletitle{
{\large nbody Compiler Median Durations}
} 

\fontsize{10.0pt}{14.4pt}\selectfont

\begin{tabular*}{\linewidth}{@{\extracolsep{\fill}}cl}
\toprule
lang & time\_as\_duration \\ 
\midrule\addlinespace[2.5pt]
go & 2s 583717322ns \\
rust & 2s 262851870ns \\
gcc & 3s 303322607ns \\
clang & 4s 927316907ns \\
checked-c & 3s 294935837ns \\
\bottomrule
\end{tabular*}
\endapptable

\par\medskip
\apptabletitle{
{\large pidigits Compiler Median Durations}
} 

\fontsize{10.0pt}{14.4pt}\selectfont

\begin{tabular*}{\linewidth}{@{\extracolsep{\fill}}cl}
\toprule
lang & time\_as\_duration \\ 
\midrule\addlinespace[2.5pt]
go & 686385733ns \\
rust & 578283342ns \\
gcc & 714030256ns \\
clang & 696200911ns \\
checked-c & 387879231ns \\
\bottomrule
\end{tabular*}
\endapptable

\par\medskip
\apptabletitle{
{\large regexredux Compiler Median Durations}
} 

\fontsize{10.0pt}{14.4pt}\selectfont

\begin{tabular*}{\linewidth}{@{\extracolsep{\fill}}cl}
\toprule
lang & time\_as\_duration \\ 
\midrule\addlinespace[2.5pt]
go & 16s 90826858ns \\
rust & 701207778ns \\
gcc & 982293121ns \\
clang & 1s 171373332ns \\
checked-c & 2s 596073753ns \\
\bottomrule
\end{tabular*}
\endapptable

\par\medskip
\apptabletitle{
{\large spectralnorm Compiler Median Durations}
} 

\fontsize{10.0pt}{14.4pt}\selectfont

\begin{tabular*}{\linewidth}{@{\extracolsep{\fill}}cl}
\toprule
lang & time\_as\_duration \\ 
\midrule\addlinespace[2.5pt]
go & 532665053ns \\
rust & 261998099ns \\
gcc & 329510217ns \\
clang & 511795787ns \\
\bottomrule
\end{tabular*}
\endapptable

\section{\label{sec:mean-tables}Mean Runtime Tables}
\subsection{\label{sec:mean-rust}By mod\_map including Rust and Go}
Mean values shown for completeness and to capture the impact of performance outliers that median values might obscure; see Section \ref{sec:median-tables} for median values which better represent non-normal distributions
\par\medskip
\apptabletitle{
{\large binarytrees Average Durations}
} 

\fontsize{10.0pt}{14.4pt}\selectfont

\begin{tabular*}{\linewidth}{@{\extracolsep{\fill}}clr}
\toprule
mod\_map & time\_as\_duration & \%\_difference \\ 
\midrule\addlinespace[2.5pt]
{[}0, 0, 0, 0, 1, 0] & 4s 685017200ns & 35.503 \\
{[}1, 1, 0, 0, 1, 0] & 4s 873680474ns & 31.666 \\
{[}1, 0, 0, 0, 0, 1] & 5s 20459841ns & 28.767 \\
{[}1, 1, 0, 0, 0, 0] & 5s 36036719ns & 28.463 \\
{[}0, 0, 0, 0, 0, 1] & 5s 36333400ns & 28.458 \\
{[}1, 0, 0, 0, 0, 0] & 5s 43961317ns & 28.309 \\
{[}1, 0, 0, 0, 1, 0] & 5s 74913229ns & 27.709 \\
{[}1, 1, 0, 0, 0, 1] & 5s 98599682ns & 27.253 \\
{[}0, 0, 0, 0, 0, 0] & 6s 707310652ns & 0.0 \\
{[}0, 0, 0, 1, 0, 0] & 8s 665667154ns & 25.478 \\
{[}0, 0, 0, 1, 0, 1] & 8s 716940187ns & 26.058 \\
{[}0, 0, 0, 1, 1, 0] & 8s 818071722ns & 27.191 \\
{[}1, 1, 0, 1, 0, 1] & 9s 65903462ns & 29.906 \\
{[}1, 1, 0, 1, 1, 0] & 9s 71700833ns & 29.969 \\
{[}1, 0, 0, 1, 1, 0] & 9s 95883050ns & 30.229 \\
{[}1, 1, 0, 1, 0, 0] & 9s 112743143ns & 30.41 \\
{[}1, 0, 0, 1, 0, 1] & 9s 115974842ns & 30.445 \\
{[}1, 0, 0, 1, 0, 0] & 9s 221413116ns & 31.567 \\
{[}0, 0, 1, 0, 0, 1] & 14s 369719556ns & 72.709 \\
{[}0, 0, 1, 0, 1, 0] & 14s 419231091ns & 73.007 \\
{[}0, 0, 1, 0, 0, 0] & 14s 459510817ns & 73.249 \\
{[}1, 1, 1, 0, 0, 1] & 14s 519146620ns & 73.605 \\
{[}1, 1, 1, 0, 1, 0] & 14s 593761701ns & 74.047 \\
{[}1, 1, 1, 0, 0, 0] & 14s 677284952ns & 74.539 \\
{[}1, 0, 1, 0, 1, 0] & 14s 867354551ns & 75.645 \\
{[}1, 0, 1, 0, 0, 1] & 14s 931657656ns & 76.014 \\
{[}1, 0, 1, 0, 0, 0] & 15s 58214189ns & 76.735 \\
\bottomrule
\end{tabular*}
\endapptable

\par\medskip
\apptabletitle{
{\large fannkuchredux Average Durations}
} 

\fontsize{10.0pt}{14.4pt}\selectfont

\begin{tabular*}{\linewidth}{@{\extracolsep{\fill}}clr}
\toprule
mod\_map & time\_as\_duration & \%\_difference \\ 
\midrule\addlinespace[2.5pt]
{[}0, 0, 0, 0, 0, 0] & 13s 696575241ns & 0.0 \\
{[}0, 0, 0, 0, 1, 0] & 19s 383956654ns & 34.385 \\
{[}0, 0, 0, 0, 0, 1] & 19s 450108956ns & 34.716 \\
{[}1, 0, 0, 0, 0, 0] & 19s 610149222ns & 35.51 \\
{[}1, 0, 0, 0, 0, 1] & 19s 638672685ns & 35.651 \\
{[}1, 0, 0, 0, 1, 0] & 19s 697723256ns & 35.941 \\
{[}1, 1, 0, 0, 0, 1] & 19s 731275513ns & 36.106 \\
{[}1, 1, 0, 0, 1, 0] & 19s 862497521ns & 36.747 \\
{[}1, 1, 0, 0, 0, 0] & 19s 988924145ns & 37.359 \\
{[}1, 1, 1, 0, 1, 0] & 26s 281667949ns & 62.96 \\
{[}1, 0, 1, 0, 0, 0] & 26s 459816181ns & 63.568 \\
{[}1, 1, 1, 0, 0, 0] & 26s 647459956ns & 64.202 \\
{[}1, 1, 1, 0, 0, 1] & 26s 716815089ns & 64.435 \\
{[}0, 0, 1, 0, 1, 0] & 26s 983284357ns & 65.323 \\
{[}0, 0, 1, 0, 0, 0] & 27s 92784966ns & 65.685 \\
{[}1, 0, 1, 0, 1, 0] & 27s 93777592ns & 65.688 \\
{[}0, 0, 1, 0, 0, 1] & 27s 369332365ns & 66.589 \\
{[}1, 0, 1, 0, 0, 1] & 27s 381995385ns & 66.63 \\
{[}0, 0, 0, 1, 0, 1] & 42s 970915867ns & 103.32 \\
{[}0, 0, 0, 1, 0, 0] & 42s 981672388ns & 103.34 \\
{[}1, 0, 0, 1, 0, 1] & 43s 66179108ns & 103.48 \\
{[}0, 0, 0, 1, 1, 0] & 43s 100639582ns & 103.54 \\
{[}1, 1, 0, 1, 0, 0] & 43s 167234945ns & 103.65 \\
{[}1, 1, 0, 1, 0, 1] & 43s 479050501ns & 104.18 \\
{[}1, 1, 0, 1, 1, 0] & 43s 484370078ns & 104.19 \\
{[}1, 0, 0, 1, 0, 0] & 44s 141222769ns & 105.28 \\
{[}1, 0, 0, 1, 1, 0] & 44s 619558539ns & 106.05 \\
\bottomrule
\end{tabular*}
\endapptable

\par\medskip
\apptabletitle{
{\large fasta Average Durations}
} 

\fontsize{10.0pt}{14.4pt}\selectfont

\begin{tabular*}{\linewidth}{@{\extracolsep{\fill}}clr}
\toprule
mod\_map & time\_as\_duration & \%\_difference \\ 
\midrule\addlinespace[2.5pt]
{[}1, 0, 0, 0, 1, 0] & 1s 956845793ns & 98.705 \\
{[}1, 0, 0, 0, 0, 0] & 1s 959510825ns & 98.602 \\
{[}0, 0, 0, 0, 0, 1] & 2s 1751097ns & 96.979 \\
{[}0, 0, 0, 0, 1, 0] & 2s 5618887ns & 96.832 \\
{[}1, 1, 0, 0, 0, 0] & 2s 8453615ns & 96.723 \\
{[}1, 1, 0, 0, 0, 1] & 2s 9876256ns & 96.669 \\
{[}1, 0, 1, 0, 1, 0] & 2s 304866298ns & 85.832 \\
{[}0, 0, 1, 0, 1, 0] & 2s 328782719ns & 84.988 \\
{[}1, 0, 1, 0, 0, 1] & 2s 344695232ns & 84.429 \\
{[}1, 1, 1, 0, 1, 0] & 2s 350664177ns & 84.22 \\
{[}1, 0, 0, 0, 0, 1] & 2s 354141890ns & 84.098 \\
{[}1, 0, 1, 0, 0, 0] & 2s 381314519ns & 83.151 \\
{[}1, 1, 1, 0, 0, 1] & 2s 382428369ns & 83.113 \\
{[}1, 1, 1, 0, 0, 0] & 2s 384970926ns & 83.024 \\
{[}0, 0, 1, 0, 0, 1] & 2s 390294µs & 82.84 \\
{[}0, 0, 1, 0, 0, 0] & 2s 434370850ns & 81.32 \\
{[}1, 1, 0, 0, 1, 0] & 2s 454421324ns & 80.634 \\
{[}0, 0, 0, 1, 0, 0] & 3s 193908211ns & 57.484 \\
{[}0, 0, 0, 1, 1, 0] & 3s 232022025ns & 56.394 \\
{[}0, 0, 0, 1, 0, 1] & 3s 256887756ns & 55.688 \\
{[}1, 0, 0, 1, 1, 0] & 3s 349352207ns & 53.096 \\
{[}1, 0, 0, 1, 0, 1] & 3s 350476576ns & 53.065 \\
{[}1, 0, 0, 1, 0, 0] & 3s 355425443ns & 52.927 \\
{[}0, 0, 0, 0, 0, 0] & 5s 770469328ns & 0.0 \\
{[}1, 1, 0, 1, 0, 1] & 15s 143566537ns & 89.635 \\
{[}1, 1, 0, 1, 1, 0] & 16s 914923863ns & 98.252 \\
{[}1, 1, 0, 1, 0, 0] & 16s 935585907ns & 98.345 \\
\bottomrule
\end{tabular*}
\endapptable

\par\medskip
\apptabletitle{
{\large knucleotide Average Durations}
} 

\fontsize{10.0pt}{14.4pt}\selectfont

\begin{tabular*}{\linewidth}{@{\extracolsep{\fill}}clr}
\toprule
mod\_map & time\_as\_duration & \%\_difference \\ 
\midrule\addlinespace[2.5pt]
{[}1, 1, 0, 0, 1, 0] & 1s 917242154ns & 63.705 \\
{[}1, 0, 0, 0, 1, 0] & 1s 918771898ns & 63.634 \\
{[}1, 0, 0, 0, 0, 1] & 1s 922099484ns & 63.478 \\
{[}0, 0, 0, 0, 0, 1] & 1s 933371271ns & 62.952 \\
{[}0, 0, 0, 0, 1, 0] & 1s 937965793ns & 62.738 \\
{[}1, 1, 0, 0, 0, 1] & 1s 946218552ns & 62.354 \\
{[}1, 1, 0, 0, 0, 0] & 2s 193290601ns & 51.373 \\
{[}1, 0, 0, 0, 0, 0] & 2s 204447820ns & 50.899 \\
{[}1, 1, 1, 0, 0, 0] & 3s 691486572ns & 0.48723 \\
{[}0, 0, 0, 0, 0, 0] & 3s 709516572ns & 0.0 \\
{[}0, 0, 1, 0, 0, 0] & 3s 724033328ns & 0.39058 \\
{[}1, 0, 1, 0, 0, 0] & 3s 762961914ns & 1.4304 \\
{[}1, 0, 1, 0, 1, 0] & 4s 6045906ns & 7.6865 \\
{[}0, 0, 1, 0, 1, 0] & 4s 23554644ns & 8.1219 \\
{[}1, 1, 1, 0, 1, 0] & 4s 124417509ns & 10.592 \\
{[}1, 1, 1, 0, 0, 1] & 4s 184478371ns & 12.034 \\
{[}0, 0, 1, 0, 0, 1] & 4s 188392395ns & 12.127 \\
{[}1, 0, 1, 0, 0, 1] & 4s 201140879ns & 12.429 \\
{[}0, 0, 0, 1, 0, 0] & 6s 760674730ns & 58.283 \\
{[}1, 1, 0, 1, 0, 0] & 6s 772429242ns & 58.442 \\
{[}1, 0, 0, 1, 0, 1] & 7s 131892522ns & 63.135 \\
{[}1, 0, 0, 1, 1, 0] & 7s 158187426ns & 63.466 \\
{[}1, 1, 0, 1, 1, 0] & 7s 173903007ns & 63.664 \\
{[}1, 0, 0, 1, 0, 0] & 7s 372111781ns & 66.102 \\
{[}1, 1, 0, 1, 0, 1] & 7s 375867317ns & 66.147 \\
{[}0, 0, 0, 1, 1, 0] & 7s 654846437ns & 69.433 \\
{[}0, 0, 0, 1, 0, 1] & 7s 741399756ns & 70.42 \\
\bottomrule
\end{tabular*}
\endapptable

\par\medskip
\apptabletitle{
{\large mandelbrot Average Durations}
} 

\fontsize{10.0pt}{14.4pt}\selectfont

\begin{tabular*}{\linewidth}{@{\extracolsep{\fill}}clr}
\toprule
mod\_map & time\_as\_duration & \%\_difference \\ 
\midrule\addlinespace[2.5pt]
{[}0, 0, 0, 1, 0, 1] & 4s 652852067ns & 13.745 \\
{[}0, 0, 0, 1, 0, 0] & 4s 686362885ns & 13.03 \\
{[}1, 0, 0, 1, 0, 0] & 4s 730226558ns & 12.102 \\
{[}1, 0, 0, 1, 0, 1] & 4s 734445014ns & 12.014 \\
{[}0, 0, 0, 1, 1, 0] & 4s 735657864ns & 11.988 \\
{[}1, 0, 0, 1, 1, 0] & 4s 746097096ns & 11.769 \\
{[}1, 1, 0, 1, 0, 1] & 4s 808975939ns & 10.457 \\
{[}1, 1, 0, 1, 1, 0] & 5s 320003945ns & 0.36711 \\
{[}0, 0, 0, 0, 0, 0] & 5s 339569940ns & 0.0 \\
{[}1, 1, 0, 1, 0, 0] & 5s 388498349ns & 0.91217 \\
{[}1, 0, 0, 0, 0, 1] & 6s 391740012ns & 17.938 \\
{[}0, 0, 0, 0, 0, 1] & 6s 472589724ns & 19.184 \\
{[}1, 0, 0, 0, 0, 0] & 6s 480713414ns & 19.308 \\
{[}1, 1, 0, 0, 0, 0] & 6s 505926533ns & 19.693 \\
{[}1, 0, 0, 0, 1, 0] & 6s 544223467ns & 20.274 \\
{[}1, 1, 0, 0, 0, 1] & 6s 621050447ns & 21.428 \\
{[}1, 1, 0, 0, 1, 0] & 6s 704873933ns & 22.671 \\
{[}0, 0, 0, 0, 1, 0] & 6s 730225241ns & 23.044 \\
{[}0, 0, 1, 0, 0, 0] & 20s 903195817ns & 118.61 \\
{[}0, 0, 1, 0, 0, 1] & 21s 366083780ns & 120.02 \\
{[}1, 0, 1, 0, 0, 0] & 21s 466326425ns & 120.32 \\
{[}1, 0, 1, 0, 1, 0] & 21s 566285920ns & 120.62 \\
{[}1, 1, 1, 0, 0, 1] & 21s 605736951ns & 120.73 \\
{[}1, 1, 1, 0, 0, 0] & 21s 673001347ns & 120.93 \\
{[}1, 0, 1, 0, 0, 1] & 21s 721173675ns & 121.07 \\
{[}1, 1, 1, 0, 1, 0] & 21s 821766098ns & 121.37 \\
{[}0, 0, 1, 0, 1, 0] & 24s 439192985ns & 128.28 \\
\bottomrule
\end{tabular*}
\endapptable

\par\medskip
\apptabletitle{
{\large nbody Average Durations}
} 

\fontsize{10.0pt}{14.4pt}\selectfont

\begin{tabular*}{\linewidth}{@{\extracolsep{\fill}}clr}
\toprule
mod\_map & time\_as\_duration & \%\_difference \\ 
\midrule\addlinespace[2.5pt]
{[}0, 0, 0, 0, 0, 0] & 2s 389480883ns & 0.0 \\
{[}0, 0, 0, 0, 1, 0] & 2s 597632345ns & 8.3476 \\
{[}1, 0, 0, 0, 0, 1] & 2s 599709530ns & 8.4274 \\
{[}0, 0, 0, 0, 0, 1] & 2s 604647976ns & 8.6168 \\
{[}1, 0, 0, 0, 0, 0] & 2s 610326306ns & 8.8342 \\
{[}1, 0, 0, 0, 1, 0] & 2s 612963243ns & 8.935 \\
{[}1, 1, 0, 0, 0, 0] & 3s 174700480ns & 28.224 \\
{[}1, 1, 0, 0, 1, 0] & 3s 199109426ns & 28.974 \\
{[}1, 1, 0, 0, 0, 1] & 3s 225555717ns & 29.78 \\
{[}1, 0, 1, 0, 0, 0] & 3s 750048517ns & 44.322 \\
{[}1, 0, 1, 0, 1, 0] & 3s 846627551ns & 46.733 \\
{[}0, 0, 1, 0, 0, 0] & 3s 869616088ns & 47.296 \\
{[}0, 0, 1, 0, 1, 0] & 3s 887819332ns & 47.738 \\
{[}1, 0, 1, 0, 0, 1] & 3s 913919119ns & 48.369 \\
{[}0, 0, 1, 0, 0, 1] & 3s 940683986ns & 49.01 \\
{[}1, 1, 1, 0, 0, 1] & 4s 446815586ns & 60.189 \\
{[}1, 1, 1, 0, 1, 0] & 4s 523887069ns & 61.747 \\
{[}1, 1, 1, 0, 0, 0] & 4s 524647484ns & 61.762 \\
{[}1, 1, 0, 1, 1, 0] & 7s 446025729ns & 102.82 \\
{[}1, 1, 0, 1, 0, 1] & 7s 475444718ns & 103.11 \\
{[}1, 1, 0, 1, 0, 0] & 7s 488405796ns & 103.24 \\
{[}1, 0, 0, 1, 1, 0] & 7s 789392015ns & 106.1 \\
{[}0, 0, 0, 1, 1, 0] & 7s 818579495ns & 106.37 \\
{[}1, 0, 0, 1, 0, 0] & 7s 818811909ns & 106.37 \\
{[}0, 0, 0, 1, 0, 1] & 7s 823060388ns & 106.41 \\
{[}1, 0, 0, 1, 0, 1] & 7s 823100815ns & 106.41 \\
{[}0, 0, 0, 1, 0, 0] & 7s 949154292ns & 107.55 \\
\bottomrule
\end{tabular*}
\endapptable

\par\medskip
\apptabletitle{
{\large pidigits Average Durations}
} 

\fontsize{10.0pt}{14.4pt}\selectfont

\begin{tabular*}{\linewidth}{@{\extracolsep{\fill}}clr}
\toprule
mod\_map & time\_as\_duration & \%\_difference \\ 
\midrule\addlinespace[2.5pt]
{[}1, 0, 0, 0, 0, 0] & 428322868ns & 52.331 \\
{[}1, 0, 0, 0, 1, 0] & 430918796ns & 51.767 \\
{[}1, 0, 0, 0, 0, 1] & 435358138ns & 50.81 \\
{[}1, 1, 0, 0, 1, 0] & 466127263ns & 44.368 \\
{[}1, 1, 0, 0, 0, 0] & 467087937ns & 44.172 \\
{[}0, 0, 0, 0, 0, 1] & 487747766ns & 40.036 \\
{[}1, 1, 0, 0, 0, 1] & 513390828ns & 35.093 \\
{[}0, 0, 0, 0, 1, 0] & 593567406ns & 20.873 \\
{[}1, 0, 1, 0, 0, 1] & 710951047ns & 2.9034 \\
{[}1, 0, 1, 0, 0, 0] & 724877297ns & 0.96377 \\
{[}1, 1, 1, 0, 1, 0] & 727375346ns & 0.61976 \\
{[}0, 0, 0, 0, 0, 0] & 731897894ns & 0.0 \\
{[}1, 0, 1, 0, 1, 0] & 735778002ns & 0.52886 \\
{[}0, 0, 1, 0, 1, 0] & 739279779ns & 1.0036 \\
{[}1, 1, 1, 0, 0, 0] & 742851035ns & 1.4855 \\
{[}0, 0, 1, 0, 0, 0] & 757120129ns & 3.3879 \\
{[}0, 0, 1, 0, 0, 1] & 768641766ns & 4.8974 \\
{[}1, 1, 1, 0, 0, 1] & 771717011ns & 5.2966 \\
{[}1, 0, 0, 1, 0, 1] & 927614729ns & 23.587 \\
{[}1, 0, 0, 1, 1, 0] & 929332275ns & 23.77 \\
{[}0, 0, 0, 1, 1, 0] & 932793907ns & 24.136 \\
{[}0, 0, 0, 1, 0, 1] & 936024777ns & 24.477 \\
{[}1, 1, 0, 1, 1, 0] & 936124283ns & 24.487 \\
{[}1, 1, 0, 1, 0, 1] & 990088905ns & 29.988 \\
{[}0, 0, 0, 1, 0, 0] & 1s 65932595ns & 37.16 \\
{[}1, 1, 0, 1, 0, 0] & 1s 92030269ns & 39.49 \\
{[}1, 0, 0, 1, 0, 0] & 1s 112508550ns & 41.272 \\
\bottomrule
\end{tabular*}
\endapptable

\par\medskip
\apptabletitle{
{\large regexredux Average Durations}
} 

\fontsize{10.0pt}{14.4pt}\selectfont

\begin{tabular*}{\linewidth}{@{\extracolsep{\fill}}clr}
\toprule
mod\_map & time\_as\_duration & \%\_difference \\ 
\midrule\addlinespace[2.5pt]
{[}0, 0, 1, 0, 0, 0] & 908626149ns & 126.29 \\
{[}1, 1, 1, 0, 0, 0] & 923210553ns & 125.33 \\
{[}1, 0, 1, 0, 0, 0] & 928924048ns & 124.95 \\
{[}1, 1, 1, 0, 1, 0] & 965021062ns & 122.6 \\
{[}1, 1, 1, 0, 0, 1] & 988720085ns & 121.07 \\
{[}1, 0, 1, 0, 1, 0] & 988806608ns & 121.07 \\
{[}0, 0, 1, 0, 1, 0] & 997552595ns & 120.51 \\
{[}0, 0, 1, 0, 0, 1] & 1s 39780290ns & 117.83 \\
{[}1, 0, 1, 0, 0, 1] & 1s 45182913ns & 117.5 \\
{[}1, 0, 0, 1, 0, 0] & 1s 73896972ns & 115.71 \\
{[}1, 0, 0, 1, 0, 1] & 1s 79047954ns & 115.39 \\
{[}1, 0, 0, 1, 1, 0] & 1s 90168290ns & 114.7 \\
{[}1, 1, 0, 1, 0, 0] & 1s 99012093ns & 114.16 \\
{[}1, 1, 0, 1, 1, 0] & 1s 127575332ns & 112.42 \\
{[}0, 0, 0, 1, 0, 0] & 1s 131861880ns & 112.16 \\
{[}1, 1, 0, 1, 0, 1] & 1s 144052616ns & 111.42 \\
{[}0, 0, 0, 1, 1, 0] & 1s 157094302ns & 110.64 \\
{[}0, 0, 0, 1, 0, 1] & 1s 161664819ns & 110.36 \\
{[}0, 0, 0, 0, 1, 0] & 1s 494943796ns & 91.613 \\
{[}1, 0, 0, 0, 1, 0] & 1s 508455494ns & 90.9 \\
{[}1, 1, 0, 0, 1, 0] & 1s 519839594ns & 90.303 \\
{[}1, 0, 0, 0, 0, 1] & 1s 521528983ns & 90.214 \\
{[}0, 0, 0, 0, 0, 1] & 1s 539629127ns & 89.27 \\
{[}1, 1, 0, 0, 0, 1] & 1s 546196341ns & 88.929 \\
{[}1, 1, 0, 0, 0, 0] & 1s 651190542ns & 83.581 \\
{[}1, 0, 0, 0, 0, 0] & 1s 671870876ns & 82.551 \\
{[}0, 0, 0, 0, 0, 0] & 4s 22096005ns & 0.0 \\
\bottomrule
\end{tabular*}
\endapptable

\par\medskip
\apptabletitle{
{\large spectralnorm Average Durations}
} 

\fontsize{10.0pt}{14.4pt}\selectfont

\begin{tabular*}{\linewidth}{@{\extracolsep{\fill}}clr}
\toprule
mod\_map & time\_as\_duration & \%\_difference \\ 
\midrule\addlinespace[2.5pt]
{[}1, 0, 1, 0, 0, 0] & 368389749ns & 99.506 \\
{[}0, 0, 1, 0, 0, 0] & 384004540ns & 96.35 \\
{[}1, 0, 1, 0, 0, 1] & 385018348ns & 96.147 \\
{[}1, 1, 1, 0, 0, 0] & 401841547ns & 92.826 \\
{[}0, 0, 1, 0, 0, 1] & 403824208ns & 92.439 \\
{[}1, 1, 1, 0, 1, 0] & 421556439ns & 89.026 \\
{[}0, 0, 0, 0, 0, 1] & 427955643ns & 87.814 \\
{[}1, 1, 0, 0, 1, 0] & 433736134ns & 86.728 \\
{[}1, 1, 0, 0, 0, 1] & 436260797ns & 86.256 \\
{[}1, 0, 0, 0, 1, 0] & 439093292ns & 85.728 \\
{[}1, 0, 0, 0, 0, 1] & 449544147ns & 83.799 \\
{[}0, 0, 0, 0, 1, 0] & 450156440ns & 83.686 \\
{[}0, 0, 1, 0, 1, 0] & 493633137ns & 75.937 \\
{[}1, 1, 1, 0, 0, 1] & 501844547ns & 74.521 \\
{[}1, 0, 1, 0, 1, 0] & 513059535ns & 72.61 \\
{[}1, 1, 0, 0, 0, 0] & 513105020ns & 72.602 \\
{[}1, 0, 0, 0, 0, 0] & 516786673ns & 71.98 \\
{[}1, 1, 0, 1, 0, 1] & 559574707ns & 64.959 \\
{[}0, 0, 0, 1, 0, 0] & 583805650ns & 61.142 \\
{[}0, 0, 0, 1, 0, 1] & 584950883ns & 60.964 \\
{[}1, 0, 0, 1, 0, 1] & 588891950ns & 60.354 \\
{[}1, 1, 0, 1, 0, 0] & 596598292ns & 59.17 \\
{[}1, 0, 0, 1, 0, 0] & 631409293ns & 53.953 \\
{[}1, 0, 0, 1, 1, 0] & 634665195ns & 53.476 \\
{[}0, 0, 0, 1, 1, 0] & 697937382ns & 44.545 \\
{[}1, 1, 0, 1, 1, 0] & 723514884ns & 41.111 \\
{[}0, 0, 0, 0, 0, 0] & 1s 97922977ns & 0.0 \\
\bottomrule
\end{tabular*}
\endapptable

\subsection{\label{sec:mean-no-rust}By mod\_map excluding Rust and Go}
\par\medskip
\apptabletitle{
{\large binarytrees Average Durations} \\
{\small Only C programs shown}
}

\fontsize{10.0pt}{14.4pt}\selectfont

\begin{tabular*}{\linewidth}{@{\extracolsep{\fill}}clr}
\toprule
mod\_map & time\_as\_duration & \%\_difference \\
\midrule\addlinespace[2.5pt]
{[}0, 0, 0, 0, 0, 0] & 4s 434497371ns & 0.0 \\
{[}0, 0, 0, 0, 1, 0] & 4s 685017200ns & 5.4942 \\
{[}1, 1, 0, 0, 1, 0] & 4s 873680474ns & 9.4365 \\
{[}1, 0, 0, 0, 0, 1] & 5s 20459841ns & 12.395 \\
{[}1, 1, 0, 0, 0, 0] & 5s 36036719ns & 12.703 \\
{[}0, 0, 0, 0, 0, 1] & 5s 36333400ns & 12.709 \\
{[}1, 0, 0, 0, 0, 0] & 5s 43961317ns & 12.86 \\
{[}1, 0, 0, 0, 1, 0] & 5s 74913229ns & 13.469 \\
{[}1, 1, 0, 0, 0, 1] & 5s 98599682ns & 13.933 \\
{[}0, 0, 0, 1, 0, 0] & 8s 665667154ns & 64.597 \\
{[}0, 0, 0, 1, 0, 1] & 8s 716940187ns & 65.125 \\
{[}0, 0, 0, 1, 1, 0] & 8s 818071722ns & 66.154 \\
{[}1, 1, 0, 1, 0, 1] & 9s 65903462ns & 68.611 \\
{[}1, 1, 0, 1, 1, 0] & 9s 71700833ns & 68.668 \\
{[}1, 0, 0, 1, 1, 0] & 9s 95883050ns & 68.903 \\
{[}1, 1, 0, 1, 0, 0] & 9s 112743143ns & 69.066 \\
{[}1, 0, 0, 1, 0, 1] & 9s 115974842ns & 69.097 \\
{[}1, 0, 0, 1, 0, 0] & 9s 221413116ns & 70.108 \\
{[}0, 0, 1, 0, 0, 1] & 14s 369719556ns & 105.67 \\
{[}0, 0, 1, 0, 1, 0] & 14s 419231091ns & 105.92 \\
{[}0, 0, 1, 0, 0, 0] & 14s 459510817ns & 106.12 \\
{[}1, 1, 1, 0, 0, 1] & 14s 519146620ns & 106.41 \\
{[}1, 1, 1, 0, 1, 0] & 14s 593761701ns & 106.78 \\
{[}1, 1, 1, 0, 0, 0] & 14s 677284952ns & 107.19 \\
{[}1, 0, 1, 0, 1, 0] & 14s 867354551ns & 108.1 \\
{[}1, 0, 1, 0, 0, 1] & 14s 931657656ns & 108.41 \\
{[}1, 0, 1, 0, 0, 0] & 15s 58214189ns & 109.0 \\
\bottomrule
\end{tabular*}
\endapptable

\par\medskip
\apptabletitle{
{\large fannkuchredux Average Durations} \\
{\small Only C programs shown}
}

\fontsize{10.0pt}{14.4pt}\selectfont

\begin{tabular*}{\linewidth}{@{\extracolsep{\fill}}clr}
\toprule
mod\_map & time\_as\_duration & \%\_difference \\
\midrule\addlinespace[2.5pt]
{[}0, 0, 0, 0, 0, 0] & 18s 714158040ns & 0.0 \\
{[}0, 0, 0, 0, 1, 0] & 19s 383956654ns & 3.5162 \\
{[}0, 0, 0, 0, 0, 1] & 19s 450108956ns & 3.8567 \\
{[}1, 0, 0, 0, 0, 0] & 19s 610149222ns & 4.6758 \\
{[}1, 0, 0, 0, 0, 1] & 19s 638672685ns & 4.8211 \\
{[}1, 0, 0, 0, 1, 0] & 19s 697723256ns & 5.1212 \\
{[}1, 1, 0, 0, 0, 1] & 19s 731275513ns & 5.2912 \\
{[}1, 1, 0, 0, 1, 0] & 19s 862497521ns & 5.9535 \\
{[}1, 1, 0, 0, 0, 0] & 19s 988924145ns & 6.5874 \\
{[}1, 1, 1, 0, 1, 0] & 26s 281667949ns & 33.636 \\
{[}1, 0, 1, 0, 0, 0] & 26s 459816181ns & 34.293 \\
{[}1, 1, 1, 0, 0, 0] & 26s 647459956ns & 34.978 \\
{[}1, 1, 1, 0, 0, 1] & 26s 716815089ns & 35.23 \\
{[}0, 0, 1, 0, 1, 0] & 26s 983284357ns & 36.191 \\
{[}0, 0, 1, 0, 0, 0] & 27s 92784966ns & 36.582 \\
{[}1, 0, 1, 0, 1, 0] & 27s 93777592ns & 36.586 \\
{[}0, 0, 1, 0, 0, 1] & 27s 369332365ns & 37.563 \\
{[}1, 0, 1, 0, 0, 1] & 27s 381995385ns & 37.608 \\
{[}0, 0, 0, 1, 0, 1] & 42s 970915867ns & 78.647 \\
{[}0, 0, 0, 1, 0, 0] & 42s 981672388ns & 78.668 \\
{[}1, 0, 0, 1, 0, 1] & 43s 66179108ns & 78.834 \\
{[}0, 0, 0, 1, 1, 0] & 43s 100639582ns & 78.902 \\
{[}1, 1, 0, 1, 0, 0] & 43s 167234945ns & 79.032 \\
{[}1, 1, 0, 1, 0, 1] & 43s 479050501ns & 79.639 \\
{[}1, 1, 0, 1, 1, 0] & 43s 484370078ns & 79.649 \\
{[}1, 0, 0, 1, 0, 0] & 44s 141222769ns & 80.907 \\
{[}1, 0, 0, 1, 1, 0] & 44s 619558539ns & 81.806 \\
\bottomrule
\end{tabular*}
\endapptable

\par\medskip
\apptabletitle{
{\large fasta Average Durations} \\
{\small Only C programs shown}
}

\fontsize{10.0pt}{14.4pt}\selectfont

\begin{tabular*}{\linewidth}{@{\extracolsep{\fill}}clr}
\toprule
mod\_map & time\_as\_duration & \%\_difference \\
\midrule\addlinespace[2.5pt]
{[}1, 0, 0, 0, 1, 0] & 1s 956845793ns & 7.4393 \\
{[}1, 0, 0, 0, 0, 0] & 1s 959510825ns & 7.3034 \\
{[}0, 0, 0, 0, 0, 1] & 2s 1751097ns & 5.1727 \\
{[}0, 0, 0, 0, 1, 0] & 2s 5618887ns & 4.9798 \\
{[}1, 1, 0, 0, 0, 0] & 2s 8453615ns & 4.8387 \\
{[}1, 1, 0, 0, 0, 1] & 2s 9876256ns & 4.7679 \\
{[}0, 0, 0, 0, 0, 0] & 2s 108045635ns & 0.0 \\
{[}1, 0, 1, 0, 1, 0] & 2s 304866298ns & 8.9202 \\
{[}0, 0, 1, 0, 1, 0] & 2s 328782719ns & 9.9502 \\
{[}1, 0, 1, 0, 0, 1] & 2s 344695232ns & 10.629 \\
{[}1, 1, 1, 0, 1, 0] & 2s 350664177ns & 10.883 \\
{[}1, 0, 0, 0, 0, 1] & 2s 354141890ns & 11.03 \\
{[}1, 0, 1, 0, 0, 0] & 2s 381314519ns & 12.174 \\
{[}1, 1, 1, 0, 0, 1] & 2s 382428369ns & 12.221 \\
{[}1, 1, 1, 0, 0, 0] & 2s 384970926ns & 12.327 \\
{[}0, 0, 1, 0, 0, 1] & 2s 390294µs & 12.549 \\
{[}0, 0, 1, 0, 0, 0] & 2s 434370850ns & 14.368 \\
{[}1, 1, 0, 0, 1, 0] & 2s 454421324ns & 15.184 \\
{[}0, 0, 0, 1, 0, 0] & 3s 193908211ns & 40.961 \\
{[}0, 0, 0, 1, 1, 0] & 3s 232022025ns & 42.096 \\
{[}0, 0, 0, 1, 0, 1] & 3s 256887756ns & 42.828 \\
{[}1, 0, 0, 1, 1, 0] & 3s 349352207ns & 45.491 \\
{[}1, 0, 0, 1, 0, 1] & 3s 350476576ns & 45.523 \\
{[}1, 0, 0, 1, 0, 0] & 3s 355425443ns & 45.663 \\
{[}1, 1, 0, 1, 0, 1] & 15s 143566537ns & 151.12 \\
{[}1, 1, 0, 1, 1, 0] & 16s 914923863ns & 155.67 \\
{[}1, 1, 0, 1, 0, 0] & 16s 935585907ns & 155.72 \\
\bottomrule
\end{tabular*}
\endapptable

\par\medskip
\apptabletitle{
{\large knucleotide Average Durations} \\
{\small Only C programs shown}
}

\fontsize{10.0pt}{14.4pt}\selectfont

%
\endapptable

\par\medskip
\apptabletitle{
{\large mandelbrot Average Durations} \\
{\small Only C programs shown}
}

\fontsize{10.0pt}{14.4pt}\selectfont

%
\endapptable

\par\medskip
\apptabletitle{
{\large nbody Average Durations} \\
{\small Only C programs shown}
}

\fontsize{10.0pt}{14.4pt}\selectfont

%
\endapptable

\par\medskip
\apptabletitle{
{\large pidigits Average Durations} \\
{\small Only C programs shown}
}

\fontsize{10.0pt}{14.4pt}\selectfont

%
\endapptable

\par\medskip
\apptabletitle{
{\large regexredux Average Durations} \\
{\small Only C programs shown}
}

\fontsize{10.0pt}{14.4pt}\selectfont

%
\endapptable

\par\medskip
\apptabletitle{
{\large spectralnorm Average Durations} \\
{\small Only C programs shown}
}

\fontsize{10.0pt}{14.4pt}\selectfont

%
\endapptable

\subsection{\label{sec:mean-compiler}By compiler}
\par\medskip
\apptabletitle{
{\large binarytrees Compiler Mean Durations}
} 

\fontsize{10.0pt}{14.4pt}\selectfont

\begin{tabular*}{\linewidth}{@{\extracolsep{\fill}}cl}
\toprule
lang & time\_as\_duration \\ 
\midrule\addlinespace[2.5pt]
go & 13s 231707860ns \\
rust & 696284628ns \\
gcc & 9s 592546578ns \\
clang & 8s 506177192ns \\
checked-c & 7s 363063525ns \\
\bottomrule
\end{tabular*}
\endapptable

\par\medskip
\apptabletitle{
{\large fannkuchredux Compiler Mean Durations}
} 

\fontsize{10.0pt}{14.4pt}\selectfont

\begin{tabular*}{\linewidth}{@{\extracolsep{\fill}}cl}
\toprule
lang & time\_as\_duration \\ 
\midrule\addlinespace[2.5pt]
go & 12s 730773182ns \\
rust & 8s 294200923ns \\
gcc & 32s 113712335ns \\
clang & 25s 93407091ns \\
checked-c & 21s 475742842ns \\
\bottomrule
\end{tabular*}
\endapptable

\par\medskip
\apptabletitle{
{\large fasta Compiler Mean Durations}
} 

\fontsize{10.0pt}{14.4pt}\selectfont

\begin{tabular*}{\linewidth}{@{\extracolsep{\fill}}cl}
\toprule
lang & time\_as\_duration \\ 
\midrule\addlinespace[2.5pt]
go & 25s 378070706ns \\
rust & 1s 622239543ns \\
gcc & 4s 221502945ns \\
clang & 1s 892301796ns \\
checked-c & 2s 145090742ns \\
\bottomrule
\end{tabular*}
\endapptable

\par\medskip
\apptabletitle{
{\large knucleotide Compiler Mean Durations}
} 

\fontsize{10.0pt}{14.4pt}\selectfont

\begin{tabular*}{\linewidth}{@{\extracolsep{\fill}}cl}
\toprule
lang & time\_as\_duration \\ 
\midrule\addlinespace[2.5pt]
go & 3s 258889358ns \\
rust & 4s 610044121ns \\
gcc & 4s 170514860ns \\
clang & 3s 556927297ns \\
\bottomrule
\end{tabular*}
\endapptable

\par\medskip
\apptabletitle{
{\large mandelbrot Compiler Mean Durations}
} 

\fontsize{10.0pt}{14.4pt}\selectfont

%
\endapptable

\par\medskip
\apptabletitle{
{\large nbody Compiler Mean Durations}
} 

\fontsize{10.0pt}{14.4pt}\selectfont

%
\endapptable

\par\medskip
\apptabletitle{
{\large pidigits Compiler Mean Durations}
} 

\fontsize{10.0pt}{14.4pt}\selectfont

%
\endapptable

\par\medskip
\apptabletitle{
{\large regexredux Compiler Mean Durations}
} 

\fontsize{10.0pt}{14.4pt}\selectfont

%
\endapptable

\par\medskip
\apptabletitle{
{\large spectralnorm Compiler Mean Durations}
} 

\fontsize{10.0pt}{14.4pt}\selectfont

%
\endapptable

\section{Minimum Runtime Tables}
\subsection{By mod\_map including Rust and Go}
\par\medskip
\apptabletitle{
{\large binarytrees Min Durations}
} 

\fontsize{10.0pt}{14.4pt}\selectfont

%
\endapptable

\par\medskip
\apptabletitle{
{\large fannkuchredux Min Durations}
} 

\fontsize{10.0pt}{14.4pt}\selectfont

%
\endapptable

\par\medskip
\apptabletitle{
{\large fasta Min Durations}
} 

\fontsize{10.0pt}{14.4pt}\selectfont

%
\endapptable

\par\medskip
\apptabletitle{
{\large knucleotide Min Durations}
} 

\fontsize{10.0pt}{14.4pt}\selectfont

%
\endapptable

\par\medskip
\apptabletitle{
{\large mandelbrot Min Durations}
} 

\fontsize{10.0pt}{14.4pt}\selectfont

%
\endapptable

\par\medskip
\apptabletitle{
{\large nbody Min Durations}
} 

\fontsize{10.0pt}{14.4pt}\selectfont

%
\endapptable

\par\medskip
\apptabletitle{
{\large pidigits Min Durations}
} 

\fontsize{10.0pt}{14.4pt}\selectfont

%
\endapptable

\par\medskip
\apptabletitle{
{\large regexredux Min Durations}
} 

\fontsize{10.0pt}{14.4pt}\selectfont

%
\endapptable

\par\medskip
\apptabletitle{
{\large spectralnorm Min Durations}
} 

\fontsize{10.0pt}{14.4pt}\selectfont

%
\endapptable

\subsection{By mod\_map excluding Rust and Go}
\par\medskip
\apptabletitle{
{\large binarytrees Min Durations} \\
{\small Only C programs shown}
}

\fontsize{10.0pt}{14.4pt}\selectfont

%
\endapptable

\par\medskip
\apptabletitle{
{\large fannkuchredux Min Durations} \\
{\small Only C programs shown}
}

\fontsize{10.0pt}{14.4pt}\selectfont

%
\endapptable

\par\medskip
\apptabletitle{
{\large fasta Min Durations} \\
{\small Only C programs shown}
}

\fontsize{10.0pt}{14.4pt}\selectfont

%
\endapptable

\par\medskip
\apptabletitle{
{\large knucleotide Min Durations} \\
{\small Only C programs shown}
}

\fontsize{10.0pt}{14.4pt}\selectfont

%
\endapptable

\par\medskip
\apptabletitle{
{\large mandelbrot Min Durations} \\
{\small Only C programs shown}
}

\fontsize{10.0pt}{14.4pt}\selectfont

%
\endapptable

\par\medskip
\apptabletitle{
{\large nbody Min Durations} \\
{\small Only C programs shown}
}

\fontsize{10.0pt}{14.4pt}\selectfont

%
\endapptable

\par\medskip
\apptabletitle{
{\large pidigits Min Durations} \\
{\small Only C programs shown}
}

\fontsize{10.0pt}{14.4pt}\selectfont

%
\endapptable

\par\medskip
\apptabletitle{
{\large regexredux Min Durations} \\
{\small Only C programs shown}
}

\fontsize{10.0pt}{14.4pt}\selectfont

%
\endapptable

\par\medskip
\apptabletitle{
{\large spectralnorm Min Durations} \\
{\small Only C programs shown}
}

\fontsize{10.0pt}{14.4pt}\selectfont

%
\endapptable

\subsection{\label{sec:compiler-min}By compiler}
\par\medskip
\apptabletitle{
{\large binarytrees Compiler Min Durations}
} 

\fontsize{10.0pt}{14.4pt}\selectfont

%
\endapptable

\par\medskip
\apptabletitle{
{\large fannkuchredux Compiler Min Durations}
} 

\fontsize{10.0pt}{14.4pt}\selectfont

%
\endapptable

\par\medskip
\apptabletitle{
{\large fasta Compiler Min Durations}
} 

\fontsize{10.0pt}{14.4pt}\selectfont

%
\endapptable

\par\medskip
\apptabletitle{
{\large knucleotide Compiler Min Durations}
} 

\fontsize{10.0pt}{14.4pt}\selectfont

%
\endapptable

\par\medskip
\apptabletitle{
{\large mandelbrot Compiler Min Durations}
} 

\fontsize{10.0pt}{14.4pt}\selectfont

%
\endapptable

\par\medskip
\apptabletitle{
{\large nbody Compiler Min Durations}
} 

\fontsize{10.0pt}{14.4pt}\selectfont

%
\endapptable

\par\medskip
\apptabletitle{
{\large pidigits Compiler Min Durations}
} 

\fontsize{10.0pt}{14.4pt}\selectfont

%
\endapptable

\par\medskip
\apptabletitle{
{\large regexredux Compiler Min Durations}
} 

\fontsize{10.0pt}{14.4pt}\selectfont

%
\endapptable

\par\medskip
\apptabletitle{
{\large spectralnorm Compiler Min Durations}
} 

\fontsize{10.0pt}{14.4pt}\selectfont

%
\endapptable


\section{Minimum Runtime Charts}
Scatterplot of minimum runtimes. X-axis: modification combinations mod\_map, Y-axis: runtime in seconds. Point colors represent different compilers and points have different protection combinations, with lower values indicating better performance.

\begin{figure*}
    \centering
    \includegraphics[width=1\textwidth]{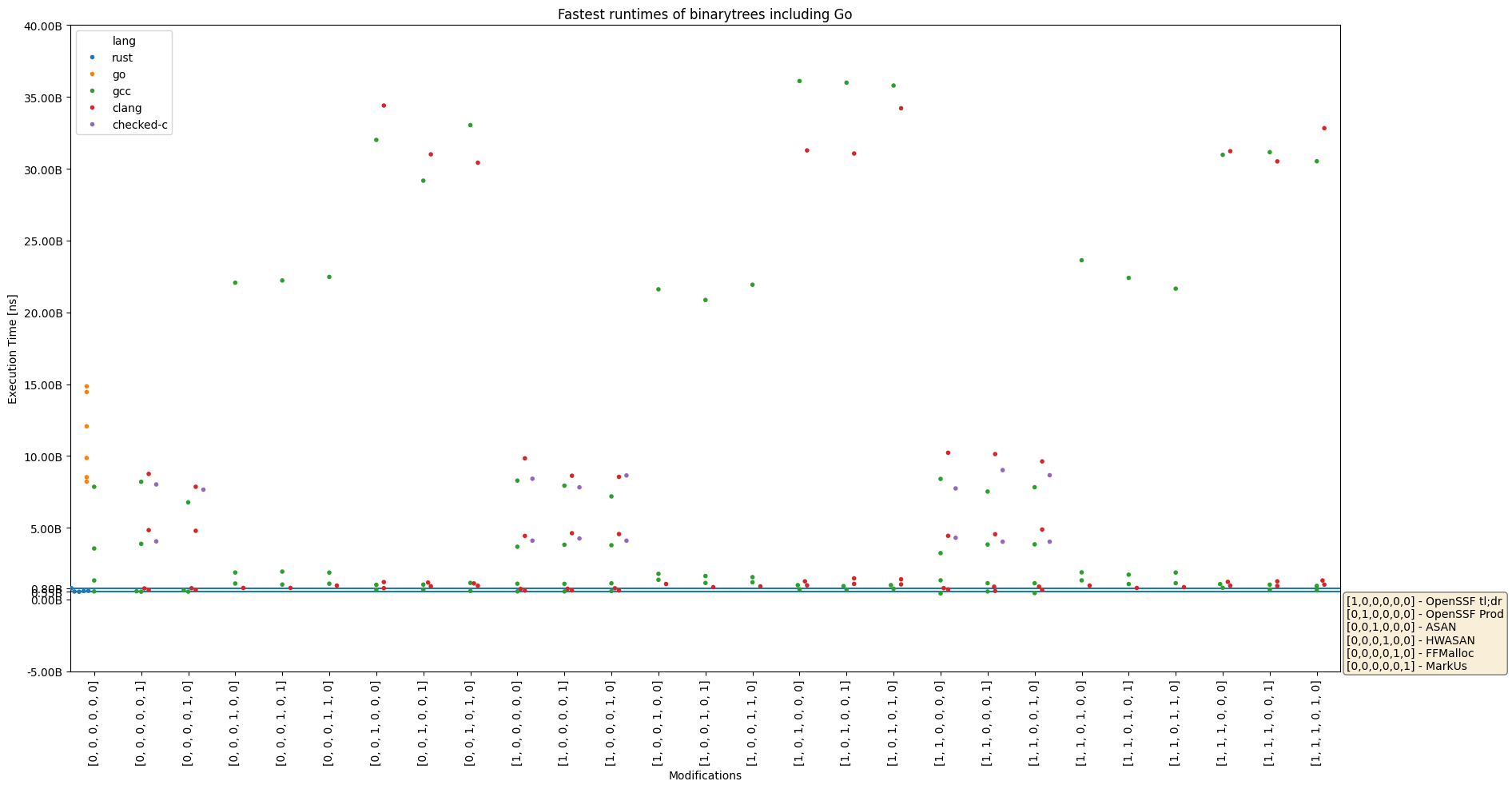}
    \caption{Scatterplot of minimum runtimes of binarytree benchmarks.}
    \label{fig:min_binarytrees}
\end{figure*}

\begin{figure*}
    \centering
    \includegraphics[width=1\textwidth]{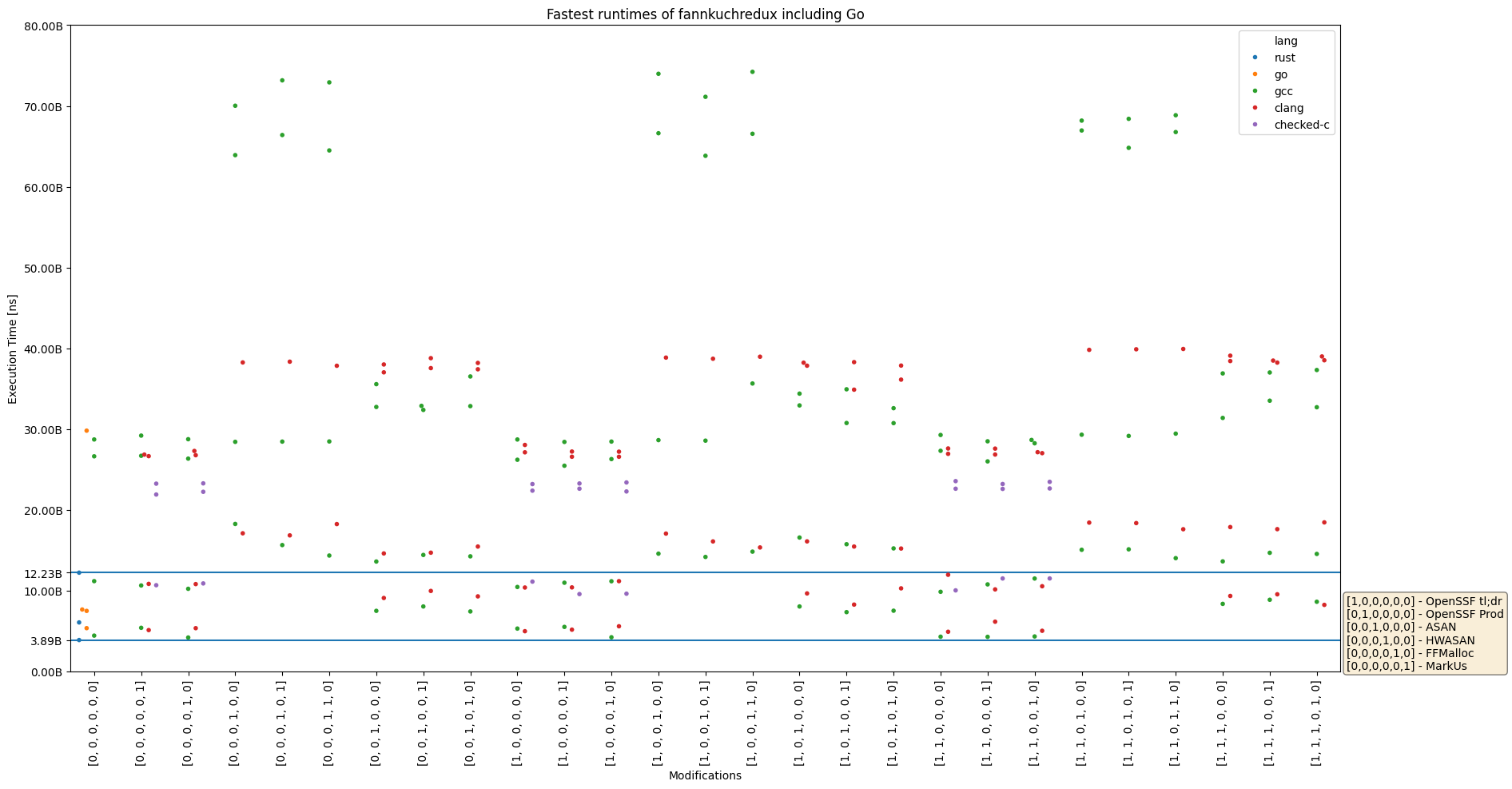}
    \caption{Scatterplot of minimum runtimes of fannkuchredux benchmarks.}
    \label{fig:min_fannkuchredux}
\end{figure*}

\begin{figure*}
    \centering
    \includegraphics[width=1\textwidth]{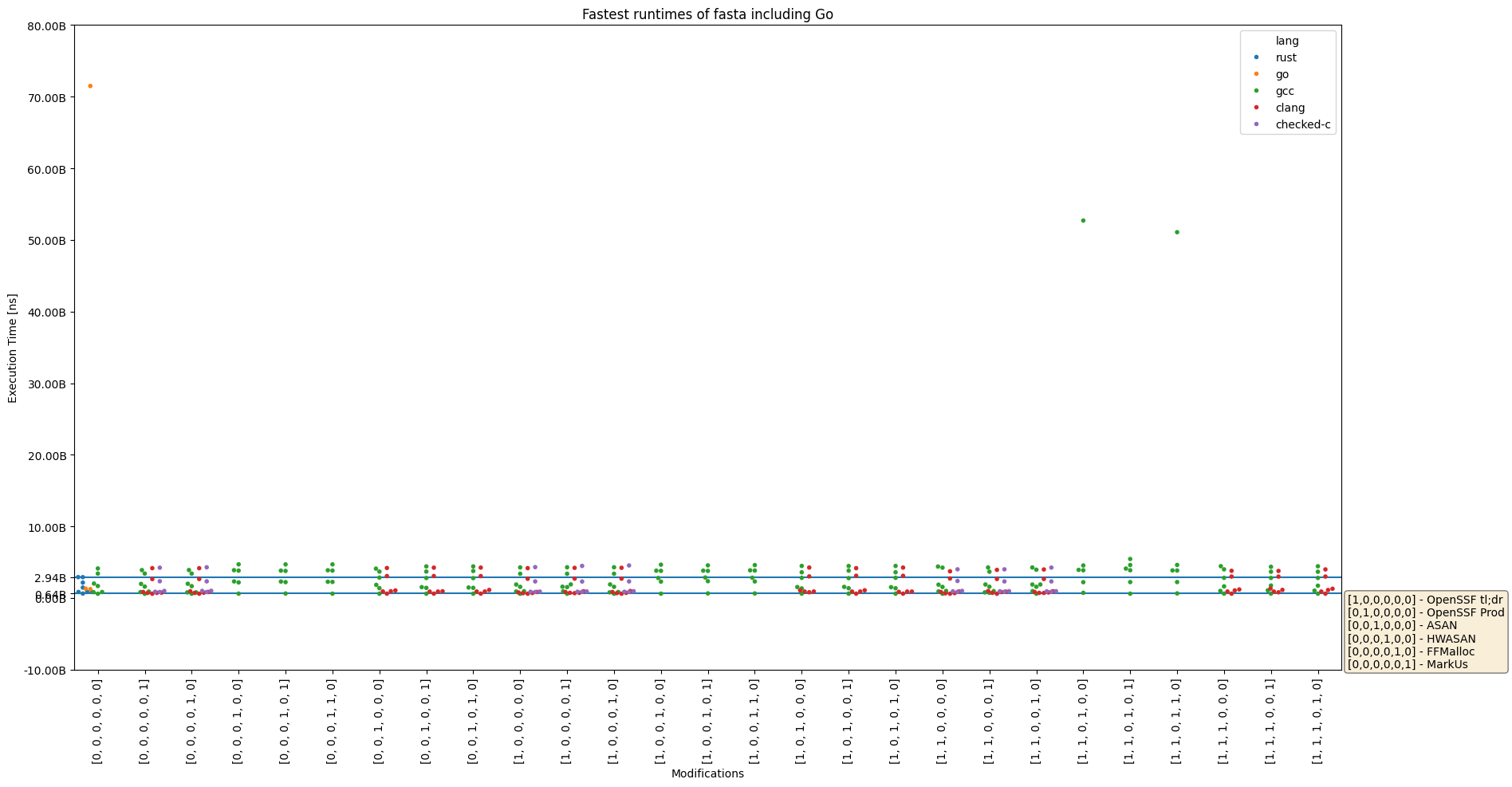}
    \caption{Scatterplot of minimum runtimes of fasta benchmarks.}
    \label{fig:min_fasta}
\end{figure*}

\begin{figure*}
    \centering
    \includegraphics[width=1\textwidth]{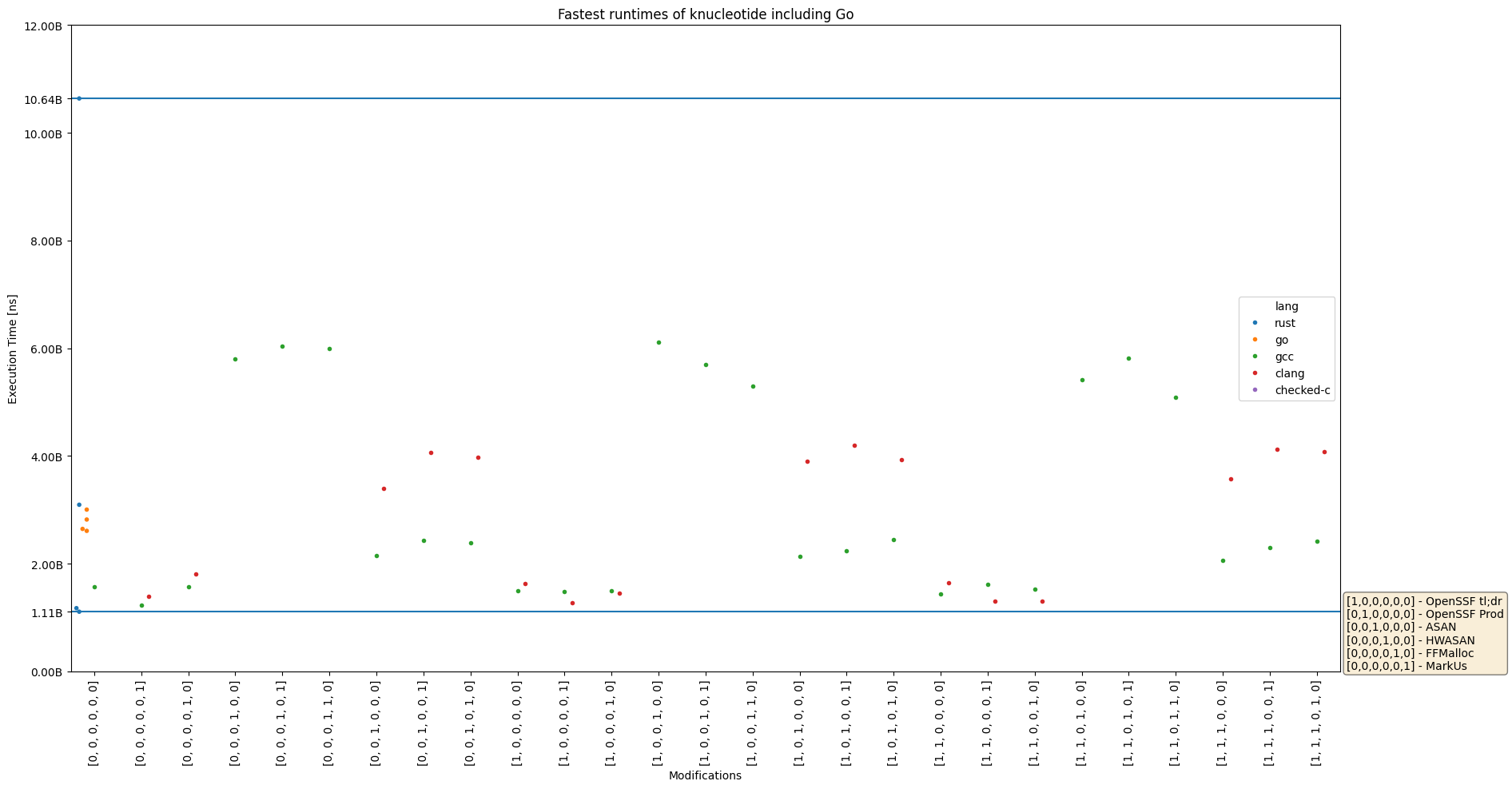}
    \caption{Scatterplot of minimum runtimes of knucleotide benchmarks.}
    \label{fig:min_knucleotide}
\end{figure*}

\begin{figure*}
    \centering
    \includegraphics[width=1\textwidth]{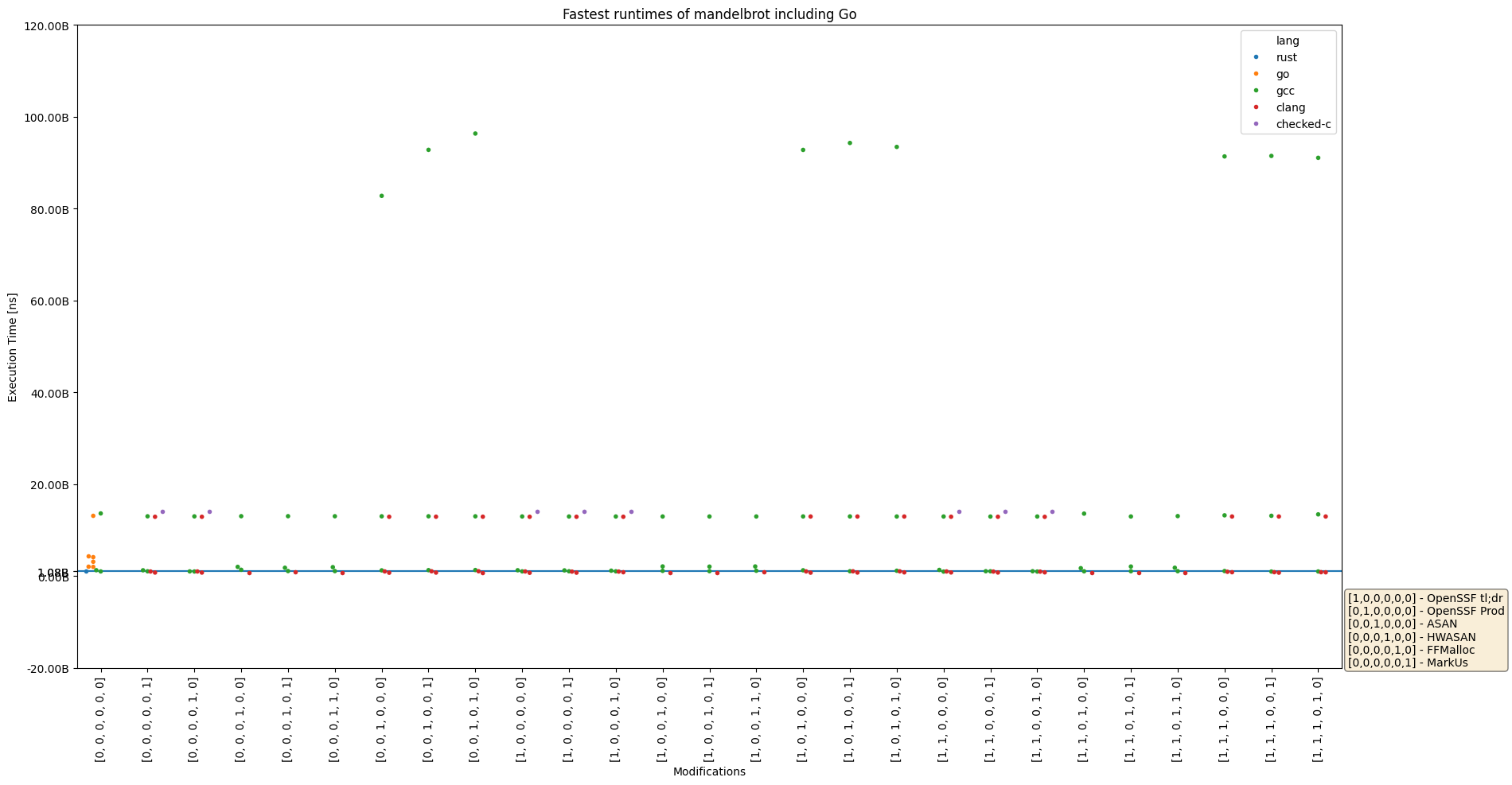}
    \caption{Scatterplot of minimum runtimes of mandelbrot benchmarks.}
    \label{fig:min_mandelbrot}
\end{figure*}

\begin{figure*}
    \centering
    \includegraphics[width=1\textwidth]{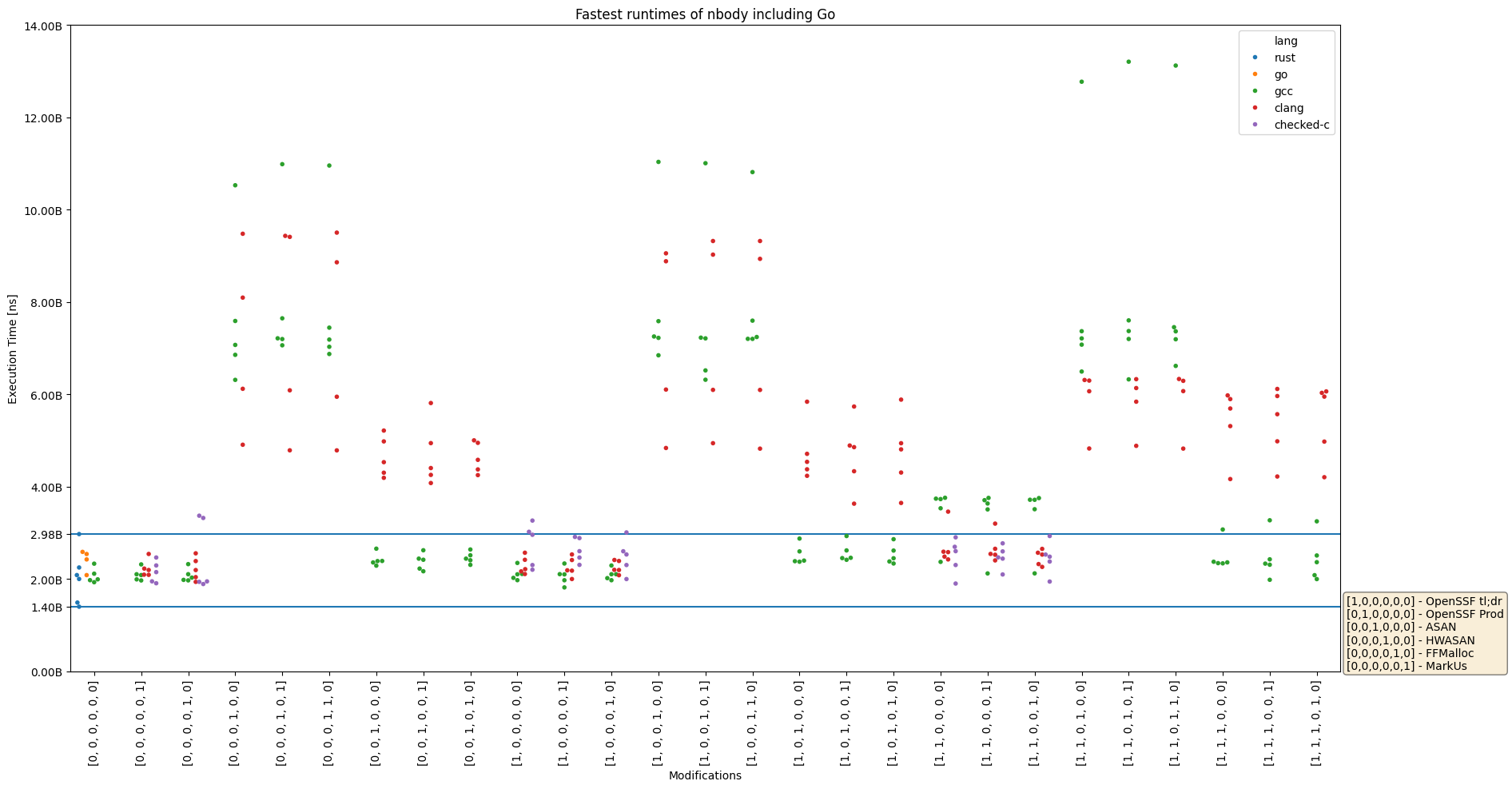}
    \caption{Scatterplot of minimum runtimes of nbody benchmarks.}
    \label{fig:min_nbody}
\end{figure*}

\begin{figure*}
    \centering
    \includegraphics[width=1\textwidth]{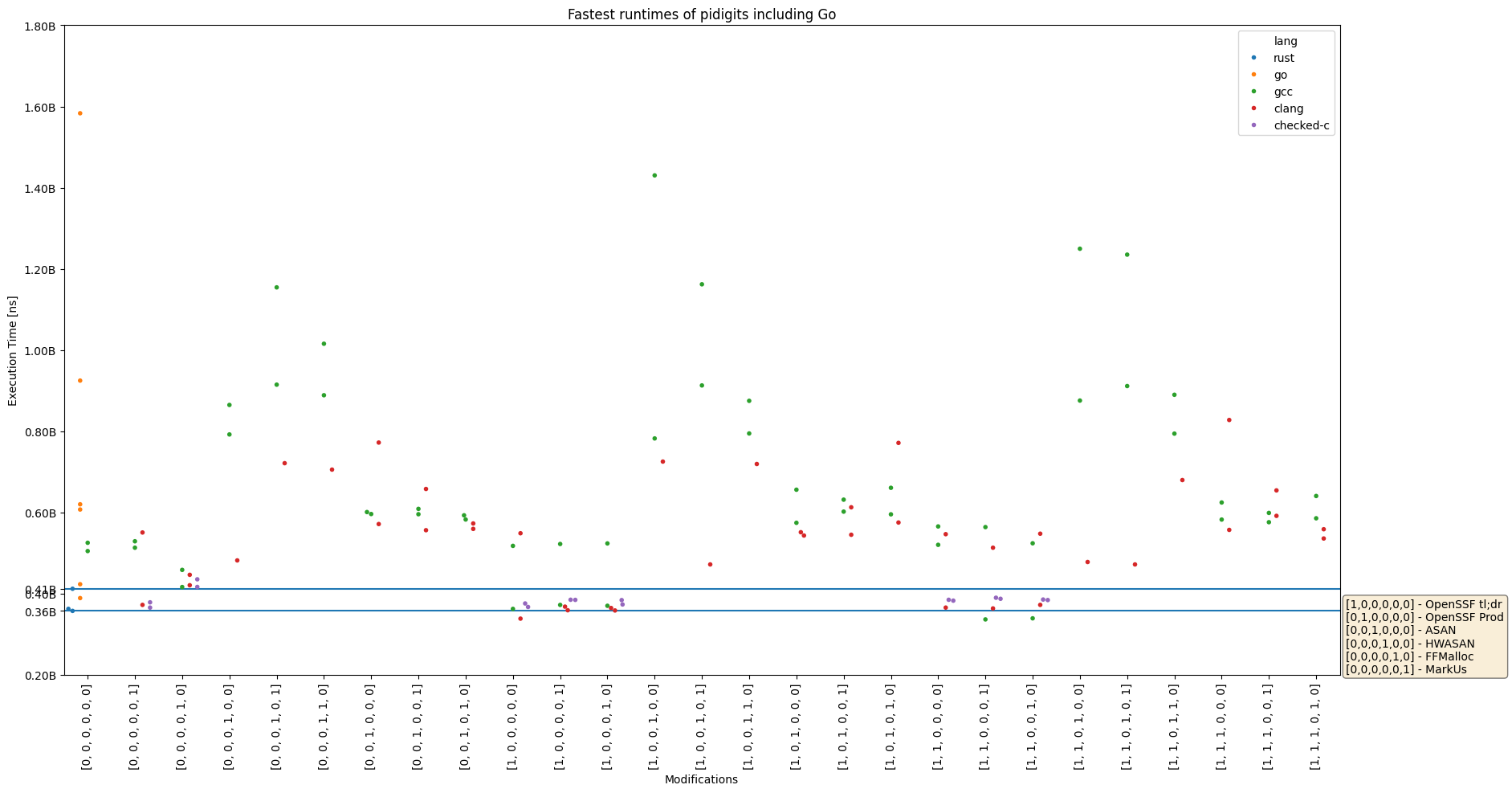}
    \caption{Scatterplot of minimum runtimes of pidigits benchmarks.}
    \label{fig:min_pidigits}
\end{figure*}

\begin{figure*}
    \centering
    \includegraphics[width=1\textwidth]{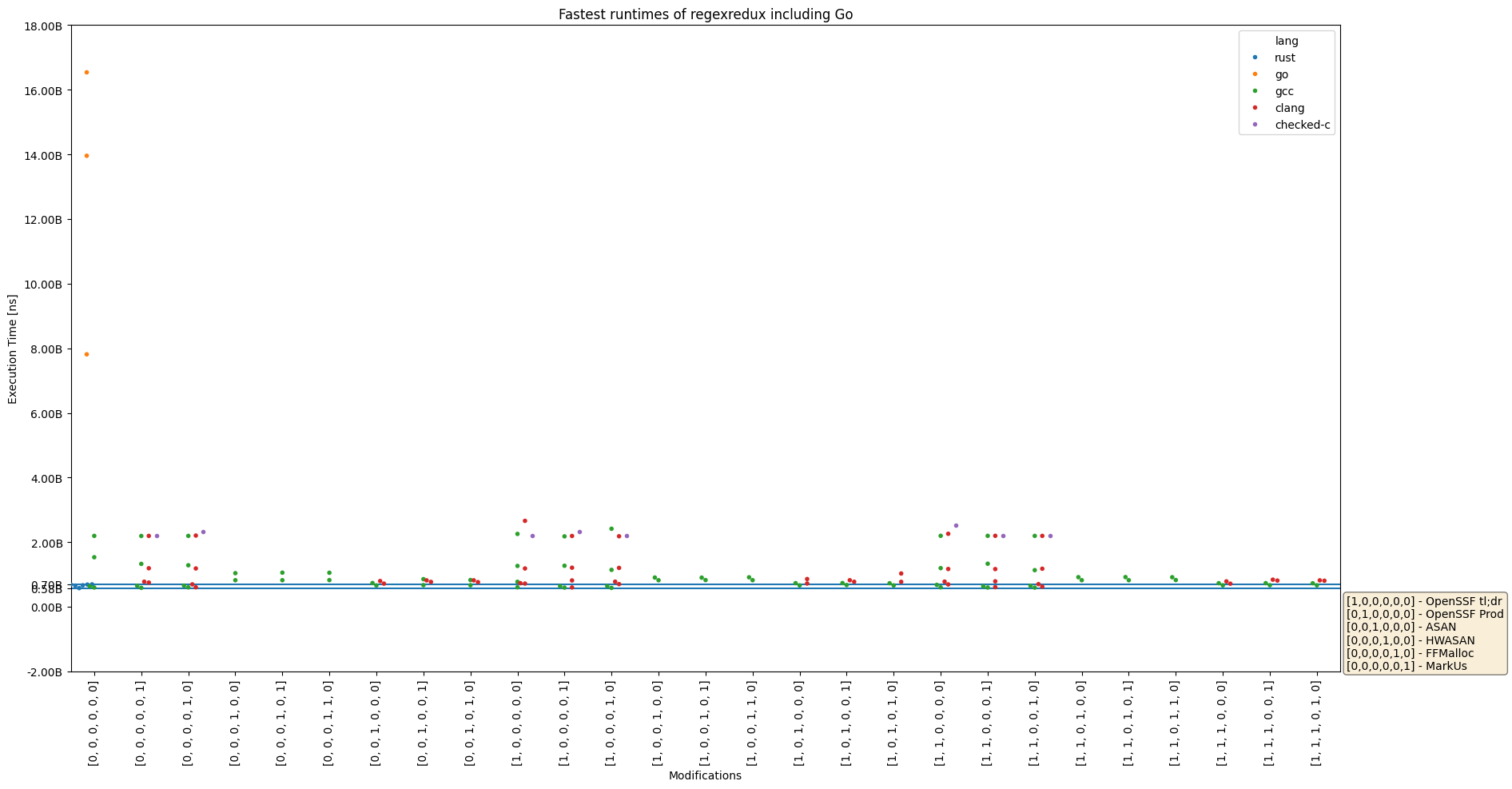}
    \caption{Scatterplot of minimum runtimes of regexredux benchmarks.}
    \label{fig:min_regexredux}
\end{figure*}

\begin{figure*}
    \centering
    \includegraphics[width=1\textwidth]{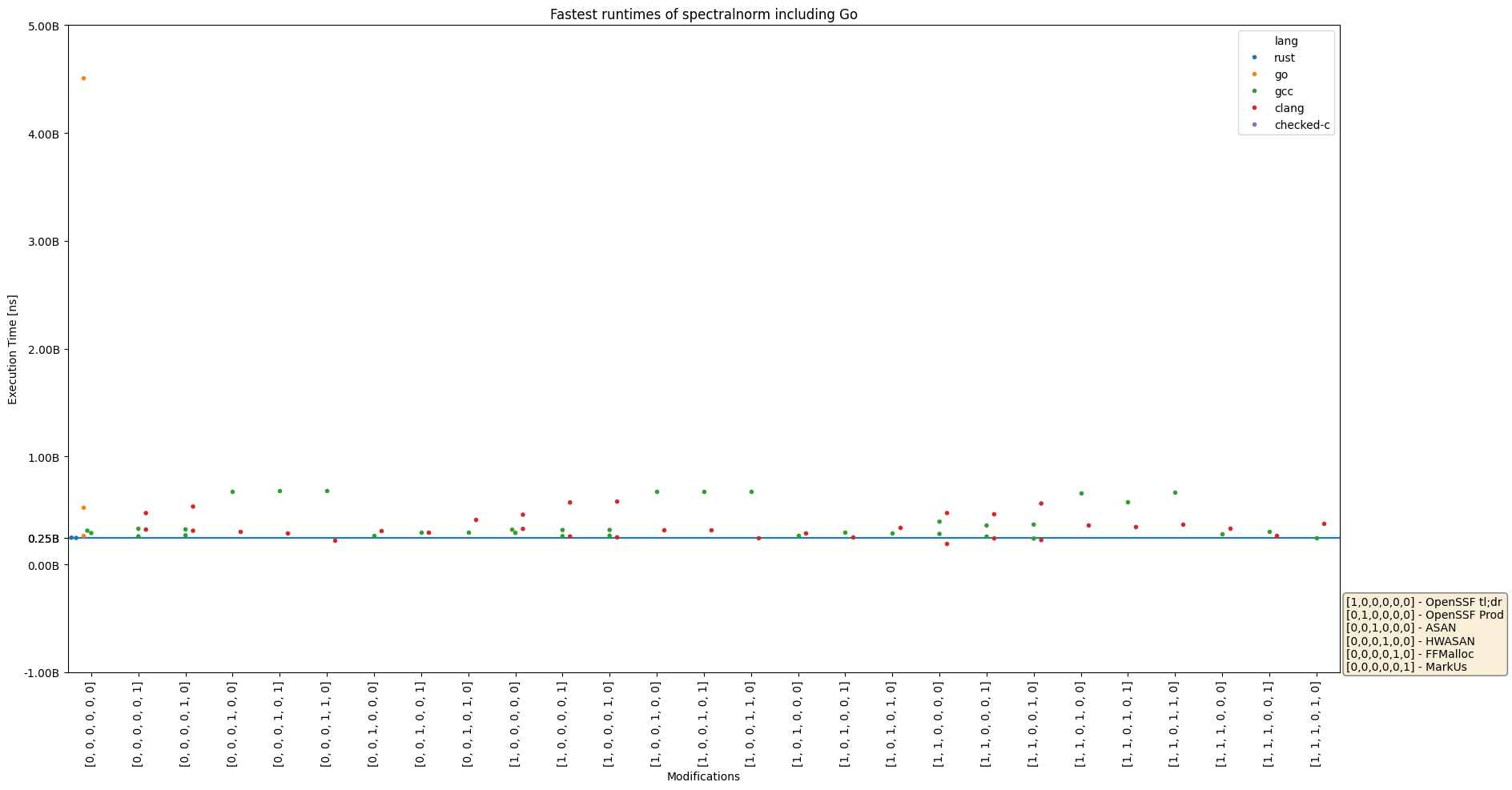}
    \caption{Scatterplot of minimum runtimes of spectralnorm benchmarks.}
    \label{fig:min_spectralnorm}
\end{figure*}

\section{Error Counts}
Error counts compared to total number of benchmarks.

\begin{figure*}
    \centering
    \includegraphics[width=1\textwidth]{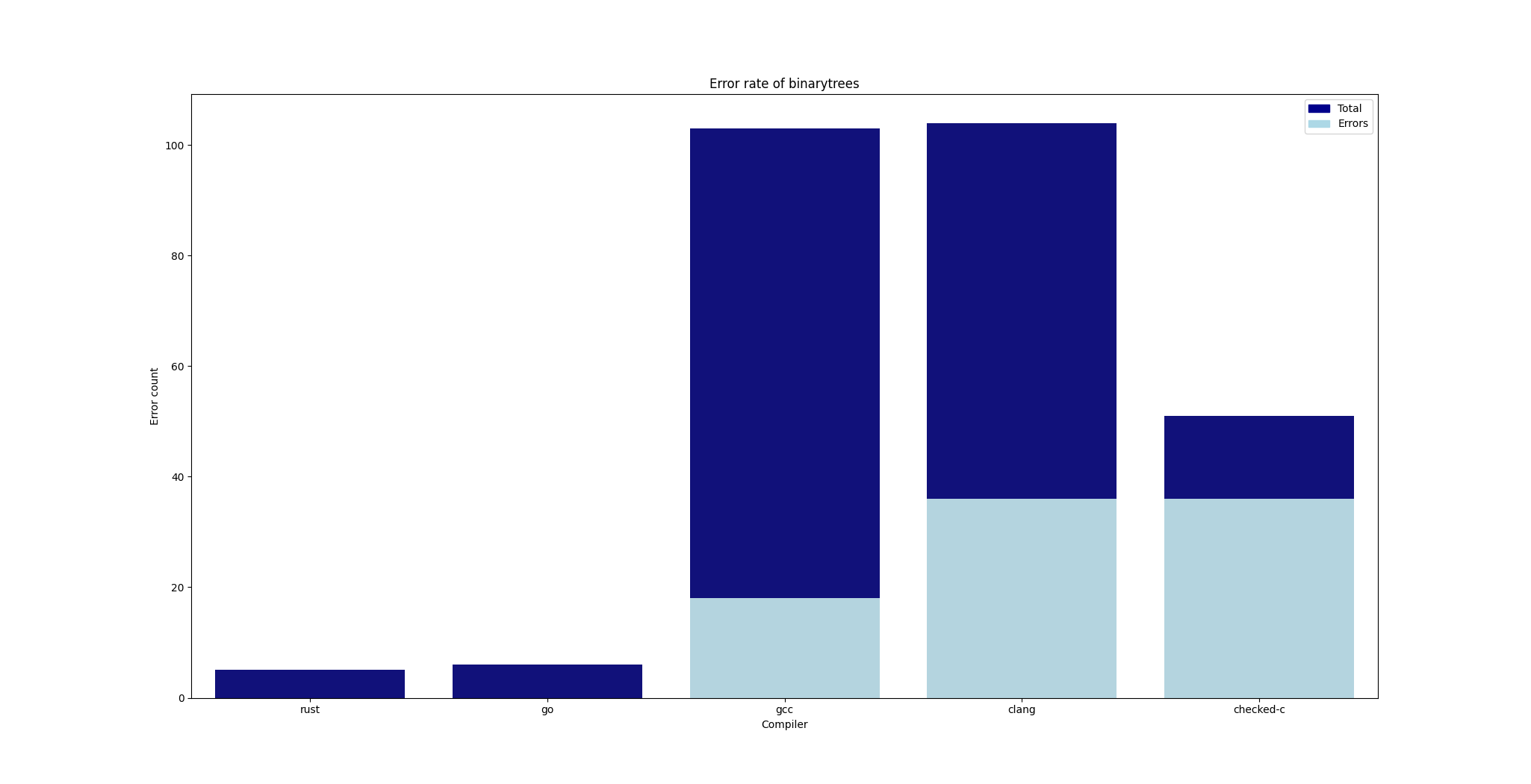}
    \caption{Error count by compiler of binarytree benchmarks.}
    \label{fig:error_count_binarytrees}
\end{figure*}

\begin{figure}
    \centering
    \includegraphics[width=1\textwidth]{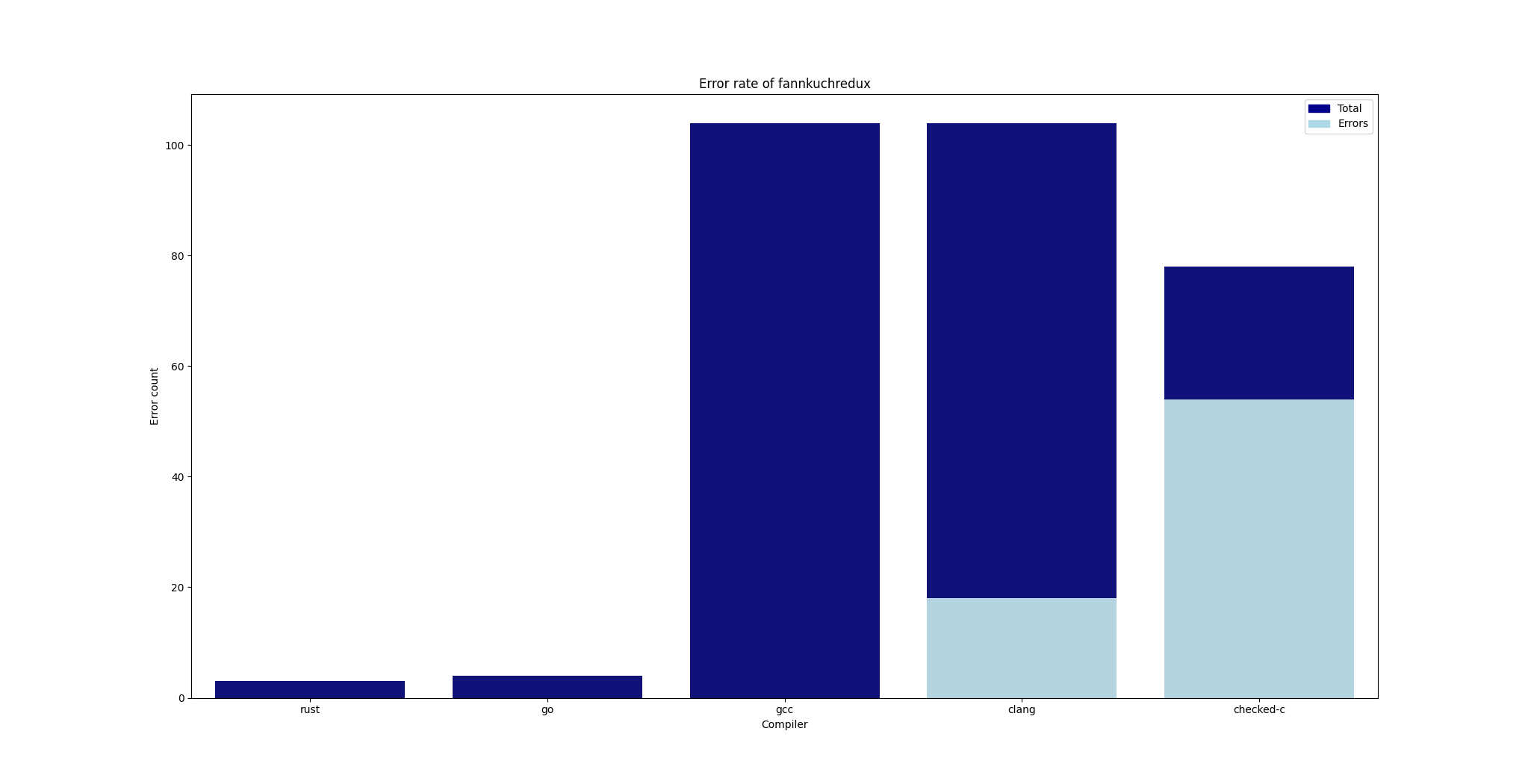}
    \caption{Error count by compiler of fannkuchredux benchmarks.}
    \label{fig:error_count_fannkuchredux}
\end{figure}

\begin{figure*}
    \centering
    \includegraphics[width=1\textwidth]{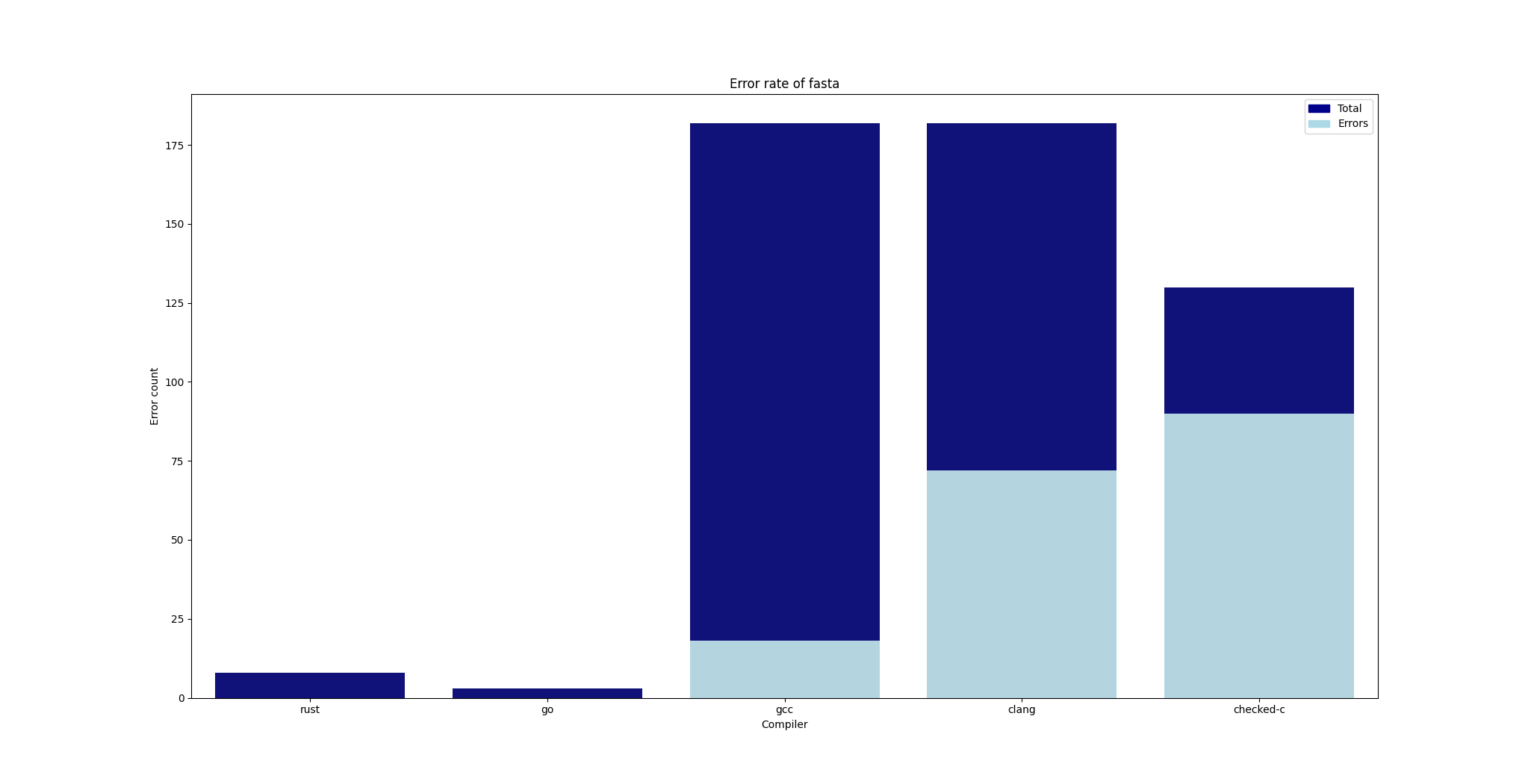}
    \caption{Error count by compiler of fasta benchmarks.}
    \label{fig:error_count_fasta}
\end{figure*}

\begin{figure*}
    \centering
    \includegraphics[width=1\textwidth]{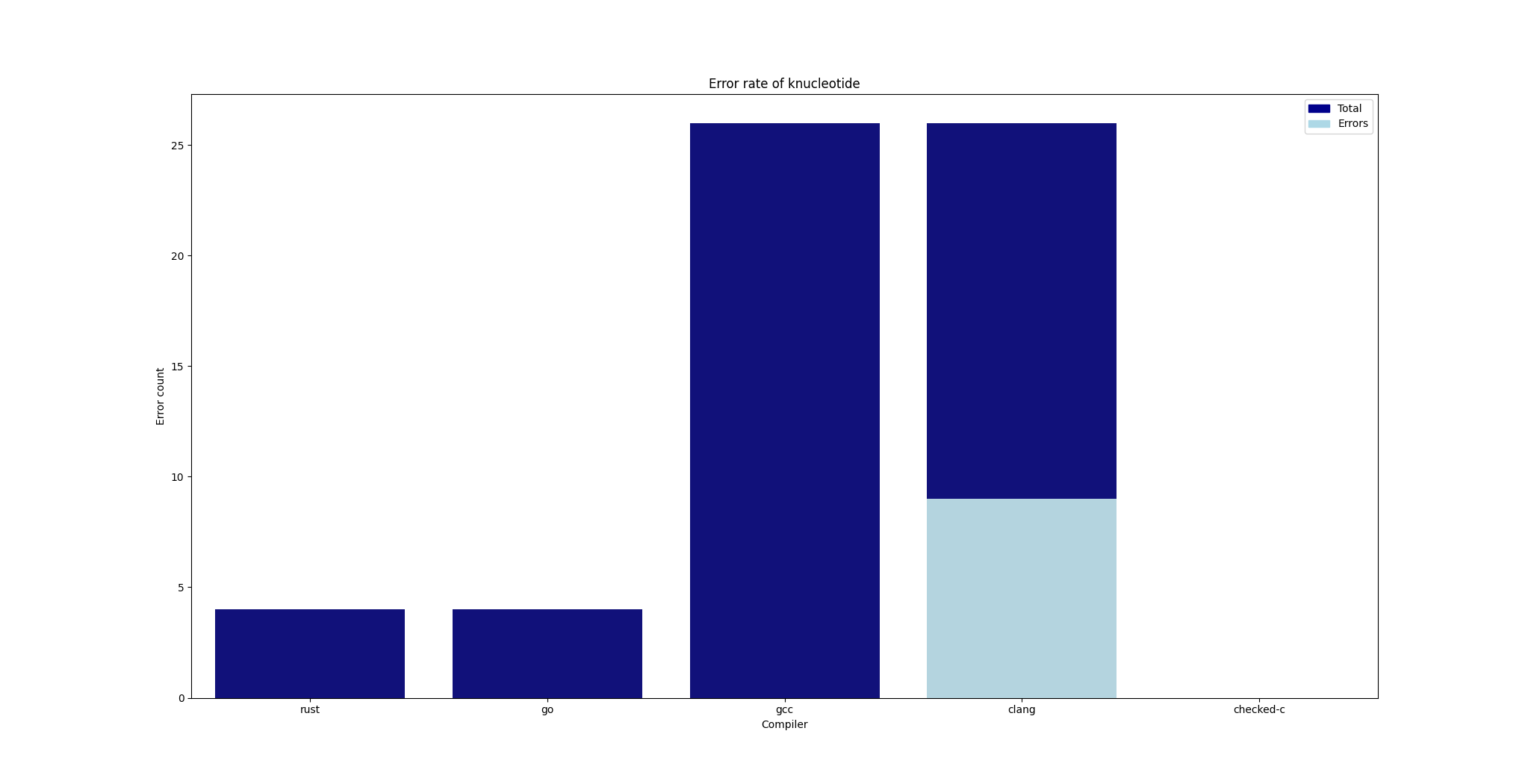}
    \caption{Error count by compiler of knucleotide benchmarks.}
    \label{fig:error_count_knucleotide}
\end{figure*}

\begin{figure*}
    \centering
    \includegraphics[width=1\textwidth]{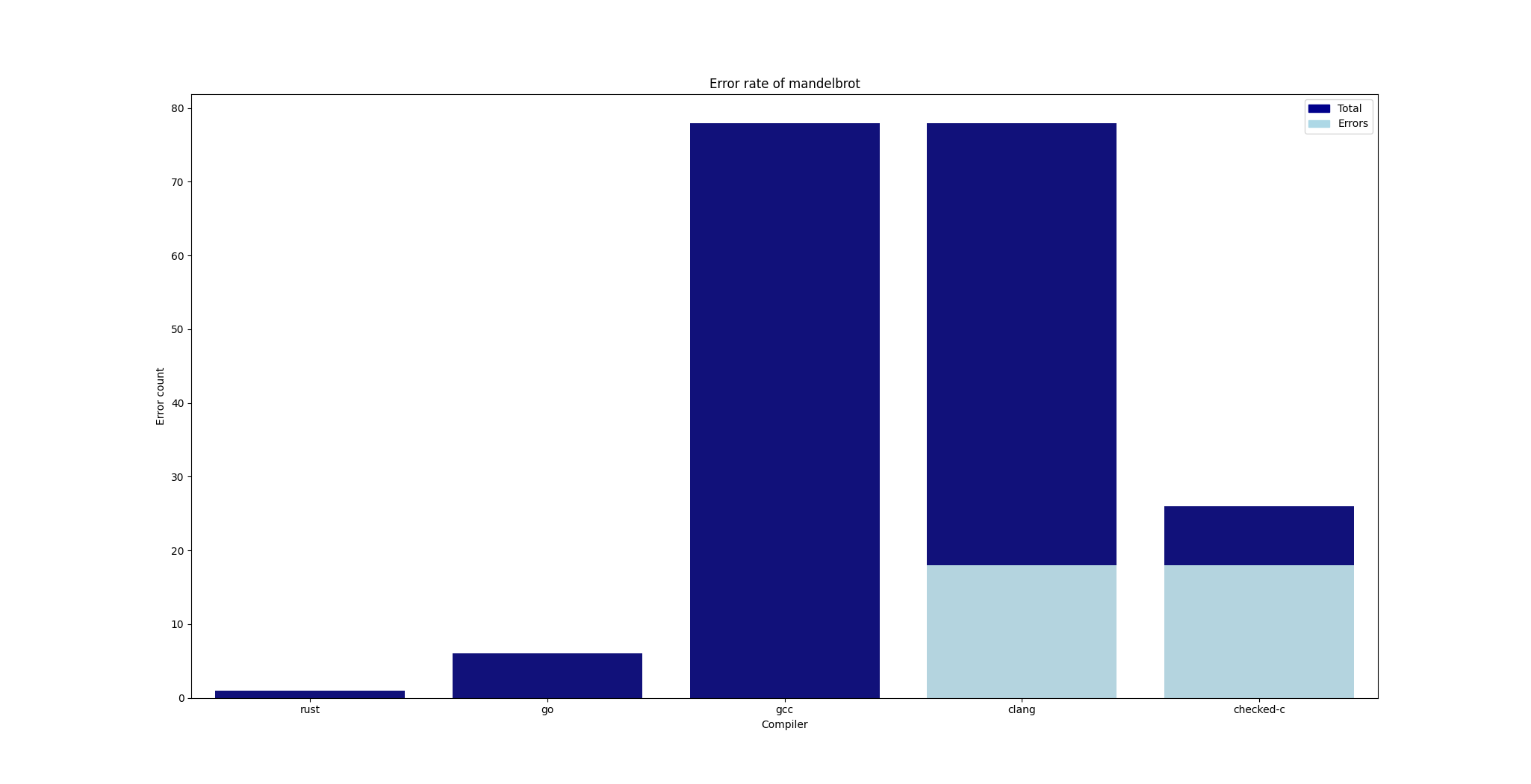}
    \caption{Error count by compiler of mandelbrot benchmarks.}
    \label{fig:error_count_mandelbrot}
\end{figure*}

\begin{figure*}
    \centering
    \includegraphics[width=1\textwidth]{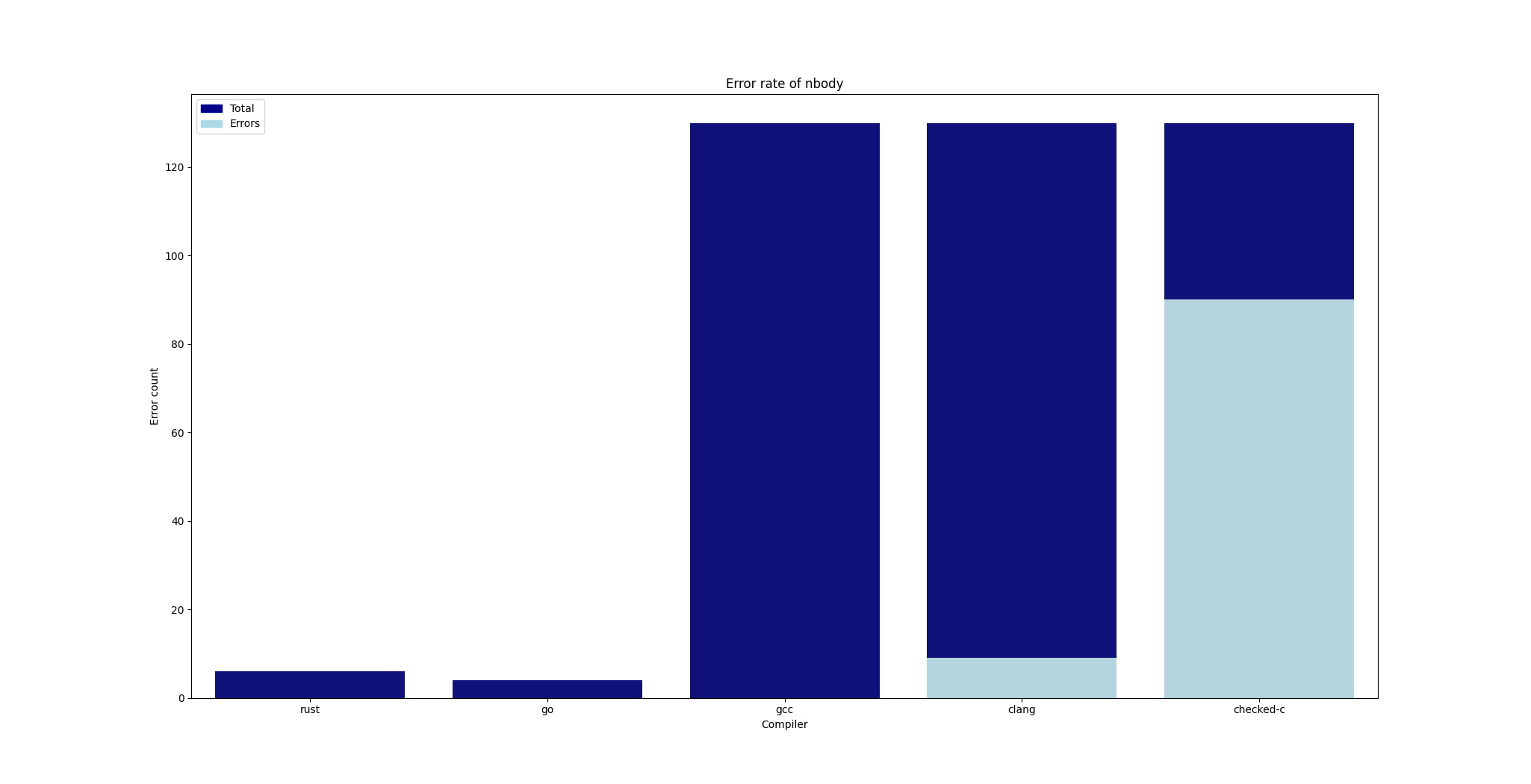}
    \caption{Error count by compiler of nbody benchmarks.}
    \label{fig:error_count_nbody}
\end{figure*}

\begin{figure*}
    \centering
    \includegraphics[width=1\textwidth]{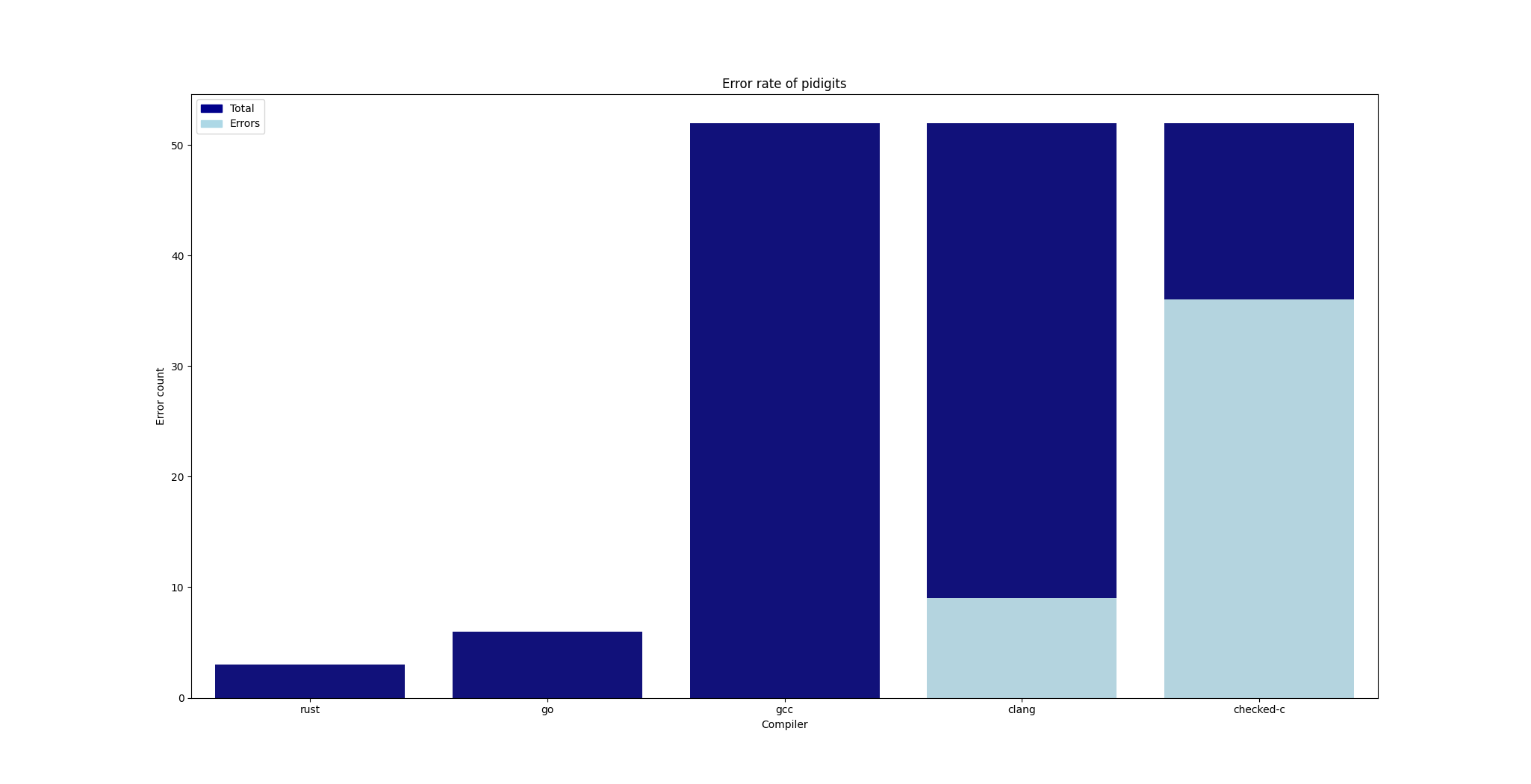}
    \caption{Error count by compiler of pidigits benchmarks.}
    \label{fig:error_count_pidigits}
\end{figure*}

\begin{figure*}
    \centering
    \includegraphics[width=1\textwidth]{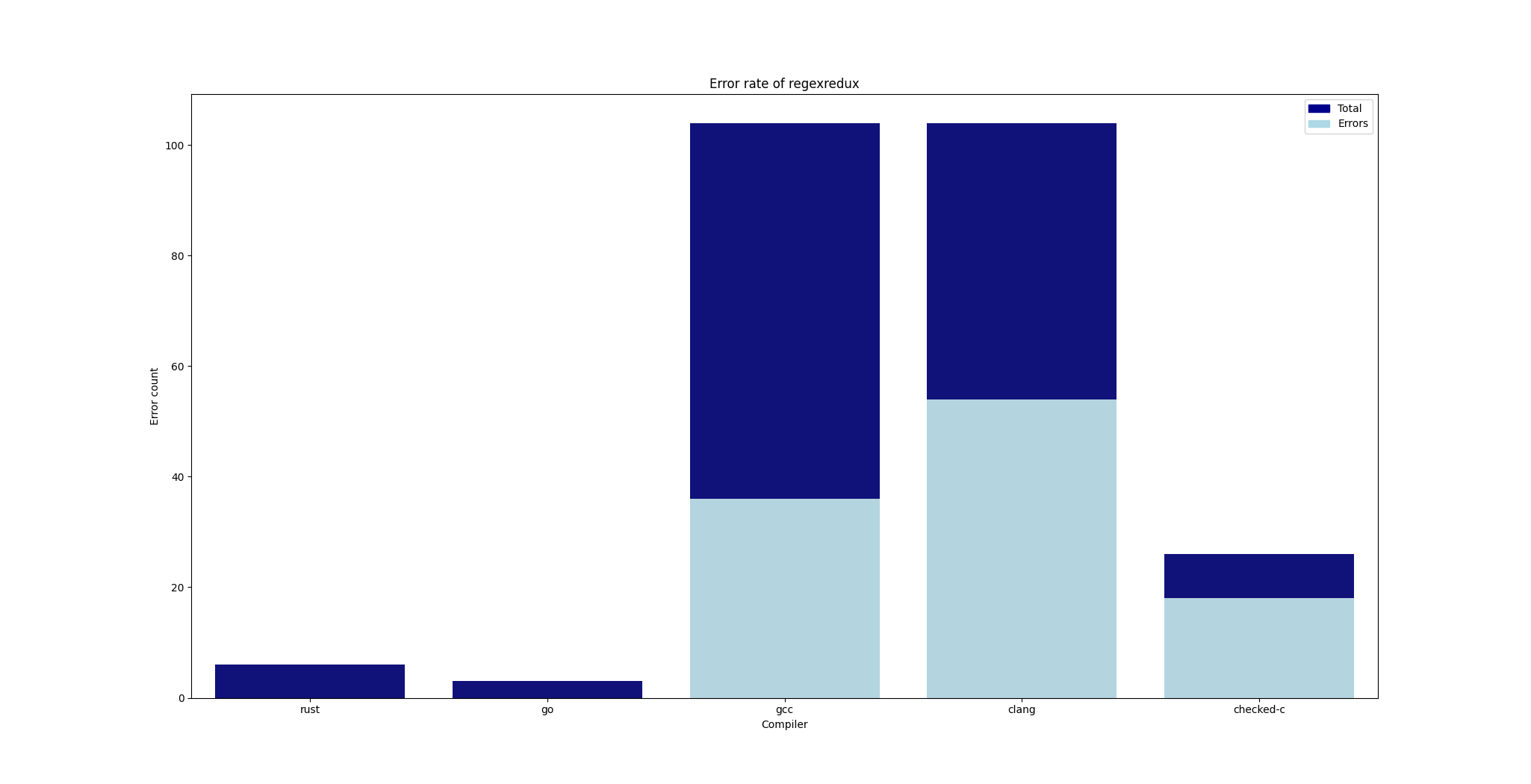}
    \caption{Error count by compiler of regexredux benchmarks.}
    \label{fig:error_count_regexredux}
\end{figure*}

\begin{figure*}[ht]
    \centering
    \includegraphics[width=1\textwidth]{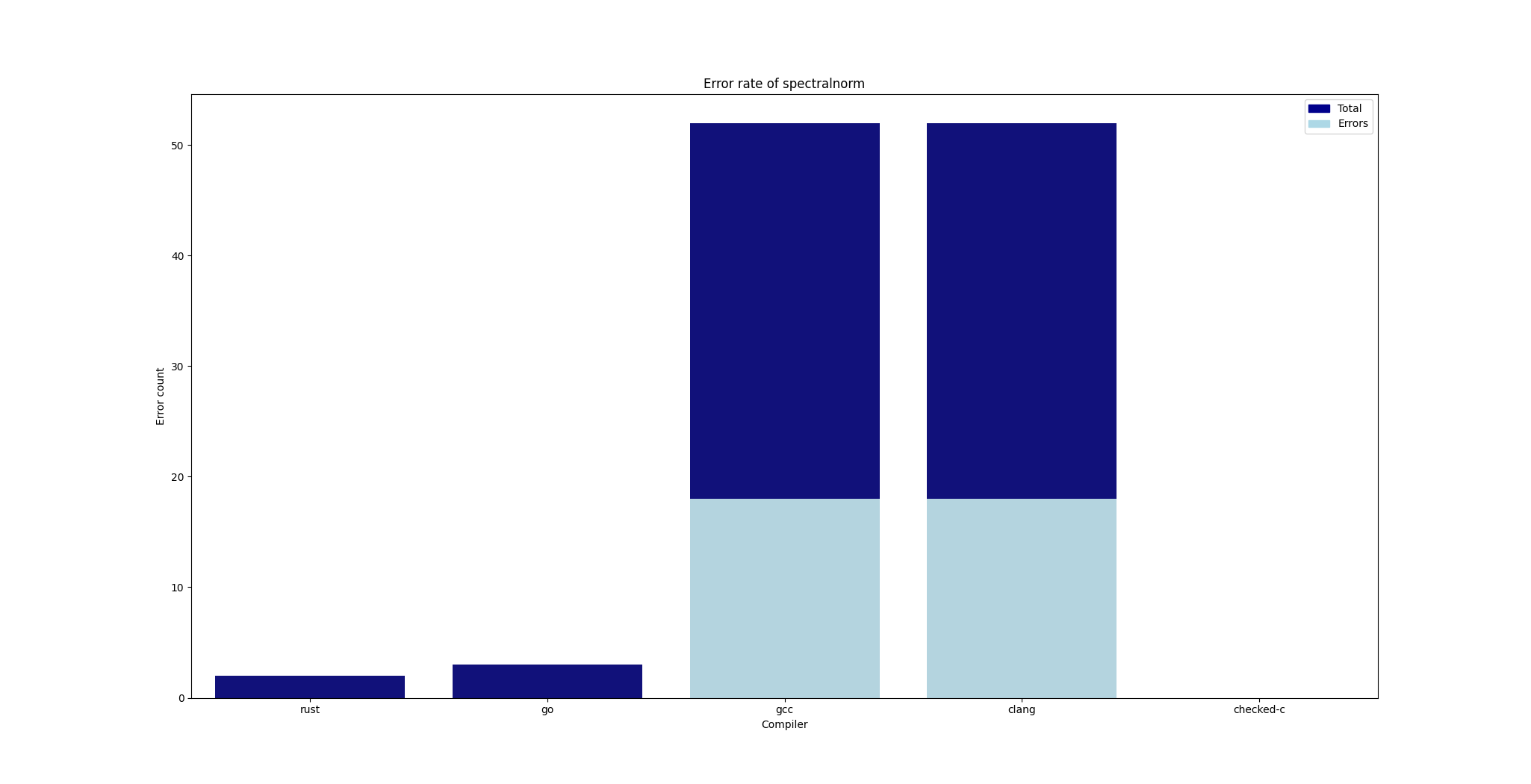}
    \caption{Error count by compiler of spectralnorm benchmarks.}
    \label{fig:error_count_spectralnorm}
\end{figure*}

\end{document}